\documentclass[11pt]{article}
\usepackage[a4paper]{geometry}
\usepackage{jcapmod}
\usepackage{amssymb}
\usepackage{bm}
\usepackage{latexsym}
\usepackage{amssymb}
\usepackage{amsmath}
\usepackage{amsfonts}
\usepackage{colortbl}
\usepackage{multirow}
\usepackage{array}
\usepackage{booktabs}
\usepackage{rotating}
\usepackage{graphicx}

\def\clap#1{\hbox to 0pt{\hss#1\hss}}

\def\mathclap{\mathpalette\mathclapinternal}

\def\mathclapinternal#1#2{\clap{$\mathsurround=0pt#1{#2}$}}

\renewcommand{\geq}{\geqslant}
\renewcommand{\leq}{\leqslant}

\newcommand{\vect}[1]{\bm{\mathrm{{#1}}}}
\newcommand{\e}[1]{\mathrm{e}^{{#1}}}
\newcommand{\im}{\mathrm{i}}
\newcommand{\EulerGamma}{\gamma_{\mathrm{E}}}

\newcommand{\ket}[1]{|{#1}\rangle}
\newcommand{\bra}[1]{\langle{#1}|}
\newcommand{\braket}[2]{\langle{{#1}}|{{#2}}\rangle}

\newcommand{\Mp}{M_{\mathrm{P}}}

\newcommand{\etal}{et al.}

\newcommand{\zterm}{z}

\newcommand{\deltamin}{\delta_{\mathrm{min}}}
\newcommand{\khard}{k_{\mathrm{UV}}}
\newcommand{\ksoft}{k_{\mathrm{IR}}}

\newcommand{\alphalambda}{\alpha}

\newcommand{\Ps}{\mathcal{P}}
\newcommand{\Pstensor}{\Ps_g}
\newcommand{\fNL}{f_{\mathrm{NL}}}

\newcommand{\nfNL}{n_{\fNL}}

\newcommand{\nt}{n_t}
\newcommand{\ntstar}{n_{t\star}}

\newcommand{\SigmaP}{\Sigma}
\newcommand{\lambdaP}{\lambda}
\newcommand{\cs}{c_s}
\newcommand{\csstar}{c_{s\star}}
\newcommand{\csdot}{\dot{c}_s}
\newcommand{\slowroll}{\mathrm{v}}
\newcommand{\epsilonSR}{\varepsilon_{\slowroll}}
\newcommand{\etaSR}{\eta_{\slowroll}}
\newcommand{\epsilonSRstar}{\varepsilon_{\slowroll\star}}
\newcommand{\etaSRstar}{\eta_{\slowroll\star}}

\newcommand{\timeorder}{\mathrm{T}}
\newcommand{\instate}{\mathrm{in}}
\newcommand{\op}{\mathcal{O}}

\newcommand{\grad}{\partial}

\DeclareMathOperator{\Or}{O}
\DeclareMathOperator{\Ei}{Ei}

\DeclareMathOperator{\sgn}{sgn}
\DeclareMathOperator{\RePart}{Re}

\renewcommand{\Re}{\RePart}

\newcommand{\complex}{\mathbb{C}}
\newcommand{\reals}{\mathbb{R}}

\DeclareMathOperator{\tr}{tr}

\renewcommand{\d}{\mathrm{d}}
\renewcommand{\vec}[1]{\bm{\mathrm{{#1}}}}

\newcommand{\para}[1]{\par\vspace{2mm}\noindent\emph{{#1}}.---}
\newcommand{\parafootnote}[2]{\par\vspace{2mm}\noindent\emph{{#1}}.\footnote{{#2}}---}

\newcommand{\tmark}[1]{\ensuremath{^{{#1}}}}

\newcolumntype{Q}{>{$\displaystyle}l<{$}}
\newcolumntype{q}{>{\columncolor[gray]{0.9}$\displaystyle}l<{$}}
\newcolumntype{R}{>{$\displaystyle}r<{$}}
\newcolumntype{S}{>{$\displaystyle}c<{$}}
\newcolumntype{s}{>{\columncolor[gray]{0.9}$\displaystyle}c<{$}}
\newcolumntype{T}{>{\columncolor[gray]{0.9}}c<{}}

\newsavebox{\tableA}
\newsavebox{\tableB}
\newlength{\tblw}

\newsavebox{\boxplot}
\newsavebox{\boxplota}
\newlength{\plotw}
\newlength{\plotwa}

\begin{document}

	\title{Large slow-roll corrections to the bispectrum of 
	noncanonical inflation}

	\author[a]{Clare Burrage,}
	\author[b]{Raquel H. Ribeiro,}
	\author[c]{David Seery}

	\affiliation[a]{D\'{e}partment de Physique Th\'{e}orique,
	Universit\'{e} de Gen\`{e}ve, \\
	24 Quai E. Ansermet, CH-1211, Gen\`{e}ve, Switzerland} 
		
	\affiliation[b]{Department of Applied Mathematics and Theoretical Physics\\
	Centre for Mathematical Sciences, Wilberforce Road \\
	Cambridge CB3 0WA, United Kingdom} 

	\affiliation[c]{Astronomy Centre, University of Sussex \\
	Falmer, Brighton BN1 9QH, United Kingdom}

	\emailAdd{Clare.Burrage@unige.ch}
	\emailAdd{R.Ribeiro@damtp.cam.ac.uk}
	\emailAdd{D.Seery@sussex.ac.uk}

	\abstract{
	Nongaussian statistics are a powerful discriminant
	between inflationary models,
	particularly those with noncanonical
	kinetic terms. Focusing on theories where the Lagrangian is an
	arbitrary Lorentz-invariant function of a scalar field and its
	first derivatives, we review and extend the calculation of the
	observable three-point function.
	We compute the ``next-order''
	slow-roll corrections to the bispectrum in closed
	form, and obtain quantitative estimates of their magnitude
	in DBI and power-law $k$-inflation.
	In the DBI case our results enable us to estimate corrections
	from the shape of the potential
	and the warp factor: these
	can be of order several tens of percent.
	We track the possible sources of large logarithms
	which can spoil ordinary perturbation theory,
	and use them to obtain a general formula for the scale dependence of the
	bispectrum.
	Our result satisfies the next-order
	version of Maldacena's consistency condition
	and an equivalent consistency condition for the scale dependence.
	We identify a new bispectrum shape available
	at next-order, which is similar to a shape encountered
	in Galileon models.
	If $\fNL$ is sufficiently large this
	shape may be independently
	detectable.}
	

	\maketitle

	\section{Introduction}
	\label{sec:introduction}
	
	In the last decade,
	our picture of the early universe has become numerically
	precise---%
	a quantitative revolution made possible by
	analysis of the cosmic microwave
	background (CMB) radiation.
	The CMB is usually interpreted as
	a relic of an earlier hot, dense, primordial era \cite{Komatsu:2009kd}.
	During this era the universe is believed to have been filled with
	an almost-smooth plasma, in which tiny perturbations were seeded
	by an unknown mechanism.
	Eventually, photons decoupled from the cooling plasma
	to form the CMB, while
	matter particles
	collapsed under gravity to generate bound
	structures.
	
	The origin of these perturbations is a matter of dispute.
	A possible candidate is the inflationary scenario, in which
	an era of quasi-de Sitter expansion smoothed the universe
	while allowing quantum fluctuations to grow and become classical
	on superhorizon scales.
	In its simplest implementation, inflation forecasts
	an approximately scale-invariant, Gaussian distribution of
	perturbations \cite{Bardeen:1983qw, Wands:2000dp},
	characterized entirely by their two-point correlations.
	Although falsifiable, these predictions are in agreement with
	present-day observations
	\cite{Jarosik:2010iu, Larson:2010gs, Komatsu:2010fb}.
	
	The increasing sophistication of CMB experiments
	has suggested that nontrivial three- and higher
	$n$-point correlations could be detected \cite{Verde:1999ij,Komatsu:2001rj},
	and
	in the medium-term future it is possible that competitive
	constraints
	will emerge from the nongaussian statistics of collapsed
	structures \cite{Bernardeau:2001qr}.
	In principle, valuable information is encoded in each
	$n$-point function. In practice, extracting information from
	the four-point function is already computationally challenging
	\cite{Smidt:2010ra,Fergusson:2010gn},
	and it is unclear whether useful constraints can be obtained
	for the five- and higher $n$-point functions. For this reason attention
	has focused on the amplitude of three-point correlations, $\fNL$
	\cite{Verde:1999ij,Komatsu:2001rj}.
	For a Gaussian field these correlations are absent and $\fNL = 0$.

	In simple models
	both the de Sitter expansion and quantum fluctuations
	originate from a single scalar field, $\phi$,
	although more complicated possibilities exist.%
		\footnote{For a review of the various generation mechanisms
		which have been proposed to date, see Lyth \& Liddle
		\cite{Lyth:2009zz}.}
	But whatever combination of ingredients we choose,
	the perturbations inherit their
	statistical properties from a
	microphysical Lagrangian
	by some mapping of the correlation functions of
	a field such as $\phi$
	\cite{Lyth:2005fi,Starobinsky:1986fxa,Sasaki:1995aw}.
	The two-point function
	for a scalar field with canonical kinetic term
	was calculated during the early days of the
	inflationary paradigm
	\cite{Bardeen:1983qw,Guth:1982ec,Hawking:1982cz,Hawking:1982my,
	Lyth:1984gv,
	Mukhanov:1985rz,Sasaki:1986hm},
	followed by partial results for the three-point function
	\cite{Falk:1992sf,Gangui:1993tt,Pyne:1995bs,Wang:1999vf}.
	The complete three-point function was eventually computed
	by Maldacena \cite{Maldacena:2002vr},
	and later extended to multiple fields
	\cite{Seery:2005gb} and four-point correlations
	\cite{Seery:2006vu,Seery:2008ax}.
	
	Maldacena's calculation showed that $\fNL$ would be unobservably
	small in the simplest model of inflation---of order the
	tensor-to-scalar ratio, $r$, which is constrained by observation to
	satisfy $r \lesssim 0.2$
	\cite{Komatsu:2010fb}.
	However, it was clear that
	information about the Lagrangian operators
	responsible for generating fluctuations in more exotic theories
	could be encoded in the three- and higher $n$-point functions.
	Which operators could be relevant?
	Creminelli argued that if the dominant kinetic operator
	for the background field was the
	canonical term
	$(\nabla \phi)^2$, where $\nabla$ is the spacetime
	covariant derivative, then
	the three-point correlations of its perturbations
	were effectively indistinguishable
	from Maldacena's simplest model
	\cite{Creminelli:2003iq}.
	However, in theories where other operators
	make significant contributions, the amplitude of three-point correlations
	can be radically different.
	This gave valuable diagnostics for more complicated scenarios such as
	``ghost inflation'' \cite{ArkaniHamed:2003uy, ArkaniHamed:2003uz}
	or models
	based on the Dirac--Born--Infeld (DBI) action \cite{Alishahiha:2004eh}.
	In these examples, and many others,
	the scalar field action controlling the dynamics
	of both background and perturbations can be
	written
	\begin{equation}
		S=\dfrac{1}{2} \int \d^4 x \;
		\sqrt{-g} \ \Big[R+ 2 P(X,\phi) \Big] ,
		\label{eq:startingaction}
	\end{equation}
	where $X \equiv - g^{ab} \nabla_a \phi \nabla_b \phi$,
	the spacetime Ricci scalar is $R$,
	and $P(X,\phi)$ is an arbitrary function.
	More recently,
	diagnostic signatures for related models based on
	``Galileon'' actions have been obtained
	\cite{Burrage:2010cu,Kobayashi:2010cm,Mizuno:2010ag,Creminelli:2010qf,
	Kamada:2010qe}.
	We describe models in which the background kinetic structure is dominated
	by $(\nabla \phi)^2$ as \emph{canonical}, and more
	general examples of the form~\eqref{eq:startingaction}
	as \emph{noncanonical}.
	Eq.~\eqref{eq:startingaction} was
	suggested by Armend\'{a}riz-Pic\'{o}n {\etal}
	\cite{ArmendarizPicon:1999rj},
	who described it as `$k$-inflation.' The corresponding two-point
	function was obtained by Garriga \& Mukhanov
	\cite{Garriga:1999vw}.
	Three-point correlations 
	were studied by Gruzinov
	in a decoupling limit where mixing with gravity could be neglected
	\cite{Gruzinov:2004jx}.
	Gravitational interactions were included
	in Ref.~\cite{Seery:2005wm}.
	The bispectrum was given in generality by Chen {\etal}
	\cite{Chen:2006nt},
	followed by extensions to multiple fields
	\cite{Langlois:2008mn,Langlois:2008qf,Arroja:2008yy}.
	The four-point function was computed by several authors
	\cite{Huang:2006eha,Arroja:2008ga,Langlois:2008wt,
	Arroja:2009pd,Mizuno:2009mv,Gao:2010xk}.
	
	In this paper we return to actions of the form~\eqref{eq:startingaction}
	and reconsider the three-point function.
	The analyses discussed above controlled their calculations
	by invoking some
	form of slow-roll approximation, typically restricting
	attention to the lowest power
	of $\varepsilon \equiv - \dot{H}/H^2$
	(or other quantities of similar magnitude)
	where $H$ is the Hubble parameter and an overdot denotes a time
	derivative.
	One may expect this procedure to yield
	estimates accurate to a fractional error of
	order $\varepsilon$,
	which in some models may be as large as $10^{-1}$ to $10^{-2}$.
	Where $\fNL$ is small, as in canonical scenarios,
	this is an excellent approximation.
	However,
	in models where $\fNL$ is numerically large,
	a fractional error of order $\varepsilon$
	may be
	comparable to
	the
	resolution of forthcoming data from the \emph{Planck} survey
	satellite,
	especially if the $\Or(\varepsilon)$ term enters with
	a relatively large coefficient.
	In the equilateral mode,
	\emph{Planck} is expected to measure
	the amplitude of three-point correlations
	with an error bar in the range $25$ -- $30$,
	and
	a future
	CMB satellite such as \emph{CMBPol} or \emph{CoRE} may
	even achieve
	$\Delta \fNL \approx 10$
	\cite{Baumann:2008aq,Core:2011ck}.
	Ideally, we would like the theoretical uncertainty
	in our predictions to fall below this threshold.

	Corrections to the power spectrum at subleading order in
	$\varepsilon$ are comparatively well-understood.
	For the two-point function of a canonical field,
	the necessary expressions
	were obtained by Stewart \& Lyth
	\cite{Stewart:1993bc}, who
	worked to second order in the slow-roll expansion.%
		\footnote{Stewart \& Lyth were obliged to assume that the
		slow-roll parameter $\varepsilon = -\dot{H}/H^2$ was small.
		Grivell \& Liddle \cite{Grivell:1996sr} gave a more general
		construction in which this assumption was not mandatory,
		but were unable to obtain analytic solutions.
		Their numerical results confirmed that the Stewart--Lyth
		formulae were valid within a small fractional error.}
	Even more precise results,
	accurate to \emph{third} order,
	were given by Gong \& Stewart \cite{Gong:2001he,Gong:2002cx}.
	Stewart \& Lyth's calculation was generalized to the
	noncanonical case by
	Chen {\etal}%
		\footnote{Earlier results had been obtained in certain special cases,
		by
		Wei {\etal} \cite{Wei:2004xx}.}~%
	\cite{Chen:2006nt}, although
	third-order results analogous to Gong \& Stewart's
	are not yet available.
	The second-order calculation was systematized
	by Lidsey {\etal} \cite{Lidsey:1995np},
	who organized their expansion into ``lowest-order''
	terms containing the fewest slow-roll parameters,
	followed by ``next-order'' terms containing a single
	extra parameter,
	``next-next-order'' terms containing
	two extra parameters, and so on.
	In what follows we
	adopt this organizational scheme.
	
	Corrections to $\fNL$ have received less attention, but are more
	likely to be required.
	The prospect of a sizeable ``theory error'' in a lowest-order
	slow-roll calculation was identified by
	Chen {\etal} \cite{Chen:2006nt}, who computed
	next-order corrections for $P(X,\phi)$ models.
	But their final expressions were given
	as quadratures which were not easy to evaluate in closed form.
	How large a correction should we expect?
	The third-order action
	derived in Ref.~\cite{Seery:2005wm}
	is perturbative in the amplitude of fluctuations,
	but exact
	in slow-roll quantities: it does not
	rely on an expansion in slow-roll parameters, although
	it mixes lowest-order and next-order terms.
	Therefore
	a subset of $\Or(\varepsilon)$ corrections to the vertices of the
	theory may be obtained without calculation,
	and can be used to estimate the size of the unknown remainder.%
		\footnote{These are the next-order components
		of the $g_i$ in Eqs.~\eqref{eq:s3tau}--\eqref{eq:gi},
		for which it is not necessary to compute corrections from the
		propagator or account for time-dependence of the interaction
		vertices.
		If the subset is
		representative of the next-order terms in sign and magnitude
		then it gives a rough estimate of the possible effect.
		In fact, we will see
		in~\S\S\ref{sec:noncanonical}--\ref{sec:conclusions}
		that
		the signs in this subset are probably systematically biased:
		when
		computed in standard scenarios the remaining terms are of
		almost exactly the same magnitude but opposite sign.
		But this could not be predicted in advance.}
	In the case of DBI inflation
	\cite{Silverstein:2003hf,Alishahiha:2004eh},
	it can be checked that these terms generate a fractional correction
	$101 \varepsilon / 7 \approx 14 \varepsilon$
	in the equilateral limit.
	For the value $\varepsilon \approx 1/20$ suggested by
	Alishahiha, Silverstein \& Tong \cite{Alishahiha:2004eh}
	this is of order 70\%.%
		\footnote{Baumann \& McAllister later suggested that
		the Lyth bound \cite{Lyth:1996im} placed a limit on $\varepsilon$
		\cite{Baumann:2006cd}.
		Lidsey \& Huston \cite{Lidsey:2007gq}
		argued that in combination with the
		large value of $\varepsilon$ implied by
		Alishahiha, Silverstein \& Tong \cite{Alishahiha:2004eh}
		(see also Ref.~\cite{Lidsey:2006ia})
		this made the ``UV'' version of the DBI model microscopically unviable.
		The UV model has other difficulties.
		Bean {\etal} \cite{Bean:2007eh}
		noted that backreaction
		could invalidate
		the probe brane approximation,
		spoiling inflation.
		For the purpose of making an estimate
		we are ignoring these details.}
	In the absence of further information one should expect the unknown
	terms to be of comparable magnitude but unknown sign,
	making
	the
	uncertainty in a purely lowest-order prediction
	of order
	10\%--20\% even if
	$\varepsilon$ is substantially smaller than that
	required for the model of Ref.~\cite{Alishahiha:2004eh}.
	In models where $\fNL$ is observable,
	but $\varepsilon$ is not negligible,
	the lowest-order prediction is likely to survive unaltered only
	if an accidental cancellation removes most of the
	$\Or(\varepsilon)$ terms.
	
	There are other reasons to pursue next-order corrections.
	In models based on~\eqref{eq:startingaction},
	only two ``shapes'' of bispectrum can be produced
	with amplitude enhanced by a low sound speed.
	At subleading order, we may expect more shapes to appear---%
	the question is only with what amplitude they are produced.
	If these shapes are distinctive, and the amplitude is
	sufficiently large, it may be possible to detect them directly.
	Were
	such a signal to be found, with amplitude appropriately correlated
	to the amplitude generated in each lowest-order shape,
	it would be a striking indication that an action
	of the form~\eqref{eq:startingaction}
	was responsible for controlling the inflationary fluctuations.

	The calculation presented in this paper enables
	a precise evaluation of the slow-roll corrections,
	resolving the large theoretical uncertainties
	and providing detailed information about their shapes.
	For canonical models where $|\fNL| \lesssim 1$
	we verify that the effect is small.
	For noncanonical models we find that
	the next-order corrections may be enhanced by a 
	numerical prefactor,
	and can be several tens of percent in interesting
	cases.
	Although next-order calculations are likely to be sufficient
	for \emph{Planck},
	this suggests that
	next-next-order results could be
	required to bring the theoretical uncertainty below
	the observational resolution of a fourth-generation
	satellite such as \emph{CMBPol} or \emph{CoRE}.
	If desired, these could be
	computed using the Green's function formalism of
	Gong \& Stewart \cite{Gong:2001he,Gong:2002cx},
	although we do not attempt this here.
	We find that $P(X,\phi)$ models can produce a distinctive---%
	but not unique---%
	bispectrum shape at next-order. A similar shape arises in
	certain Galileon models.
	
	\para{Outline}%
	This paper is organized as follows.
	In~\S\ref{sec:singlefieldinflation} we discuss the background model
	and specify our version of the slow-roll approximation.
	In~\S\ref{subsec:second} we give a short account
	of the two-point function,
	which is required to study the squeezed limit of the bispectrum.
	We review the derivation of the third-order action in~\S\ref{sec:third},
	and give an extremely brief introduction
	to the ``in--in'' formulation for
	expectation values in \S\ref{subsec:inin}.
	We give a careful discussion of boundary terms
	in~\S\ref{subsec:boundaryterms}.

	Next-order corrections to the bispectrum are calculated
	in~\S\ref{subsec:bispectrum}. Readers familiar with existing calculations
	of the three-point function may wish to skip directly to this section,
	and refer to \S\S\ref{sec:singlefieldinflation}--\ref{subsec:boundaryterms}
	for our notation and definitions.
	We translate
	the next-order bispectrum to a
	prediction for $\fNL$ in \S\ref{subsec:fnl},
	and study its shape and scale dependence
	in \S\S\ref{sec:shapes}--\ref{sec:running}.
	We identify a new bispectrum shape orthogonal to the
	lowest-order possibilities,
	and give cosines for the
	next-order shapes with standard templates.
	In \S\ref{sec:running}
	we obtain a general formula for the overall running of $\fNL$ with
	scale.
	
	In~\S\ref{sec:tensor} we briefly recapitulate the next-order
	computation for tensor fluctuations. This extends the range of
	observables to include the tensor-to-scalar ratio, $r$,
	and the tensor spectral index $n_t$,
	which are helpful in
	assessing observational signatures.
	In~\S\S\ref{sec:vanilla}--\ref{sec:noncanonical} we apply our
	results to a selection of concrete models.
	\S\ref{sec:vanilla} concentrates on canonical single-field
	inflation with an arbitrary potential.
	The next-order corrections are small, but we are able to
	verify Maldacena's consistency condition to second-order in
	slow-roll parameters.
	In~\S\ref{sec:noncanonical} we discuss models which can imprint
	a significant nongaussian signature, focusing on DBI inflation
	and $k$-inflation.
	For DBI inflation one can choose
	a late-time power-law attractor solution,
	or a generic quasi-de Sitter background which
	depends on the shape of the potential and warp factor.
	In~\S\ref{sec:dbi} we calculate the next-order corrections
	for both cases, obtaining corrections to $\fNL$ arising from the
	shape of the potential and warp factor for the first time.
	We compare our results to known exact solutions.
	In~\S\ref{sec:k-inflation} we apply our
	calculation to the power-law
	solution of $k$-inflation.
	Finally, we summarize our findings and conclusions
	in~\S\ref{sec:conclusions}.

	A number of appendices collect extra material.
	We briefly recount the derivation of next-order corrections
	to the propagator in Appendix~\ref{appendix:propagator}.
	Appendices~\ref{appendix:a}--\ref{appendix:b}
	give mathematical details
	of certain integrals and discuss the role of analytic continuation in
	obtaining expressions valid for arbitrary momentum shapes.
	In Appendix~\ref{appendix:compare}
	we provide explicit formulae for the quadratures
	used in Ref.~\cite{Chen:2006nt}.
	
	We choose units in which $c = \hbar = 1$.
	We use the reduced Planck mass, $\Mp = (8 \pi G)^{-1/2}$,
	usually setting $\Mp = 1$. Spacetime tensors are labelled by Latin indices
	$\{ a, b, \ldots \}$, whereas purely spatial tensors
	are labelled by indices $\{ i, j, \ldots \}$.

	\section{Single-field inflation: an overview}
	\label{sec:singlefieldinflation}

	Consider the theory~\eqref{eq:startingaction}, with
	a homogeneous background solution given by the
	Robertson--Walker metric, 
	\begin{equation}
		\d s^2 = - \d t^2 + a^2(t) \d \vec{x}^2 .
		\label{eq:frw}
	\end{equation} 
	The corresponding Friedmann equations are
	\begin{equation}
 		3 H^2= \left(2X P_{,X}-P \right) \quad
 		\textrm{and} \quad
 		2 \dot{H}+3H^2=-P ,
		\label{eq:friedmann}
	\end{equation}
	where $H\equiv \dot{a}/a$ is the Hubble parameter,
	an overdot denotes differentiation with respect to cosmological time,
	$t$,
	and a
	comma denotes a partial derivative.
	The nontrivial kinetic structure of $P$
	causes fluctuations of the scalar field $\phi$ to
	propagate with phase velocity---which we will
	refer to as the sound speed, $c_s$---different to unity
	\cite{Christopherson:2008ry},
	\begin{equation}
		\cs^2 
		=
		\dfrac{P_{,X}}{P_{,X}+2XP_{,XX}} .
		\label{eq:speedofsound}
	\end{equation}
	In the special case of a canonical field,
	we have $P=X/2-V(\phi)$ and the fluctuations in $\phi$
	propagate at the speed of light.

	\subsection{Fluctuations}
	During inflation the universe rapidly isotropizes
	\cite{Wald:1983ky}, making $\phi$
	spatially homogeneous to a good approximation. At the same time,
	quantum fluctuations generate small perturbations whose statistics
	it is our intention to calculate. Since $\phi$ dominates the
	energy density of the universe by assumption, its fluctuations
	must be communicated to the metric.

	Our freedom to make coordinate redefinitions allows these
	metric and field fluctuations to be studied in a variety of
	gauges \cite{Kodama:1985bj,Malik:2008im}.
	Without commitment to any particular gauge,
	a perturbed metric can be written in terms
	of the Arnowitt--Deser--Misner (ADM) lapse and shift functions
	\cite{Arnowitt:1962hi},
	respectively
	denoted $N$ and $N_i$,
	\begin{equation}
		\d s^2 = -N^2 \d t^2 + h_{ij} (\d x^i + N^i \d t)(\d x^j + N^j \d t) .
		\label{eq:adm}
	\end{equation}
	In the absence of perturbations,
	Eq.~\eqref{eq:adm} reduces to the background metric~\eqref{eq:frw}.
	The utility of this formulation is that $N$ and $N_i$
	do not support propagating modes,
	which are restricted to $\phi$ and the spatial
	metric $h_{ij}$. There are three scalar
	and two tensor modes,
	of which one scalar can be gauged away by choice of spatial
	coordinates and another by choice of time.
	At quadratic order
	the tensor modes decouple from the remaining scalar.
	The tensors contribute at tree-level
	for $n$-point functions with
	$n \geq 4$
	\cite{Arroja:2008ga,Seery:2008ax},
	but only at loop-level%
		\footnote{The loop expansion in question is generated
		by insertion of vertices with extra factors of
		the fluctuations, and so is effectively an expansion in
		powers of $H^2 \approx 10^{-10}$.
		Conditions under which the ``classical'' tree graphs
		dominate loops
		were discussed by
		Weinberg \cite{Weinberg:2008mc,Weinberg:2005vy,Weinberg:2006ac}
		and van der Meulen \& Smit
		\cite{vanderMeulen:2007ah}.}
	to the three-point function.
	In what follows
	we work to tree-level
	and discard tensor modes.
	Whether or not they are retained, however,
	one uses constraint equations (discussed below)
	to express the lapse and shift algebraically
	in terms of the
	propagating modes.%
		\footnote{If desired, the constraints need not be solved
		explicitly but can be enforced via the use of auxiliary
		fields, yielding an ``off-shell'' formulation
		\cite{Prokopec:2010be}.}
	
	Where only a single scalar field is present, it is convenient
	to work on spatial slices of uniform field value. In this gauge
	the propagating scalar mode is carried by $h_{ij}$, and expresses
	local modulations in the expansion $a(t)$,
	\begin{equation}
		h_{ij} = a^2(t) \e{2\zeta} \delta_{ij} .
		\label{eq:comoving-curvature-perturbation}
	\end{equation}
	Writing $\ln a(t) = \int^t \d N$, where
	$\d N = H \, \d t$ and
	$N(t) - N(t_0)$ expresses
	the number of e-folds of expansion between times $t$ and $t_0$,
	it follows that $\zeta = \delta N$.%
		\footnote{Our notation is chosen to coincide with
		the recent literature,
		including Refs.~\cite{Salopek:1990jq,Maldacena:2002vr,Chen:2006nt,
		Cheung:2007st,
		Creminelli:2010qf,Baumann:2011su}.
		However,
		the uniform-$\phi$ slicing corresponds
		to comoving gauge---in which the perturbation in
		Eq.~\eqref{eq:comoving-curvature-perturbation}
		has traditionally been denoted $\mathcal{R}$
		\cite{Liddle:1993fq,Lyth:1998xn,Wands:2000dp}.
		The symbol $\zeta$ has often been reserved for
		the corresponding quantity in uniform density gauge.
		The comoving and uniform density slicings agree
		for adiabatic, superhorizon-scale
		perturbations in single-field inflation
		\cite{Wands:2000dp},
		and therefore $\zeta = \mathcal{R}$ up to an irrelevant
		sign convention.}
	
	\para{Slow-variation parameters}%
	We work perturbatively in $\zeta$.
	During inflation the universe is typically well-approximated by
	a de Sitter epoch, in which the Hubble parameter is constant.
	To express the action~\eqref{eq:startingaction}
	in terms of $\zeta$, it is convenient to introduce dimensionless
	parameters which measure the variation of background quantities
	per Hubble time.
	We define
	\begin{equation}
		\varepsilon \equiv - \frac{\d \ln H}{\d N}
		= - \frac{\dot{H}}{H^2} ,
		\quad
		\eta \equiv \frac{\d \ln \varepsilon}{\d N}
		= \frac{\dot{\varepsilon}}{H\varepsilon} ,
		\quad
		\text{and}
		\quad
		s \equiv \frac{\d \ln \cs}{\d N}
		= \frac{\csdot}{H \cs} \ .
		\label{eq:slowvariation}
	\end{equation}
	Note that $\varepsilon \geq 0$, provided $H$ decreases with time.
	There is no requirement of principle that any of
	$\varepsilon$, $\eta$ or $s$
	are small, although whenever background quantities are
	slowly varying we may expect
	$\varepsilon, |\eta|, |s| \ll 1$.

	\para{Action and constraints}%
	In ADM variables, the action~\eqref{eq:startingaction} can be
	written
	\begin{equation}
		S = \dfrac{1}{2} \int \d^4 x \; \sqrt{h} \,
			N \Big[
				{R}^{(3)} +2P (X, \phi) 
			\Big]
			+
			\dfrac{1}{2} \int \d^4 x \; \sqrt{h} \,
			N^{-1} \Big[
				E_{ij} E^{ij} - E^2
			\Big] ,
		\label{eq:admaction}
	\end{equation}
	where $E_{ij}$ satisfies
	\begin{equation}
		E_{ij} = \frac{1}{2} \Big( \dot{h}_{ij} - N_{(i \mid j)} \Big) ,
	\end{equation}
	where $\mid$ denotes a covariant derivative compatible with $h_{ij}$,
	and indices enclosed in brackets $(\cdots)$ are to be symmetrized.
	The extrinsic curvature of spatial slices, $K_{ij}$,
	can be written $K_{ij} = N^{-1} E_{ij}$.

	Varying the action with respect to $N$ and $N_i$ yields constraint
	equations \cite{Seery:2005wm},
	\begin{subequations}
	\begin{equation}
		{R}^{(3)} + 2P - 4P_{,X}
			\left( X+h^{ij} \partial_i \phi \partial_j \phi	\right)
		- \dfrac{1}{N^2} \left(E_{ij}E^{ij}-E^2 \right)
		=
		0
		\label{eq:shift-constraint}
	\end{equation}
	and
	\begin{equation}
		\nabla^{j} \bigg[
			\dfrac{1}{N}\left( E_{ij}-Eh_{ij} \right)
		\bigg]
		=
		\dfrac{2 P_{,X}}{N} \bigg(
			\dot{\phi}\partial_i \phi
			- N^{j}\partial_i \phi \partial_j \phi
		\bigg) .
		\label{eq:lapse-constraint}
	\end{equation}
	\end{subequations}
	Eqs.~\eqref{eq:shift-constraint}--\eqref{eq:lapse-constraint}
	can be solved order-by-order. We write
	$N = 1 + \alpha$.
	Likewise,
	the shift vector can be decomposed into its
	irrotational and solenoidal components,
	$N_i = \partial_i \theta + \beta_i$,
	where $\partial_i \beta_i = 0$.
	We expand $\alpha$, $\beta$ and $\theta$ perturbatively
	in powers of $\zeta$, writing the terms of $n^{\mathrm{th}}$ order
	as $\alpha_n$, $\beta_n$ and $\theta_n$.
	To study the three-point correlations it is only necessary
	to solve the constraints to first order
	\cite{Maldacena:2002vr,Chen:2006nt}.
	One finds
	\begin{equation}
		\alpha_1 = \dfrac{\dot{\zeta}}{H},
		\quad
		\beta_{1i} = 0,
		\quad
		\text{and}
		\quad
		\theta_{1} = - \dfrac{\zeta}{H} + 
			\dfrac{a^2 \SigmaP}{H^2} \partial^{-2} \dot{\zeta} .
		\label{eq:order1constraints}
	\end{equation}
	The quantity $\SigmaP$ measures the second $X$-derivative of $P$.
	In what follows
	it will be
	helpful to define an analogous quantity for the third derivative,
	denoted $\lambdaP$ \cite{Seery:2005wm},
	\begin{subequations}
	\begin{align}
		\SigmaP
			& \equiv
			X P_{,X} + 2 X^2 P_{,XX}
			=
			\dfrac{\varepsilon H^2}{\cs^2} .
			\label{eq:sigmadef}
		\\
		\lambdaP
			& \equiv
			X^2P_{,XX}+\dfrac{2}{3}X^3P_{,XXX} .
			\label{eq:lambdadef}
	\end{align}
	\end{subequations}

	\subsection{Two-point correlations}
	\label{subsec:second}
	
	To obtain the two-point statistics of $\zeta$ one must
	compute the action~\eqref{eq:admaction} to second order,
	which was first accomplished by Garriga \& Mukhanov
	\cite{Garriga:1999vw}.
	Defining a conformal time variable by
	$\tau = \int_\infty^{t} \d t / a(t)$, one finds
	\begin{equation}
		S_2
		=
		\int \d^3x \, \d\tau \;
		a^2 \zterm
		\big[
			(\zeta')^2 - \cs^2 (\partial \zeta)^2
		\big],
		\label{eq:eom2conformaltime}
	\end{equation}
	where $\zterm \equiv \varepsilon / \cs^2$,
	a prime denotes differentiation with respect to $\tau$,
	and $\partial$ represents a spatial derivative.
	Although our primary interest lies with $P(X,\phi)$-type models
	of the form~\eqref{eq:startingaction}, we leave intermediate
	expressions in terms of $\zterm$ and its corresponding slow-variation
	parameter,
	\begin{equation}
		v \equiv \frac{\d \ln \zterm}{\d N} = \frac{\dot{\zterm}}{H \zterm}
		= \eta - 2 s .
	\label{eq:defv}
	\end{equation}
	The last equality applies in a $P(X,\phi)$ model.
	When calculating three-point functions in
	\S\ref{sec:third}, we will do so
	for an arbitrary $\zterm$.
	In order that the fluctuations are healthy and not ghostlike
	we must choose $\zterm$ positive.
	
	To simplify certain intermediate expressions,
	it will be necessary to have an expression for the
	variation $\delta S_2 / \delta \zeta$,
	\begin{equation}
		\dot{\chi}
		=
		\varepsilon \zeta
		- H \chi
		- \dfrac{1}{2a}\partial^{-2} \dfrac{\delta S_2}{\delta \zeta} ,
		\label{eq:eom2zeta}
	\end{equation} 
	where we have introduced a variable $\chi$, which satisfies
	\begin{equation}
		\partial^2 \chi
		=
		\frac{\varepsilon a^2}{\cs^2} \dot{\zeta} .
		\label{eq:chieq}
	\end{equation}
	The equation of motion follows by setting $\delta S_2 / \delta \zeta = 0$
	in~\eqref{eq:eom2zeta}.
	
	\para{Slow-variation approximation}%
	Although Eqs.~\eqref{eq:eom2conformaltime}
	and \eqref{eq:eom2zeta}--\eqref{eq:chieq}
	are exact at linear order in $\zeta$,
	it is not known how to solve the equation of motion~\eqref{eq:eom2zeta}
	for arbitrary backgrounds.
	Lidsey {\etal} \cite{Lidsey:1995np}
	noted that the time derivative of each slow-variation parameter
	is proportional to a sum of products of slow-variation parameters,
	and therefore if we assume
	\begin{equation}
		0 < \varepsilon \ll 1,
		\quad
		|\eta| \ll 1,
		\quad
		|s| \ll 1
		\quad
		\text{and}
		\quad
		|v| \ll 1 ,
		\label{eq:slow-variation-approximation}
	\end{equation}
	and work to first order in these quantities, we may formally treat them as
	constants.
	
	Expanding a general background quantity such as $H(t)$
	around a reference time $t_\star$, one finds
	\begin{equation}
		H(t) \approx H(t_\star) \left[ 1 + \varepsilon_{\star}
			\Delta N_\star(t) + \cdots \right] ,
		\label{eq:time-dependence}
	\end{equation}
	where $\Delta N_\star(t) = N(t) - N(t_\star)$.
	Similar formulae apply for the slow-variation parameters $\varepsilon$,
	$\eta$, $s$ and $v$ themselves, and their derivatives.
	Physical quantities do not depend on the arbitrary scale $t_\star$,
	which plays a similar role to the arbitrary renormalization scale
	in quantum field theory.
		Eq.~\eqref{eq:time-dependence}
	yields an approximation to the full time evolution whenever
	$|\varepsilon_\star \Delta N_\star(t) | \ll 1$
	\cite{Gong:2001he,Gong:2002cx,Choe:2004zg},
	but fails
	no later than $\sim 1/\varepsilon_\star$ e-folds after the
	reference time $t_\star$.
	In order that solutions of~\eqref{eq:eom2zeta} are sufficiently
	accurate for their intended use---to compute correlation functions
	for $t \sim t_\star$---we must demand that $\Or(\varepsilon)$
	quantities are sufficiently small that~\eqref{eq:time-dependence}
	applies for at least a few e-folds around the time of horizon crossing,
	but
	it is not usually necessary to impose more stringent
	restrictions. This approach will fail if any slow-variation quantity
	becomes temporarily large around the time of horizon exit,
	which may happen in ``feature'' models
	\cite{Chen:2006xjb,Chen:2008wn,Chen:2010bka,Leblond:2010yq,Adshead:2011bw}.
	
	Working in an arbitrary model,
	results valid many e-folds after horizon exit
	typically require an improved formulation of perturbation theory
	obtained by resumming powers of $\Delta N$
	\cite{Seery:2007wf,Burgess:2009bs,Seery:2010kh},
	for which various formalisms are in use
	\cite{Lyth:2005fi,Mulryne:2009kh,Mulryne:2010rp,Peterson:2010mv}.
	However, as is well-known (and we shall see below),
	this difficulty does not arise
	for single-field inflation.
		
	\para{Two-point function}%
	The time-ordered two-point correlation function is
	the Feynman propagator,
	$\langle \timeorder \, \zeta(\tau, \vect{x}_1) \zeta(\tau', \vect{x}_2)
	\rangle = G(\tau, \tau'; |\vec{x}_1 - \vec{x}_2|)$,
	which depends on the 3-dimensional invariant
	$|\vec{x}_1 - \vec{x}_2|$.
	In Fourier space $G = \int \d^3 q \, (2\pi)^{-3} G_q(\tau, \tau')
	\e{\im \vec{q} \cdot ( \vec{x}_1 - \vec{x}_2 )}$,
	which implies
	\begin{equation}
		\langle
			\timeorder \,
			\zeta(\vec{k}_1, \tau)
			\zeta(\vec{k}_2, \tau')
		\rangle
		=
		(2\pi)^3
		\delta(\vec{k}_1+\vec{k}_2)
		G_k(\tau, \tau') .
	\end{equation}
	The $\delta$-function enforces conservation of
	three-momentum, and
	$k = |\vec{k}_1| = |\vec{k}_2|$.
	One finds
	\begin{equation}
		G_k (\tau, \tau') = \left\{
		\begin{array}{l@{\hspace{5mm}}l}
			u_k(\tau) u_k^\ast (\tau')  & \text{if $\tau < \tau'$} \\
			u_k^\ast (\tau) u_k(\tau') & \text{if $\tau' < \tau$}
		\end{array}
		\right. .
		\label{eq:propagator}
	\end{equation}
	The mode function $u_k$ is a positive frequency solution
	of~\eqref{eq:eom2zeta} with $\delta S_2 / \delta \zeta = 0$.
	Invoking~\eqref{eq:slow-variation-approximation}
	and working to first-order in each slow-variation parameter,
	we find
	\begin{equation}
		u_k(\tau)
		=
		\dfrac{\sqrt{\pi}}{2\sqrt{2}}
		\dfrac{1}{a(\tau)}
		\sqrt{\dfrac{-(1+s) \tau}{\zterm(\tau)}}
		H_{\frac{3}{2} + \varpi}^{(2)}\left[-k \cs (1+s) \tau \right] ,
		\label{eq:wavefunction}
	\end{equation}
	in which $H_{\nu}^{(2)}$ is the Hankel function of the second kind
	of order $\nu$, and $\varpi \equiv \varepsilon+v/2+3s/2$.
	At sufficiently early times, for which $|k \cs \tau| \gg 1$,
	an oscillator of comoving wavenumber $k$ cannot explore the curvature
	of spacetime and feels itself to be in Minkowski space.
	In this limit~\eqref{eq:wavefunction} approaches the corresponding
	Minkowski wavefunction \cite{Bunch:1978yq}.

	\para{Power spectrum}%
	When
	evaluated at equal times, the two-point function defines
	a power spectrum $P(k,\tau)$ by the rule
	\begin{equation}
		P(k, \tau) = G_k(\tau, \tau) .
		\label{eq:power-spectrum-def}
	\end{equation}
	In principle, the power spectrum depends on time.
	It is conventionally denoted $P(k)$ and should not
	be confused with the Lagrangian $P(X,\phi)$.
	Using~\eqref{eq:propagator} for $\tau' \rightarrow \tau$,
	working in the limit $|k \cs \tau| \rightarrow 0$
	and expanding uniformly around a reference time $\tau_\star$, one
	finds
	\begin{equation}
		P(k) = \frac{H_\star^2}{4 \zterm_\star \csstar^3}
			\frac{1}{k^3} \left[
				1 + 2 \left\{
					\varpi_\star (2 - \EulerGamma - \ln \frac{2k}{k_\star})
					- \varepsilon_\star - s_\star
				\right\}
			\right] .
		 \label{eq:powerspectrum}
	\end{equation}
	The Euler--Mascheroni
	constant is
	$\EulerGamma \approx 0.577$.
	We have introduced a quantity
	$k_\star$ satisfying $|k_\star c_{s\star}\tau_\star| = 1$,
	where we recall that
	$\tau_\star$ is arbitrary and need not be determined by
	$k$.
	Since $|k_\star c_{s\star} \tau_\star| \approx
	|k_\star c_{s\star}/ a_\star H_\star|$,
	we describe $\tau_\star$ as the horizon-crossing time associated with
	the wavenumber $k_\star$. Inclusion of next-order effects
	marginally shifts the time of horizon exit,
	leading to a small mismatch between
	$|k_\star c_{s\star} \tau_\star|$ and
	$|k_\star c_{s\star} / a_\star H_\star|$.%
		\footnote{Chen {\etal} \cite{Chen:2006nt} adopted a definition
		in terms of $|k_\star / a_\star H_\star|$, making
		some intermediate
		expressions different in appearance but identical in content.}
	In the terminology of Lidsey {\etal} \cite{Lidsey:1995np},
	the ``lowest-order'' result is the coefficient of the square
	bracket $[ \cdots ]$, and the ``next-order'' correction arises from the
	term it contains of first-order in slow-variation parameters.
	Therefore the lowest-order result can be recovered by setting the
	square bracket to unity.
	This convention
	was introduced in Ref.~\cite{Lidsey:1995np}, and
	when writing explicit expressions
	we adopt it
	in the remainder of this paper.
	
	Although expanded around some reference time $\tau_\star$,
	Eq.~\eqref{eq:powerspectrum}
	does not depend on $\Delta N_\star$
	and therefore becomes time-independent 
	once the scale of wavenumber $k$ has passed outside the horizon.
	This is a special property of single-field inflation.
	Working in classical perturbation theory
	it is known that $\zeta$ becomes constant
	in the superhorizon limit provided the fluctuations are adiabatic
	and the background solution is an attractor
	\cite{Rigopoulos:2003ak,Lyth:2004gb,Naruko:2011zk}.%
		\footnote{For canonical inflation this conclusion can be reached
		without use of the Einstein equations. Recently, Naruko \&
		Sasaki argued that the Einstein equations may be required for
		some types of noncanonical models \cite{Naruko:2011zk}.}
	We are not aware of
	a corresponding theorem for
	the correlation functions of $\zeta$, computed according to the rules
	of quantum field theory.
	However, it appears that in all
	examples compatible with the classical conservation laws
	discussed in Refs.~\cite{Rigopoulos:2003ak,Lyth:2004gb,Naruko:2011zk},
	a time-independent limit is reached. We will return to this issue
	in \S\ref{subsec:3pf} below.
	
	\para{Scale dependence}%
	The logarithmic term in $P(k)$
	indicates that
	the power spectrum varies weakly with scale $k$,
	making~\eqref{eq:powerspectrum} quantitatively
	reliable only if the reference time $k_\star$ is chosen sufficiently
	close to $k$ that $|\ln(2k/k_\star)| \lesssim 1$.
	Defining a ``dimensionless'' power spectrum
	$\Ps$ by the rule $\Ps = k^3 P(k) / 2\pi^2$,
	the variation of $\Ps(k)$ with scale is conventionally described in terms
	of a spectral index,
	\begin{equation}
		n_s - 1 = \frac{\d \ln \Ps}{\d \ln k} = -2 \varpi_\star ,
		\label{eq:rge}
	\end{equation}
	which is valid to lowest-order provided $k_\star \approx k$.
	Eq.~\eqref{eq:rge} is a renormalization group equation describing
	the flow of $\Ps$ with $k$, where $\beta_{\Ps} \equiv (n_s - 1) \Ps$ plays
	the role of the $\beta$-function.
	An expression for $n_s$ valid to next-order can be obtained
	by setting $k = k_\star$,
	making `$\star$' the time of horizon exit of wavenumber $k$.
	Having made this choice, the
	$k$-dependence
	in~\eqref{eq:powerspectrum} appears only through the time of
	evaluation `$\star$' and is accurate to next-order
	\cite{Stewart:1993bc}.
	We define additional slow-variation parameters,
	\begin{equation}
		\xi \equiv \frac{\dot{\eta}}{H \eta} ,
		\quad
		t \equiv \frac{\dot{s}}{H s} ,
		\quad
		\text{and}
		\quad
		w \equiv \frac{\dot{v}}{H v} .
	\end{equation}
	and quote results in Table~\ref{table:spectral-index} for $n_s - 1$
	at lowest-order and next-order.
	
	\begin{table}

	\heavyrulewidth=.08em
	\lightrulewidth=.05em
	\cmidrulewidth=.03em
	\belowrulesep=.65ex
	\belowbottomsep=0pt
	\aboverulesep=.4ex
	\abovetopsep=0pt
	\cmidrulesep=\doublerulesep
	\cmidrulekern=.5em
	\defaultaddspace=.5em
	\renewcommand{\arraystretch}{1.6}

	\begin{center}
		\small
		\begin{tabular}{lll}

			\toprule
		
			model & lowest-order & next-order \\
			\midrule
		
			\rowcolor[gray]{0.9}
				arbitrary &
				$-2\epsilon_\star - v_\star - 3 s_\star$ &
				$\displaystyle
				-2\epsilon_\star^2 + \epsilon_\star \eta_\star
				\big(2 - 2 \EulerGamma - 2 \ln \frac{2k}{k_\star}\big)
				+ s_\star t_\star
				\big( 4 - 3 \EulerGamma - 3 \ln \frac{2k}{k_\star} \big)$ \\
			\rowcolor[gray]{0.9}
				& & $\displaystyle
				\mbox{} - 5 \epsilon_\star s_\star - 3 s_\star^2
				- v_\star( \epsilon_\star + s_\star)
				+ v_\star w_\star \big(
				2 - \EulerGamma - \ln \frac{2k}{k_\star} \big)$
				\\[2mm]

				canonical &
				$-2 \varepsilon_\star - \eta_\star$ &
				$\displaystyle
				-2\epsilon_\star^2 + \epsilon_\star \eta_\star
				\big(1 - 2 \EulerGamma - 2 \ln \frac{2k}{k_\star} \big)
				+ \eta_\star \xi_\star
				\big( 2 - \EulerGamma - \ln \frac{2k}{k_\star} \big)$
				\\[2mm]

			\rowcolor[gray]{0.9}
				$P(X,\phi)$ &
				$-2\epsilon_\star - \eta_\star - s_\star$ &
				$\displaystyle
				-2\epsilon_\star^2 + \epsilon_\star \eta_\star
				\big(1 - 2 \EulerGamma - 2 \ln \frac{2k}{k_\star} \big)
				- s_\star t_\star
				\big( \EulerGamma + \ln \frac{2k}{k_\star} \big)$ \\
			\rowcolor[gray]{0.9}
				& & $\displaystyle
				\mbox{} + \eta_\star \xi_\star
				\big( 2 - \EulerGamma - \ln \frac{2k}{k_\star} \big)
				- s_\star^2 - 3 \epsilon_\star s_\star - s_\star \eta_\star$ \\

 			\bottomrule
	
		\end{tabular}
	\end{center}
	\caption{$n_s - 1$ at lowest-order and next-order.
		The first row applies for arbitrary positive, smooth $\zterm$,
		as explained below Eq.~\eqref{eq:defv}.
	\label{table:spectral-index}}
	\end{table}	

	\section{Three-point correlations}
	\label{sec:third}
	
	\subsection{Third-order action}
	\label{sec:third-order-action}
	
	Three-point statistics can be obtained from the third-order
	action. This calculation was first given in
	Ref.~\cite{Seery:2005wm}, where the three-point
	function was obtained under certain hypotheses.
	Chen {\etal} later computed the full three-point function
	\cite{Chen:2006nt}.
	After integration by parts,
	using both the background equations of motion~\eqref{eq:friedmann}
	and the solutions of the
	constraints given in~\eqref{eq:order1constraints},
	we find
	\begin{equation}
		\begin{split}
		S_3 \supseteq & \int_{\partial} \d^3 x \; a^3
		\bigg\{
			-9 H \zeta^3 + \frac{1}{a^2 H} \zeta (\partial \zeta)^2
		\bigg\} \\
		& \mbox{}
		+ \dfrac{1}{2} \int \d^3 x \, \d t \; a^3
		\bigg\{
			- 2 \frac{\varepsilon}{a^2} \zeta (\partial \zeta)^2
			+ 6 \frac{\SigmaP}{H^2} \zeta \dot{\zeta}^2
			- 2 \frac{\SigmaP +2\lambdaP}{H^3} \dot{\zeta}^3
			- \dfrac{4}{a^4} \partial^2 \theta_{1} \partial_j \theta_{1}
				\partial_j \zeta
		\\ &
		\hspace{3.05cm} \mbox{}
			+ \dfrac{1}{a^4}
				\bigg(
					\dfrac{\dot{\zeta}}{H} - 3\zeta
				\bigg)
				\partial^2 \theta_{1}
				\partial^2 \theta_{1}
			- \dfrac{1}{a^4}
				\big(
					\dfrac{\dot{\zeta}}{H} - 3\zeta
				\big)
				\partial_i \partial_j \theta_{1}
				\partial_i \partial_j \theta_{1}
		\bigg\} ,
		\end{split}
		\label{eq:firstS3}
	\end{equation}
	where $\int_\partial$ denotes an integral over a formal boundary,
	whose role we will discuss in more detail below.
	We have temporarily reverted to cosmic time $t$, rather than
	the conformal time $\tau$.
	The parameters $\SigmaP$ and $\lambdaP$ were defined in
	Eqs.~\eqref{eq:sigmadef} and~\eqref{eq:lambdadef}.
	After further integration by parts, and combining~\eqref{eq:eom2zeta}
	and~\eqref{eq:order1constraints} with~\eqref{eq:chieq}, one finds%
		\footnote{In Refs.~\cite{Seery:2005wm,Chen:2006nt},
		a further transformation
		was made to rewrite the terms proportional to $\eta$.
		Using the field equation~\eqref{eq:eom2zeta} and integrating by parts,
		these can be consolidated into the coefficient of a new operator
		$\zeta^2 \dot{\zeta}$---which does not appear
		in~\eqref{eq:shiftaction}---%
		together with corresponding new contributions to $f$ and the boundary
		term.
		In this paper, we will leave the action as in~\eqref{eq:shiftaction}
		for the following reasons.
		First, when computing
		the three-point correlation function, the contribution of each operator
		must be obtained separately. Therefore nothing is gained by introducing
		an extra operator, whose contribution we can avoid calculating
		by working with~\eqref{eq:shiftaction}.
		Second, after making the transformation,
		the contribution from the boundary term is nonzero
		and must be accommodated by making a field redefinition.
		This redefinition must eventually be reversed to obtain the
		correlation functions of the physical field $\zeta$.
		If we leave the action as in~\eqref{eq:shiftaction} then
		it transpires that no field redefinition is necessary.}
	\begin{equation}
		\begin{split}
			S_3 \supseteq \mbox{}
			& \frac{1}{2} \int_{\partial} \d^3 x \; a^3
			\bigg\{
				- 18 H^3 \zeta^3
				+ \frac{2}{a^2 H} \bigg( 1 - \frac{\varepsilon}{\cs^2} \bigg)
					\zeta (\partial \zeta)^2
				- \frac{1}{2a^4 H^3} \partial^2 \zeta (\partial \zeta)^2
				- \frac{2 \varepsilon}{H \cs^4} \zeta \dot{\zeta}^2
				\\ & \hspace{2.15cm} \mbox{}
				- \frac{1}{a^4 H} \partial^2 \chi
					\partial_j \chi \partial_j \zeta
				- \frac{1}{2a^4 H} \partial^2 \zeta (\partial \chi)^2
				+ \frac{1}{a^4 H^2} \partial^2 \zeta
					\partial_j \chi \partial_j \zeta
				\\ & \hspace{2.15cm} \mbox{}
				+ \frac{1}{2a^4 H^2} \partial^2 \chi (\partial \zeta)^2
			\bigg\}
			\\
			& + \frac{1}{2} \int \d^3 x \, \d t \; a^3
			\bigg\{
				\frac{2}{\cs^2 a^2}
					\{
						\varepsilon ( 1 - \cs^2) + \eta \varepsilon
						+ \varepsilon^2 + \varepsilon \eta - 2 \varepsilon s
					\}
					\zeta (\partial \zeta)^2
				\\ & \hspace{3cm} \mbox{}
				+ \frac{1}{\cs^4}
					\{
						6 \varepsilon (\cs^2 - 1) + 2 \varepsilon^2
						- 2 \varepsilon \eta
					\}
					\zeta \dot{\zeta}^2
				\\ & \hspace{3cm} \mbox{}
				+ \frac{1}{H}
					\bigg(
						2 \frac{\varepsilon}{\cs^4} ( 1 - \cs^2 )
						- 4 \frac{\lambdaP}{H^2}
					\bigg)
					\dot{\zeta}^3
				\\ & \hspace{3cm} \mbox{}
				+ \frac{\varepsilon}{2a^4}
					\partial^2 \zeta (\partial \chi)^2
				+ \frac{\varepsilon-4}{a^4}
					\partial^2 \chi
					\partial_j \zeta \partial_j \chi
				+ \frac{2 f}{a^3} \frac{\delta S_2}{\delta \zeta}
			\bigg\}
		\end{split}
		\label{eq:shiftaction}
	\end{equation}
	where $f$ is defined by
	\begin{equation}
		\begin{split}
		f \equiv &
			- \frac{1}{H \cs^2} \zeta \dot{\zeta}^2
			+ \frac{1}{4a^2 H^2} ( \partial \zeta )^2
			- \frac{1}{4 a^2 H^2} \partial_j \zeta \partial_j \chi
			- \frac{1}{4 a^2 H^2} \partial^{-2}
				\big\{
					\partial_i \partial_j ( \partial_i \zeta \partial_j \zeta )
				\big\}
			\\ & \mbox{}
			+ \frac{1}{4a^2 H} \partial^{-2} \partial_j
				\big\{ \partial^2 \zeta \partial_j \chi
					+ \partial^2 \chi \partial_j \zeta
				\big\}\;.
		\end{split}
		\label{eq:secondS3}
	\end{equation}

	\para{Boundary terms}%
	The boundary terms in Eqs.~\eqref{eq:firstS3}--\eqref{eq:secondS3}
	arise from integration by parts with respect to time,
	and were not quoted
	for the original calculations reported in Refs.~\cite{Maldacena:2002vr,
	Seery:2005wm,Chen:2006nt}.
	Adopting a procedure initially used by Maldacena,
	these calculations discarded all boundary terms,
	retaining only contributions proportional to $\delta S_2 / \delta \zeta$
	in the bulk component of~\eqref{eq:shiftaction}.
	The $\delta S_2 / \delta \zeta$ terms were subtracted by making
	a field redefinition.

	This procedure can be misleading.
	The terms proportional to $\delta S_2 / \delta \zeta$
	give no contribution to any Feynman graph at any order in
	perturbation theory, because $\delta S_2 / \delta \zeta$ is
	zero by construction when evaluated on a propagator.
	Therefore these terms
	give nothing whether they are subtracted or not.
	On the other hand,
	a field redefinition may certainly shift the
	three-point correlation function.
	Therefore, in general,
	the subtraction procedure will yield correct answers only if
	this nonzero shift reproduces the contribution of the boundary
	component in~\eqref{eq:shiftaction},
	which need not be related to $f$.
	This argument was first given in 
	Ref.~\cite{Seery:2005gb},
	and later in more detail in Refs.~\cite{Seery:2006tq,Seery:2010kh},
	but was
	applied to the third-order action
	for field fluctuations in the spatially flat gauge.
	In this gauge only a few
	integrations by parts are required. The boundary term is
	not complicated and the subtraction procedure works as intended.
	In the present case, however,
	it appears impossible that the subtraction
	procedure could be correct, because the boundary
	term contains operators such as $\zeta^3$ which are not present in
	$f$.
	Indeed, because
	$\zeta$ approaches a constant at late times, the $\zeta^3$ term
	apparently leads to a catastrophic divergence
	which should manifest itself as a rapidly evolving
	contribution to the
	three-point function outside the horizon.

	This potential problem can be seen most clearly after
	making the redefinition $\zeta \rightarrow \pi - f$
	under which the quadratic action transforms according to
	\begin{equation}
		S_2[\zeta] \rightarrow
		S_2[\pi]
			- 2 \int_\partial \d^3 x \; a^3 \dfrac{\varepsilon }{\cs^2}
				\dot{\pi} f
			- \int{d^3 x \, \d \tau \; f
				\dfrac{\delta S_2}{\delta \zeta}} .
		\label{eq:field-redefinition}
	\end{equation}
	The bulk term proportional to $\delta S_2 / \delta \zeta$ disappears
	by construction.
	After the transformation, the boundary term becomes
	\begin{equation}
		\begin{split}
		S_3 \supseteq \dfrac{1}{2} \int_{\partial} \d^3 x \; a^3
			\bigg\{ &
				- 18 H \pi^3
				+ \frac{2}{a^2 H} \bigg( 1 - \frac{\varepsilon}{\cs^2} \bigg)
					\pi (\grad \pi)^2
				- \frac{1}{2 a^4 H^3} \partial^2 \pi (\partial \pi)^2
				\\ & \mbox{}
				+ \frac{2 \varepsilon}{H \cs^4} \pi \dot{\pi}^2
				+ \frac{1}{aH} \partial^2 \pi (\partial \chi)^2
				- \frac{1}{a H} \partial_i \partial_j \pi
					\partial_i \chi \partial_j \chi
			\bigg\} ,
		\end{split}
		\label{eq:boundaryaction}
	\end{equation}
	in which $\chi$ is to be interpreted as a function of $\pi$
	[cf. Eq. (\ref{eq:chieq})].

	Eq.~\eqref{eq:boundaryaction} is not zero.
	To satisfy ourselves that it
	does not spoil the conclusions of
	Refs.~\cite{Maldacena:2002vr,Seery:2005wm,Chen:2006nt},
	we must determine how it contributes to
	the three-point correlation function.
	Before doing so, we briefly describe the \emph{in--in}
	formalism which is required. Readers familiar with this
	technique may wish to skip to \S\ref{subsec:boundaryterms}.

	\subsubsection{Schwinger's \textit{in--in} formalism}
	\label{subsec:inin}

	The correlation functions of interest are
	equal time expectation values taken in the state
	corresponding to the vacuum at past infinity.
	At later times, persistent
	nontrivial correlations exist owing to
	gravitational effects associated with the time-dependent background
	of de Sitter.
	
	Feynman's path integral computes the overlap between
	two states separated by a finite time interval,
	which is taken to infinity in scattering calculations.
	Schwinger obtained expectation values
	at a finite time $t_\ast$
	by inserting a complete
	set of states $\ket{i, t_f}$ at an arbitrary time $t_f \geq t_\ast$,
	\begin{equation}
		\bra{\instate} \op(t_\ast) \ket{\instate}
		= 
		\sum_i
		\braket{\instate}{i, t_f}
		\bra{i, t_f} \op(t_\ast) \ket{\instate} .
		\label{eq:schwinger-prototype}
	\end{equation}
	where
	$\ket{\mathrm{in}}$ is the ``in'' vacuum in which one wishes
	to compute the expectation value
	and $\mathcal{O}$ is an arbitrary local functional.
	Our choice of $t_f$ is irrelevant. Choosing
	a basis of energy eigenstates,
	$\ket{i, t_f} = \e{-\im E_i (t_f - t_\ast)} \ket{i, t_\ast}$
	where $E_i$ is the energy of the state $\ket{i}$.
	This phase cancels in~\eqref{eq:schwinger-prototype}.%
		\footnote{One could just as well insert a complete set of
		states at an arbitrary time $t < t_\ast$, but the resulting
		overlap would not be expressible in terms of a path integral.
		Had we retained $t_f > t_\ast$, the resulting contributions
		would have yielded only cancelling phases in~\eqref{eq:schwinger}.}
	Expressing each overlap as a Feynman path integral, we obtain
	\cite{Schwinger:1960qe, Bakshi:1962dv, Bakshi:1963bn, Keldysh:1964ud}
	\begin{equation}
		\bra{\instate} \op(t_\ast) \ket{\instate}
		=
		\int [ \d \phi_+ \, \d \phi_- ] \; \mathcal{O}(t_\ast) \,
		\e{\im S[\phi_+] - \im S[\phi_-]}
		\delta[
			\phi_+(t_\ast) - \phi_-(t_\ast)
		] .
		\label{eq:schwinger}
	\end{equation}
	The $\delta$-function restricts the domain of integration to
	fields $\phi_+$ and $\phi_-$ which agree at time $t_\ast$,
	but are unrestricted at past infinity.
	The requisite overlap of an arbitrary field configuration with the vacuum
	is obtained by deforming the contour of time integration.
	Cosmological applications of Schwinger's
	formulation were considered by Jordan \cite{Jordan:1986ug}
	and by Calzetta \& Hu \cite{Calzetta:1986ey},
	to which we refer for further details.
	Applications to inflationary correlation functions were
	discussed by Weinberg \cite{Weinberg:2005vy,Weinberg:2006ac}
	and have been reviewed elsewhere \cite{Seery:2007we,Koyama:2010xj}.
	
	\subsubsection{Removal of boundary terms}
	\label{subsec:boundaryterms}

	If boundary terms are present they appear in~\eqref{eq:schwinger}
	as part of the action $S$ with
	support at past infinity and $t = t_\ast$.
	The deformed contour of integration
	kills any contribution from past infinity,
	leaving a boundary term evaluated
	precisely at $t_\ast$,
	where the $\delta$-function constrains the fields to agree.
	Therefore,
	at least for $\op$ containing fields but not
	time derivatives of fields, any
	boundary operators which do not involve \emph{time} derivatives
	produce only a phase
	which cancels
	between the $+$ and $-$ contours.%
		\footnote{Spatial derivatives play no role in this argument,
		which therefore applies to the entire first line
		of~\eqref{eq:secondS3}.}
	This cancellation is a special property of the in--in formulation:
	it would not occur when calculating in--out amplitudes, for which
	the uncancelled boundary term would diverge near future infinity.
	Rapid oscillations of $\e{\im S}$ induced by this divergence
	would damp the path integral, yielding an
	amplitude for any scattering process which is formally zero.
	This can be understood as
	a consequence of the lack of an S-matrix in de Sitter
	space \cite{Witten:2001kn}.
	
	In virtue of this cancellation
	we may disregard the first three operators in the
	boundary part of~\eqref{eq:boundaryaction}.
	However, the $\delta$-function in~\eqref{eq:schwinger}
	in no way requires that time derivatives
	of the $+$ and $-$ fields are related at $t = t_\ast$.
	Therefore
	operators involving time derivatives need not
	reduce to cancelling phases.
	To understand their significance we subtract them using
	a further field redefinition.

	Inspection of Eq.~\eqref{eq:boundaryaction} shows that
	the time-derivative terms are of the schematic form
	$\pi \dot{\pi}^2$,
	and therefore lead to a field redefinition of the form
	$\zeta \rightarrow \pi + \pi \dot{\pi}$.
	We now argue that boundary operators with two or more time derivatives
	are irrelevant on superhorizon scales.
	Using the schematic field redefinition, the three-point correlation
	functions of $\zeta$ and $\pi$ are related by
	$\langle \zeta^3 \rangle
	= \langle \pi^3 \rangle + 3 \langle \pi^2 \rangle \langle \pi \dot{\pi}
	\rangle$
	plus higher-order contributions. However,
	Eq.~\eqref{eq:powerspectrum}
	implies
	$\langle \pi \dot{\pi} \rangle \rightarrow 0$
	on superhorizon scales,
	and therefore $\langle \zeta^3 \rangle = \langle \pi^3 \rangle$
	up to a decaying mode.
	This field redefinition will inevitably produce bulk
	terms proportional to
	$\delta S_2 / \delta \zeta$, but we have already seen that these
	do not contribute to Feynman diagrams at any order in
	perturbation theory.
	Therefore, on superhorizon scales, the correlation functions of the
	original and redefined fields agree.
	It follows that after subtraction by a field redefinition,
	the unwanted boundary terms in~\eqref{eq:boundaryaction}
	are irrelevant and can be ignored.
	Equivalently, one may confirm
	this conclusion by checking that operators with two or more
	time derivatives
	give convergent contributions to the boundary action at late times.
	Similar arguments apply for any higher-derivative combination.

	We conclude that the only non-negligible
	field redefinitions are of the schematic form
	$\zeta \rightarrow \pi + \pi^2$, which arise from boundary
	operators containing a \emph{single} time derivative.
	Eq.~\eqref{eq:boundaryaction} contains no such operators.
	However,
	had we made a further transformation
	to consolidate the $\eta$ dependence,
	a contribution of this form would have appeared in $f$
	and the boundary action.
	It can be checked that this single-derivative term would be
	correctly subtracted by Maldacena's procedure,
	and in this case the subtraction method applied
	in Refs.~\cite{Maldacena:2002vr,Seery:2005wm,Chen:2006nt}
	yields the correct answer. However,
	in theories where
	single-derivative terms already appear in the boundary component
	of~\eqref{eq:firstS3}
	there seems no guarantee it will continue to do so.

	\subsection{The bispectrum beyond lowest-order}
	\label{subsec:bispectrum}

	We define the bispectrum, $B$, in terms of the three-point function,
	\begin{equation}
		\langle
			\zeta(\vec{k}_1) \zeta(\vec{k}_2) \zeta(\vec{k}_3)
		\rangle
		=
		(2\pi)^3 \delta(\vec{k}_1+\vec{k}_2+\vec{k}_3) B({k}_1, {k}_2, {k}_3) .
		\label{eq:bispectrum-def}
	\end{equation}
	Observational constraints are typically quoted in terms of
	the reduced bispectrum, $\fNL$, which satisfies
	\cite{Komatsu:2001rj,Lyth:2005fi}
	\begin{equation}
		\fNL \equiv \dfrac{5}{6}
		\dfrac{B({k}_1, {k}_2, {k}_3)}{P(k_1)P(k_2) + P(k_1) P(k_3) +
		P(k_2) P(k_3)} .
		\label{eq:deffnl}
	\end{equation}
	Current constraints on $\fNL$ in the simplest inflationary models
	have been discussed by Senatore {\etal} \cite{Senatore:2009gt}.
	
	\para{Next-order corrections}%
	We are now in a position to compute the three-point function
	of~\eqref{eq:shiftaction}
	to next-order.
	The arguments of the previous section show that the
	boundary action and $\delta S_2 / \delta \zeta$ contributions
	can be discarded. Next-order terms arise in the remaining operators
	from a variety of sources.
	First, the coefficients of each vertex in~\eqref{eq:shiftaction}
	contain a mixture of lowest-order and next-order contributions.
	Second, the lowest-order part of each vertex is a time-dependent
	quantity which must be expanded around a reference time, as in
	\eqref{eq:time-dependence}, producing next-order terms.
	Third, there are next-order corrections to the propagator,
	obtained by expanding
	\eqref{eq:propagator}--\eqref{eq:wavefunction}
	in the neighbourhood of the chosen reference time.
	Propagator corrections appear on both the external
	and internal legs of the diagram.
	
	\para{Reference time, factorization scale}%
	To proceed, we must choose a reference point
	$\tau_\star$ around which to expand
	time-dependent quantities.
	Consider an arbitrary correlation function of fields $\zeta(\vec{k}_i)$.
	Whatever our choice of $\tau_\star$, the
	result~\eqref{eq:powerspectrum} for the power spectrum shows that
	we must expect logarithms of the form $\ln k_i/k_\star$
	which account for the difference in time of horizon exit between
	the mode $k_i$ and the reference wavenumber $k_\star$.
	To obtain a reliable answer we should attempt to minimize
	these logarithms.
	
	If all fields participating in the correlation function carry
	momenta of
	approximately common magnitude $k_i \sim k$---%
	described as the
	``equilateral limit''---the logarithm will
	be small when $k_\star \sim k$.
	In this case, na\"{\i}ve perturbation theory is not spoiled
	by the appearance of large logarithms.
	In the opposite limit,
	one or more fields have ``soft'' momenta of order
	$\ksoft$ which are much smaller than
	the remaining ``hard'' momenta of order $\khard$.
	When $\ksoft/\khard \rightarrow 0$ it will not be possible
	to find a choice of $k_\star$ which keeps
	all logarithms small
	and
	the calculation passes outside the validity of
	ordinary perturbation theory. We have encountered
	the problem
	of large logarithms which led to the
	renormalization
	group of Gell-Mann \& Low
	\cite{GellMann:1954fq}.
	
	In the study of inflationary correlation functions, configurations
	mixing hard and soft momenta with $\ksoft \ll \khard$ are
	referred to as ``squeezed,''
	and are of significance because they
	dominate the bispectrum for canonical inflation
	\cite{Maldacena:2002vr}.
	In principle one could study the behaviour of a correlation
	function as its momenta are squeezed
	by setting up an appropriate renormalization group analysis
	\cite{Seery:2009hs}.
	But this is more complicated
	than necessary.
	Maldacena argued that, as the momentum carried by one operator
	becomes soft, the three-point function would factorize:
	it can be written as
	a ``hard subprocess,'' described by the two-point correlation
	between the remaining hard operators
	on a background created by the soft operator
	\cite{Maldacena:2002vr}.
	Factorization of this kind is typical in the infrared dynamics
	of gauge theories such as QCD, where it plays an important role
	in extracting observational predictions.
	The various factorization theorems for QCD correlation functions
	have been
	comprehensively reviewed by Collins,
	Soper \& Sterman \cite{Collins:1989gx}.%
		\footnote{The background created by soft modes is typically
		described
		by some version of the DGLAP (or Altarelli--Parisi) equation.
		A similar phenomenon seems to occur in the inflationary case
		\cite{Seery:2009hs}.
		Equally, the separate universe method can be thought of
		as a factorization
		theorem for secular time-dependent logarithms
		$\sim \ln | k \cs \tau |$.
		The $\delta N$ rules which translate
		correlation functions of the field perturbations
		into correlation functions of $\zeta$ are an important special case.
		In this sense, factorization is as important in extracting
		observable quantities for inflation as it is for QCD.}
	Maldacena's argument was later generalized by
	Creminelli {\etal} \cite{Creminelli:2004yq}.
	The factorization property can be exhibited by an explicit decomposition
	of the field into hard and soft modes
	\cite{Ganc:2010ff,RenauxPetel:2010ty}.
	
	Because the squeezed limit can be described by Maldacena's
	method,
	the outcome of this discussion is that the reference scale
	should usually be chosen
	to minimize the logarithms when all momenta are
	comparable.
	In the remainder of this paper we quote results for arbitrary
	$k_\star$,
	but frequently adopt the symmetric choice
	$k_\star = k_1 + k_2 + k_3$ where numerical results are required.%
		\footnote{In the gauge theory language discussed above,
		the scale $k_\star$ can be thought of as the factorization
		scale. Operators carrying momentum $k \ll k_\star$
		should not be included as part of the hard subprocess,
		but factorized into the background.}
	Having done so, we will be formally unable
	to describe the squeezed limit.
	Nevertheless, because there is no other scale in the problem,
	our results must be compatible with the onset of factorization in
	appropriate circumstances---a property usually referred to as
	Maldacena's \emph{consistency relation}.
	We will see below that this constitutes a nontrivial check on the
	correctness of our calculation; see also
	Renaux-Petel \cite{RenauxPetel:2010ty} for a recent discussion of
	Maldacena's condition in the case of $P(X,\phi)$ models.

	\para{Operators}%
	To simplify our notation, we rewrite
	the cubic action~\eqref{eq:shiftaction} as
	\begin{equation}
		\begin{split}
			S_3
			=
			\int \d^3 x \, \d \tau \; a^2
			\bigg\{
				\frac{g_1}{a} \zeta'^3
				+ g_2 \zeta \zeta'^2
				+ g_3 \zeta (\partial \zeta)^2
				+ g_4 \zeta' \partial_j \zeta \partial_j \partial^{-2} \zeta'
				+ g_5 \partial^2 \zeta
					(\partial_j \partial^{-2} \zeta')
					(\partial_j \partial^{-2} \zeta')
			\bigg\} .
		\end{split}
		\label{eq:s3tau}
	\end{equation}
	In a $P(X,\phi)$ model
	the interaction vertices are
	\begin{equation}
		\begin{aligned}
		g_1 &
		=
		\dfrac{\varepsilon}{H \cs^4}
			\left(
				1-\cs^2-2 \dfrac{\lambdaP \cs^2}{\SigmaP}
			\right)
		&
		g_2 &
		=
		\dfrac{\varepsilon}{\cs^4}
			\left[
				-3(1-\cs^2)+\varepsilon-\eta
			\right]
		&
		&
		\\
		g_3 &
		=
		\dfrac{\varepsilon}{\cs^2}
			\left[
				(1-\cs^2) +\varepsilon+ \eta -2s
			\right]
		&
		g_4 &
		=
		\dfrac{\varepsilon^2}{2 \cs^4}(\varepsilon-4)
		&
		g_5
		&
		=
		\dfrac{\varepsilon^3}{4 \cs^4} ,
		\end{aligned}
		\quad\Bigg\}
		\label{eq:gi}
	\end{equation}
	but our calculation will apply for arbitrary $g_i$.
	Although $\zeta$ is dimensionless, it is helpful for power-counting
	purposes to think of it as a field of
	engineering dimension $[\text{mass}]$,
	obtained after division by the Hubble rate $H$.
	In this counting scheme, the $\zeta'^3$ operator is dimension-6, whereas
	the remaining four operators are dimension-5.
	At low energies one would na\"{\i}vely expect
	the dimension-6 operator to be irrelevant in comparison to those
	of dimension-5.
	However, the dimension-5 operators are suppressed by the scale $H$
	making all contributions equally relevant.
	This manifests itself as an extra power
	of $H$ in the denominator of $g_1$.

	The vertex factors $g_i$ are themselves time-dependent background
	quantities.
	We define slow-variation parameters $h_i$ which measure their
	rate of change per e-fold,
	\begin{equation}
		h_i \equiv \dfrac{\dot{g}_i}{H g_i} ,
	\end{equation}
	and take these to be $\Or(\varepsilon)$ in the slow-variation approximation.

	\subsection{Three-point correlations}
	\label{subsec:3pf}

	We use these conventions to compute the next-order bispectra
	for each operator in~\eqref{eq:s3tau}.
	The resulting three-point functions
	are complicated objects, and when quoting their values
	it is helpful to adopt an organizing principle.
	We divide the possible contributions into broadly similar
	classes.
	In the first class, labelled `$a$,' we collect
	(i) the lowest-order bispectrum;
	(ii) effects arising
	from corrections to the wavefunctions associated with external
	lines; and (iii) effects arising from the vertex corrections.
	In the second class, labelled `$b$,' we restrict attention to
	effects arising from wavefunctions associated with internal
	lines.
	These are qualitatively different in character because
	wavefunctions associated with the internal lines
	are integrated over time.
	Adapting terminology from particle physics,
	we occasionally refer to the lowest-order bispectrum as the ``LO''
	part, and the next-order piece as the ``NLO'' part.
	
	\para{Large logarithms, infrared singularities}%
	The computation of inflationary $n$-point functions has been 
	reviewed by Chen~\cite{Chen:2010xk} and Koyama~\cite{Koyama:2010xj}. 
	At least three species of large logarithms appear, disrupting 
	ordinary perturbation theory. We carefully track the contribution from 
	each species. The most familiar types---already encountered
	in the two-point function---%
	measure time- and
	scale-dependence. A third type of large logarithm
	is associated with the far infrared limit
	$\ksoft / \khard \rightarrow 0$
	discussed in \S\ref{subsec:bispectrum}.
	This is Maldacena's ``squeezed'' limit, discussed
	in \S\ref{subsec:bispectrum}, in which the behaviour of the
	three-point function obeys a factorization principle.
	We will show that the various large logarithms
	arrange themselves in such a way that they can be absorbed into the
	scale-dependence of background quantities.
	
	Time-dependent logarithms appear
	after expanding background quantities near a fixed reference
	scale, as in~\eqref{eq:time-dependence},
	where at conformal time $\tau$ we have
	$N_\star = \ln | k_\star \cs \tau |$.
	In \S\ref{subsec:second} we explained
	that the correlation functions of $\zeta$ are expected to become
	time-independent outside the horizon.
	Therefore one should expect all $\ln \tau$ dependence
	to disappear.
	Some $N_\star$-type
	logarithms cancel among themselves
	but others cancel with time-dependent logarithms
	arising from wavefunction corrections associated with internal lines.
	The internal lines are aware only of
	the intrinsic geometrical scale $k_t \equiv k_1 + k_2 + k_3$
	and cannot depend on the
	arbitrary reference scale $k_\star$,
	so the outcome of such a cancellation
	leaves a residue of the form $\ln k_t/k_\star$.
	These are scaling logarithms,
	entirely analogous to the logarithm of~\eqref{eq:powerspectrum},
	describing variation of the three-point function
	with the geometrical scale $k_t$.
	Scale logarithms can also occur in the form $\ln k_i / k_\star$.
		
	The
	third species of logarithm takes the form $\ln k_i / k_t$.
	Each side of the triangle must scale linearly with
	the perimeter, so
	despite appearances
	these have no dependence on $k_t$---they are unaffected by rigid
	rescalings of the momentum triangle
	(cf. Eqs.~\eqref{eq:fergusson-shellard-a}--\eqref{eq:fergusson-shellard-c}
	below).
	We describe them as `purely' shape dependent.
	The `pure' shape logarithms become large in the squeezed
	limit $k_i / k_t \rightarrow 0$.

	\para{$a$-type bispectrum}%
	Collecting the $a$-type contributions to the bispectrum, we find
	\begin{equation}
		\begin{split}
		B^{a} = \mbox{} &
			\frac{H_{\star}^4}{2^4 \csstar^6}
			\frac{g_{i\star}}{\zterm_\star^3}
			\frac{T^{a}(k_1)}{k_t^2\prod_i{k_i^3}}
			\bigg\{
				-\varpi_{\star} U^{a}(k_1)
					\ln\frac{k_1k_2k_3}{k_\star^3}
				\\
				& \hspace{3.2cm} \mbox{}
					+ 2V^{a}(k_1)\varepsilon_{\star}\ln\frac{k_t}{k_\star}
					+ W^{a}(k_1) h_{i\star}\ln\frac{k_t}{k_\star}
				\\
				& \hspace{3.2cm} \mbox{}
				+ X^{a}(k_1) (1 + 3E_{\star})
				+ 2Y^{a}(k_1)\varepsilon_{\star}
				+ Z^{a}(k_1) h_{i\star}	
			\bigg\}
		\\
		& \mbox{} + \text{cyclic permutations} .
		\end{split}
		\label{eq:Ba}
	\end{equation}
	The coefficients $T^{a}(k_1)$, $U^{a}(k_1)$, $V^{a}(k_1)$,
	$W^{a}(k_1)$, $X^{a}(k_1)$, $Y^{a}(k_1)$ and $Z^{a}(k_1)$ are
	functions of all three momenta $\vect{k}_i$
	and are symmetric under the exchange $k_2 \leftrightarrow k_3$.
	We adopt the convention, used through the remainder of this paper,
	of writing only the asymmetric momentum explicitly.
	The notation `cyclic permutations' denotes
	addition of the preceding term under
	cyclic permutations of the $\vect{k}_i$.
	The result is symmetric under interchange of any two momenta.	

	We give explicit expressions for the coefficient functions in
	Table~\ref{table:Blead}. The quantity
	$E$ is a combination of slow-variation parameters,
	$E = \varpi (2- \EulerGamma -\ln 2) - \varepsilon - s$,
	and
	also appears in the power
	spectrum~\eqref{eq:powerspectrum}.
	The term proportional to $X^a(k_1)$ includes the
	entire lowest-order
	bispectrum.

	\begin{table}[htp]

	\heavyrulewidth=.08em
	\lightrulewidth=.05em
	\cmidrulewidth=.03em
	\belowrulesep=.65ex
	\belowbottomsep=0pt
	\aboverulesep=.4ex
	\abovetopsep=0pt
	\cmidrulesep=\doublerulesep
	\cmidrulekern=.5em
	\defaultaddspace=.5em
	\renewcommand{\arraystretch}{1.6}
	\begin{center}
		\small
		\begin{tabular}{QqQqQq}

			\toprule
			
			&
			\multicolumn{5}{c}{operator}
			\\
			\cmidrule(l){2-6}
		
		 	& 
		 	\multicolumn{1}{c}{$\zeta'^3$} &
		 	\multicolumn{1}{c}{$\zeta \zeta'^2$} &
		 	\multicolumn{1}{c}{$\zeta ( \partial \zeta)^2$} &
		 	\multicolumn{1}{c}{$\zeta' \partial_j \zeta \partial_j \partial^{-2} \zeta'$} &
		 	\multicolumn{1}{c}{$\partial^2 \zeta (\partial_j \partial^{-2} \zeta')^2$}
		 	\\
			\midrule

			T^{a}(k_1) &
			6H_{\star}\frac{k_1^2k_2^2k_3^2}{k_t} &
			k_2^2k_3^2(k_1+k_t) & 
			\frac{k_t}{\csstar^4}(\vec{k}_2 \cdot \vec{k}_3) &  
			\frac{k_1^2}{2}(\vec{k}_2 \cdot \vec{k}_3) &
			\begin{array}{l}
				k_1^2(\vec{k}_2 \cdot \vec{k}_3) \\
				\; \mbox{} \times (k_1 + k_t)
			\end{array}
			\\[2mm]

			\cmidrule{2-6}

			U^{a}(k_1) & 
			1 &
			1 & 
			\csstar^2\left(K^2 - k_t^2 + \frac{k_1 k_2 k_3}{k_t} \right)	& 
			3k_t-k_1 &  
			1
			\\[2mm]

			\cmidrule{2-6}

			V^{a}(k_1) &
			1 &
			1 &
			\begin{array}{l}
				\displaystyle -\csstar^2
				\bigg(
					k_t^2-K^2-\frac{k_1k_2k_3}{k_t}
				\bigg)
			\end{array} &
			3k_t-k_1 &
			1
			\\[2mm]

			\cmidrule{2-6}
			
			W^{a}(k_1) &
			1 &
			1 &
			3\csstar^2\left(K^2 - k_t^2 + \frac{k_1 k_2 k_3}{k_t} \right) &
			3k_t-k_1 &
			1
			\\[2mm]
			
			\cmidrule{2-6}
			
			X^{a}(k_1) &	 
			1 &
			1 & 
			\csstar^2\left(K^2 - k_t^2 + \frac{k_1 k_2 k_3}{k_t} \right) &
			3k_t-k_1 &
			1
			\\[2mm]
 
			\cmidrule{2-6}

			Y^{a}(k_1) &
			\EulerGamma -\frac{1}{2} &
			\EulerGamma + \frac{k_t}{k_1+k_t} &
			\begin{array}{l}
				\displaystyle \csstar^2\bigg[
					K^2 \\
				\displaystyle \; \mbox{} -\EulerGamma
				\bigg(
					k_t^2-K^2-\frac{k_1k_2k_3}{k_t}
				\bigg)
				\bigg]
			\end{array} &
			\begin{array}{l}
				\displaystyle (3k_t-k_1)\EulerGamma \\
				\displaystyle \; \mbox{} + 2k_t
			\end{array}&
			\EulerGamma + \frac{k_t}{k_1+k_t}
			\\[2mm]

			\cmidrule{2-6}

			Z^{a}(k_1) &
			\EulerGamma-\frac{3}{2} &
			\EulerGamma - \frac{k_1}{k_1+k_t} &
			\begin{array}{l}
				\displaystyle 3\csstar^2\bigg[
					\EulerGamma K^2 \\
				\displaystyle \; \mbox{} +
				(1-\EulerGamma)
				\bigg(
					k_t^2-\frac{k_1k_2k_3}{k_t}
				\bigg)
				\bigg]
			\end{array} &
			\begin{array}{l}
				\displaystyle (3k_t-k_1) \EulerGamma \\
				\displaystyle \; \mbox{} + k_1 - k_t
			\end{array}
			&
			\EulerGamma - \frac{k_1}{k_1+k_t}
			\\[2mm]
			
 			\bottomrule
	
		\end{tabular}
	\end{center}
	\caption{Coefficients of the leading order bispectrum.
			 $K^2=k_1k_2+k_1k_3+k_2k_3$.
	\label{table:Blead}}
	\end{table}	

	\para{$b$-type bispectrum}%
	The $b$-type bispectrum must be added to the
	$a$-type terms.
	It has no lowest-order contributions,
	and can be written
	\begin{equation}
		\begin{split}
			B^{b} = \mbox{} &
				\frac{H_{\star}^4}{2^4 \csstar^6}
				\frac{g_{i\star}}{\zterm_\star^3}
				\frac{T^{b}(k_1)}{k_t^2\prod_i{k_i^3}}	
				\bigg\{
					\varpi_{\star}
					\sum^3_{i=1}
						\left(
							k_t U^{b}(k_i)J_0(k_i)
							+ V^{b}_i(k_1)J_1(k_i)
							+ k_t^2 W^{b}(k_i) \ln \frac{2k_i}{k_\star}
						\right)
					\\
					& \hspace{3.2cm} \mbox{}
					+ \varpi_{\star}
						\left(
							X^{b} J_2(k_1)
							+ Y^{b} k_t^3 \ln\frac{k_t}{k_\star}
						\right)
					+ Z^{b}
					+ \csstar k_t^2 \Re\left( \mathcal{J}_\star \right)
				\bigg\}
				\\
				& \mbox{} + \text{cyclic permutations} .
		\end{split}
		\label{eq:Bb}
	\end{equation}
	The same convention applies to the arguments of
	the coefficient functions
	$T^{b}(k_1)$, $U^{b}(k_i)$, $V^{b}_i(k_1)$, $W^{b}(k_i)$, $X^{b}$,
	$Y^{b}$ and $Z^{b}$.
	We give explicit expressions in Table~\ref{table:Bsub}.
	
	Eq.~\eqref{eq:Bb} depends on three logarithmic functions $J_i$
	(which are \emph{not} Bessel functions)
	defined by
	\begin{subequations}
		\begin{align}
			\vartheta_i J_0(k_i) & =
				\ln \frac{2k_i}{k_t} , \\
			\vartheta_i^2 J_1(k_i) & =
				\vartheta_i + \ln \frac{2k_i}{k_t} , \\
			\vartheta_i^3 J_2(k_i) & =
				\vartheta_i (2+\vartheta_i) + 2\ln \frac{2k_i}{k_t} ,
		\end{align}
		\label{eq:clare-integrals}
	\end{subequations}
	where $\vartheta_i =1-2k_i/k_t$.
	These exhaust the `pure' shape logarithms of the form $\ln k_i / k_t$,
	discussed above, which appear only in the $J_i$.
	There is an obvious logarithmic divergence in the squeezed limit
	$k_i \rightarrow 0$, which we will show to be responsible for
	factorization of the correlation function.
	There is potentially
	a power-law divergence in the limit
	$k_t \rightarrow 2 k_i$. This is \emph{also} a squeezed limit---in which
	the $i^{\mathrm{th}}$ side stays fixed while a different momentum
	goes to zero. In this limit $\vartheta_i \rightarrow 0$,
	making
	the $J_i$ na\"{\i}vely divergent.
	If present, such power-law divergences would be puzzling.
	However, it can be checked that---in combination
	with the logarithm---each $J_i$ is finite.
	This infrared-safe behaviour relies on
	a resummation procedure which is
	discussed in more detail in Appendix~\ref{appendix:a}.

	The function $\mathcal{J}_\star$ satisfies
	\begin{equation}
		\begin{split}
			\mathcal{J}_\star
			=
			\frac{1}{k_t \csstar}
			\bigg[
				&
				\gamma_0
				- \frac{\gamma_1 + \delta_1}{k_t}
				- \dfrac{2\gamma_2+3\delta_2}{k_t^2}
				+ \dfrac{6\gamma_3+11\delta_3}{k_t^3}
				+ \dfrac{24\gamma_4+50\delta_4}{k_t^4}
				\\
				& \mbox{}
				- \left(
					\EulerGamma
					+ \ln \frac{k_t}{k_\star}
					+ \im \dfrac{\pi}{2}
				\right)
				\left(
					\delta_0
					- \dfrac{\delta_1}{k_t}
					- 2 \dfrac{\delta_2}{k_t^2}
					+ 6 \dfrac{\delta_3}{k_t^3}
					+ 24\dfrac{\delta_4}{k_t^4}
				\right)
			\bigg] .
		\end{split}
	\end{equation}
	This function
	is discussed in Appendix \ref{appendix:b}.
	The coefficients $\gamma_0$, $\gamma_1$, $\gamma_2$, $\gamma_3$,
	$\delta_0$, $\delta_1$, $\delta_2$ and $\delta_3$
	depend on the operator under consideration.
	We quote values for each operator in Table~\ref{table:Jgamma}.

	\begin{table}[htp]

	\heavyrulewidth=.08em
	\lightrulewidth=.05em
	\cmidrulewidth=.03em
	\belowrulesep=.65ex
	\belowbottomsep=0pt
	\aboverulesep=.4ex
	\abovetopsep=0pt
	\cmidrulesep=\doublerulesep
	\cmidrulekern=.5em
	\defaultaddspace=.5em
	\renewcommand{\arraystretch}{1.6}
	\begin{center}
		\small
		\begin{tabular}{QqQqQq}

			\toprule
			
			&
			\multicolumn{5}{c}{operator}
			\\
			\cmidrule(l){2-6}
		
		 	& 
		 	\multicolumn{1}{c}{$\zeta'^3$} &
		 	\multicolumn{1}{c}{$\zeta \zeta'^2$} &
		 	\multicolumn{1}{c}{$\zeta ( \partial \zeta)^2$} &
		 	\multicolumn{1}{c}{$\zeta' \partial_j \zeta \partial_j \partial^{-2} \zeta'$} &
		 	\multicolumn{1}{c}{$\partial^2 \zeta (\partial_j \partial^{-2} \zeta')^2$}
		 	\\
			\midrule

			T^{b}(k_1)	& 
			-\frac{3}{2}H_{\star}\csstar^2k_1^2k_2^2k_3^2 &
			k_1^2 & 
			\frac{1}{\csstar^2}(\vec{k}_2 \cdot \vec{k}_3) & 
			k_1^2(\vec{k}_2 \cdot \vec{k}_3) &  
			k_1^2(\vec{k}_2 \cdot \vec{k}_3)
	  		\\[2mm]

			\cmidrule{2-6}

			U^{b}(k_i) &
			&
			-1 & 
			2k_ik_t-2k_i^2 -K^2& 
			\csstar k_t & 
			\csstar k_t
	 		\\[2mm]

			\cmidrule{2-6}

			V^{b}_1(k_1) &
			&
			k_1 & 
			k_1k_2k_3 & 
			-\frac{1}{2}(k_2+k_3) & 
			k_1
		 	\\[2mm]

			\cmidrule{2-6}

			V^{b}_2(k_1) &
			&
			-k_1 & 
			k_1k_2k_3 & 
			\frac{1}{2}(k_2-k_3) & 
			-k_1
			\\[2mm]

			\cmidrule{2-6}

			V^{b}_3(k_2) & 
			&
			-k_1 & 
			k_1k_2k_3 &  
			\frac{1}{2}(k_3-k_2) & 
			-k_1
		 	\\[2mm]
	  
	  		\cmidrule{2-6}
			
			W^{b}(k_i) &
			&
			&
			k_t-2k_i & 
			& 
			\\[2mm]
	  
			\cmidrule{2-6}

  			X^{b} & 
			\frac{1}{\csstar^2 k_t} & 
  			& 
  			&	
  			&
			\\[2mm]

			\cmidrule{2-6}
			
			Y^{(b)} &
			&
			&
			2 & 
			& 
			\\[2mm]

			\cmidrule{2-6}

			Z^{b} &
			&
			&
			k_t^3[2\varpi_{1\star}-3\Re(\mu_{0\star})
				+3\gamma_E\varpi_{1\star}] & 
			& 
			\\[2mm]
						
 			\bottomrule
	
		\end{tabular}
	\end{center}
	\caption{Coefficients of the subleading corrections to the bispectrum.
			 $K^2=k_1k_2+k_1k_3+k_2k_3$.
	\label{table:Bsub}}
	\end{table}	

	The operators $\zeta \zeta'^2$,
	$\zeta' \partial_j \zeta \partial^{-2} \partial_j \zeta$
	and $\partial^2 \zeta (\partial^{-2} \partial_j \zeta')^2$
	are all dimension-5, and
	differ only in the arrangement of spatial gradients.
	For arbitrary shapes their three-point functions will not coincide,
	but for equilateral triangles the arrangement of gradients
	is irrelevant
	and the resulting $\fNL$ should agree.
	This will represent a minimal check of our expressions.
	We will carry out further checks in
	\S\ref{sec:running}
	and \S\S\ref{sec:vanilla}--\ref{sec:noncanonical}.

	\begin{table}

	\heavyrulewidth=.08em
	\lightrulewidth=.05em
	\cmidrulewidth=.03em
	\belowrulesep=.65ex
	\belowbottomsep=0pt
	\aboverulesep=.4ex
	\abovetopsep=0pt
	\cmidrulesep=\doublerulesep
	\cmidrulekern=.5em
	\defaultaddspace=.5em
	\renewcommand{\arraystretch}{1.6}
	\begin{center}
		\small
		\begin{tabular}{QqQqQq}

			\toprule
			
			&
			\multicolumn{5}{c}{operator}
			\\
			\cmidrule(l){2-6}
		
		 	& 
		 	\multicolumn{1}{c}{$\zeta'^3$} &
		 	\multicolumn{1}{c}{$\zeta \zeta'^2$} &
		 	\multicolumn{1}{c}{$\zeta ( \partial \zeta)^2$} &
		 	\multicolumn{1}{c}{$\zeta' \partial_j \zeta \partial_j \partial^{-2} \zeta'$} &
		 	\multicolumn{1}{c}{$\partial^2 \zeta (\partial_j \partial^{-2} \zeta')^2$} \\
			\midrule

			\gamma_0 &
			&
			\mu_{0\star}+2s_{\star} -2\mu_{1\star} & 
			\begin{array}{l}
				\displaystyle s_{\star}k_1^2+k_1 \mu_{1\star} (k_2+k_3) \\
				\displaystyle \; \mbox{} -\mu_{0\star}k_2 k_3
			\end{array} &
			\mu_{0\star}+2s_{\star} -2\mu_{1\star} & 
			\mu_{0\star}+2s_{\star} -2\mu_{1\star}
			\\[2mm]

			\cmidrule{2-6}

			\gamma_1 & 
			& 
			\begin{array}{l}
				\displaystyle 3k_1\mu_{1\star} +k_t s_{\star} \\
				\displaystyle \; \mbox{} - 3k_1 s_{\star}
			\end{array} &
			\begin{array}{l}
				\displaystyle -s_{\star} k_{1}^2 (k_2+k_3) \\
				\displaystyle \; \mbox{} - \mu_{1\star}k_1 k_2 k_3
			\end{array} &
			\begin{array}{l}
				\displaystyle k_t s_{\star}-3k_2 s_{\star} \\
				\displaystyle \; \mbox{} +3 \mu_{1\star} k_2
			\end{array} & 
			\begin{array}{l}
				\displaystyle k_t s_{\star}-3k_1 s_{\star} \\
				\displaystyle \; \mbox{} +3 \mu_{1\star} k_1
			\end{array}
			\\[2mm]

			\cmidrule{2-6}

			\gamma_2 & 
			\frac{s_{\star}-\mu_{1 \star}}{\csstar^2} &
			k_1 k_t s_{\star} &	
			-s_{\star} k_1^2 k_2 k_3 & 
			k_2 k_t s_{\star} &	
			k_1 k_t s_{\star} 
			\\[2mm]

			\cmidrule{2-6}

			\gamma_3 &	
			k_1 \frac{s_{\star}}{\csstar^2} &
			&	 
			& 
			& 
			
			\\[2mm]

			\cmidrule{2-6}

			\delta_0 & 
			& 
			3\varpi_{1\star}-4s_{\star} & 
			-s_{\star} k_1^2-K^2\varpi_{1\star} &	
			3\varpi_{1\star}-4s_{\star} & 
			3\varpi_{1\star}-4s_{\star}
			\\[2mm]
	
			\cmidrule{2-6}

			\delta_1 &
			&
			\begin{array}{l}
				\displaystyle -k_t s_{\star} +5k_1 s_{\star} \\
				\displaystyle \; \mbox{} -3k_1 \varpi_{1\star}
			\end{array} &
			\begin{array}{l}
				\displaystyle s_{\star} k_1^2(k_2+k_3) \\
				\displaystyle \; \mbox{} +\varpi_{1\star} k_1k_2k_3 
			\end{array} & 
			\begin{array}{l}
				\displaystyle -k_t s_{\star} +5k_2 s_{\star} \\
				\displaystyle \; \mbox{} -3k_2 \varpi_{1\star}
			\end{array} & 
			\begin{array}{l}
				\displaystyle -k_t s_{\star} +5k_1 s_{\star} \\
				\displaystyle \; \mbox{} -3k_1 \varpi_{1\star}
			\end{array}
			\\[2mm]

			\cmidrule{2-6}

			\delta_2 & 
			\frac{\varpi_{1\star}-2s_{ \star}}{\csstar^2} & 
			-s_{\star} k_1 k_t	&  
			s_{\star} k_1^2 k_2k_3 &	 
			-k_2 k_t s_{\star} & 
			-k_1 k_t s_{\star} 
			\\[2mm]

			\cmidrule{2-6}

			\delta_3 & 
			-k_1 \frac{s_{\star}}{\csstar^2} &		 
			& 
			&	  
			& 
			
			\\

 			\bottomrule
	
		\end{tabular}
	\end{center}
	\caption{Coefficients appearing in the function $\mathcal{J}$
	for each operators. Note that the $\gamma_i$ contain complex numbers.
	The imaginary part is cancelled on addition of the $+$ and $-$
	Feynman diagrams, and
	only the real part of these coefficients contribute.
	In an intermediate step for the
	three-point function of
	$\zeta (\partial \zeta)^2$,
	the cancellation of power-law divergences
	in the conformal time $\tau$
	(which is required by Weinberg's theorem \cite{Weinberg:2005vy})
	depends on a real contribution generated
	from the product of two imaginary terms.
	\label{table:Jgamma}}
	\end{table}	
	
	A subset of these terms were calculated by Chen {\etal}
	\cite{Chen:2006nt}.
	Our calculations exhibit two principal differences.
	First, Chen {\etal} worked to fixed order in slow-roll quantities,
	keeping terms of $\Or(\epsilon)$ only. In a model where $\cs \ll 1$
	this gives the next-order corrections.
	However, in a model where $\cs \sim 1$ the leading terms are
	themselves $\Or(\epsilon)$
	and the formulae of Chen {\etal} reduce to these
	leading contributions.
	In our calculation, we work uniformly to next-order rather than
	a fixed order in powers of $\epsilon$.
	When $\cs \ll 1$ our next-order corrections are $\Or(\epsilon)$,
	and we have verified that they agree with those computed in
	Ref.~\cite{Chen:2006nt}.
	(We give more details of the relation between our calculations
	in Appendix~\ref{appendix:compare}.)
	When $\cs \sim 1$ the next-order corrections are $\Or(\epsilon^2)$.
	These were not included in the formulae of Ref.~\cite{Chen:2006nt}.

	Second, we retain a floating reference scale $k_\star$.
	In Appendix~B%
		\footnote{Slightly different conventions for $k_\star$ were used
		elsewhere in Ref.~\cite{Chen:2006nt}.}
	of Ref.~\cite{Chen:2006nt}
	this was chosen to be $k_\star = k_t$.
	Retaining this scale allows us to extract the scale- and
	shape-dependence of
	$\fNL$ (\S\S\ref{sec:shapes}--\ref{sec:running}). 

	\subsection{Formulae for $\fNL$}
	\label{subsec:fnl}

	The individual bispectra, with their detailed shape-dependence,
	are the principal observable objects.
	However, for simple model comparisons it is helpful to have
	an explicit expression for the nonlinearity parameter $\fNL$
	defined in Eq.~\eqref{eq:deffnl}.
	Accounting for scale-dependent logarithms present in the power
	spectrum, one finds
	\begin{equation}
		\fNL
		=
		\dfrac{5}{6}
		\left(
			\dfrac{4 z_{\star} \csstar^3}{H_{\star}^2}
		\right)^2
		\frac{B(k_1, k_2, k_3) \prod_i k_i^3}
			{\sum_i k_i^3
			\big(
				1
				+ 4E_{\star}
				- 2\varpi_{1\star}
				\ln \big\{
					k_i^{-1} k_t^{-2} \prod_j k_j
				\big\}
			\big)}
		.
		\label{eq:fnl-logarithms}
	\end{equation} 
	This expression is to be expanded uniformly to
	$\Or(\varepsilon)$ in slow-variation parameters.
	
	There is
	another reason to study $\fNL$.
	We have explained that
	large logarithms of the form
	$\ln k_i / k_\star$ or $\ln k_i / k_t$
	are to be expected
	in the squeezed limit $k_i \rightarrow 0$,
	describing variation of the bispectrum with shape.
	The power spectrum $P(k)$ contains similar large logarithms.
	Since copies of the power spectrum must
	be factored out to obtain $\fNL$, one may expect it to be
	more regular in the squeezed limit.
	Indeed, a stronger statement is possible.
	Partitioning the momenta into a single soft mode of order $\ksoft$
	and two hard modes of order $\khard$,
	Maldacena's consistency condition requires
	\cite{Maldacena:2002vr}
	\begin{equation}
		\fNL \rightarrow - \frac{5}{12} ( n_s - 1 )|_{\khard},
		\label{eq:maldacena-condition}
	\end{equation}
	as $\ksoft \rightarrow 0$,
	where the right-hand side is to be evaluated at horizon
	exit for the mode of wavenumber $\khard$.
	Eq.~\eqref{eq:maldacena-condition} is finite
	and independent of any logarithms associated with the
	limit $\ksoft \rightarrow 0$,
	which is why this behaviour is described as factorization.
	It imposes the nontrivial requirement that all large logarithms
	can be absorbed
	into $P(\ksoft)$.
	Such logarithms are subtracted by the
	denominator of~\eqref{eq:fnl-logarithms},
	making $\fNL$ finite.
	
	For each operator $i$, we write the corresponding $\fNL$
	as ${\fNL}_i$
	and quote it in the form
	\begin{equation}
		{\fNL}_i = \left. \fNL \right|_{i0} \big[
			1
			+ \kappa_{h|i} h_{i\star}
			+ \kappa_{v|i} v_{\star}
			+ \kappa_{s|i} s_{\star}
			+ \kappa_{\varepsilon|i} \varepsilon_{\star}
		\big] .
		\label{eq:next-order-fnl}
	\end{equation}
	In Tables~\ref{table:fNLequilateral}
	and~\ref{table:fNLsqueezed}
	we give explicit expressions for the coefficient functions
	$\left. \fNL \right|_{i0}$
	and $\kappa_i$
	in the case of equilateral%
		\footnote{The quoted quantity is $\fNL(k,k,k)$, which is
		not the
		same object as $\fNL^{\mathrm{equi}}$ for which constraints
		are typically derived from data
		\cite{Senatore:2009gt,Komatsu:2010fb}.
		To obtain $\fNL^{\mathrm{equi}}$, one should
		take an appropriately normalized inner product
		(see~\S\ref{sec:shapes} for a simple example)
		between the full next-order
		bispectrum and the equilateral template
		\cite{Creminelli:2005hu,Creminelli:2006rz}.
		At this level of precision, it may even be desirable
		to include experiment-dependent information in the
		inner product.
		\label{footnote:equilateral}}
	and squeezed triangles.
 	Table~\ref{table:fNLsqueezed} confirms
 	that~\eqref{eq:next-order-fnl} is finite in the squeezed limit,
 	as required.
 	In the equilateral case,
	we find that the operators $\zeta \zeta'^2$,
	$\zeta' \partial_j \zeta \partial^{-2} \partial_j \zeta$
	and
	$\partial^2 \zeta ( \partial^{-2} \partial_j \zeta' )^2$
	agree, for the reasons explained above.
 
	\begin{sidewaystable}
	\small
	\heavyrulewidth=.08em
	\lightrulewidth=.05em
	\cmidrulewidth=.03em
	\belowrulesep=.65ex
	\belowbottomsep=0pt
	\aboverulesep=.4ex
	\abovetopsep=0pt
	\cmidrulesep=\doublerulesep
	\cmidrulekern=.5em
	\defaultaddspace=.5em
	\renewcommand{\arraystretch}{1.6}
	
	\sbox{\tableA}{%
		\begin{tabular}{QqQqQq}
		   	\toprule

			&
			\multicolumn{5}{c}{operator}
			\\

			\cmidrule(l){2-6}

		 	& 
		 	\multicolumn{1}{c}{$\zeta'^3$} &
		 	\multicolumn{1}{c}{$\zeta \zeta'^2$} &
		 	\multicolumn{1}{c}{$\zeta ( \partial \zeta)^2$} &		 	\multicolumn{1}{c}{$\zeta' \partial_j \zeta \partial_j \partial^{-2} \zeta'$} &
		 	\multicolumn{1}{c}{$\partial^2 \zeta (\partial_j \partial^{-2} \zeta')^2$}
		 	\\

			\midrule

			\left. \fNL \right|_{i0} &

			\frac{5}{81} \frac{g_{1\star} H_\star}{\zterm_\star} &
			\frac{10}{27} \frac{g_{2\star}}{\zterm_\star} & 
			\frac{85}{108} \frac{g_{3\star}}{\zterm_\star \csstar^2} & 			-\frac{5}{27} \frac{g_{4\star}}{\zterm_\star} &
			\frac{10}{27} \frac{g_{5\star}}{\zterm_\star}
			\\[2mm]

			\cmidrule{2-6}

			\multirow{2}*{$\kappa_{h|i}$} &
			\EulerGamma - \frac{3}{2} + \ln \frac{3k}{k_\star} &
			\EulerGamma - \frac{1}{4} + \ln \frac{3k}{k_\star} &
			\EulerGamma - \frac{26}{17} + \ln \frac{3k}{k_\star} &
			\EulerGamma - \frac{1}{4} + \ln \frac{3k}{k_\star} &
			\EulerGamma - \frac{1}{4} + \ln \frac{3k}{k_\star}
			\\

			&
			-0.922784\tmark{a} &
			0.327216\tmark{a} &
			-0.952196\tmark{a} &
			0.327216\tmark{a} &
			0.327216\tmark{a}
			\\

			\cmidrule{2-6}

			\multirow{2}*{$\kappa_{v|i}$} &
			-\EulerGamma - \frac{29}{2} + 78\omega - \ln \frac{2k}{k_\star} &
			-\EulerGamma - \frac{1}{4} + 6\omega - \ln \frac{2k}{k_\star} &
			-\EulerGamma + \frac{22}{17} + \frac{30}{17}\omega
				- \ln \frac{2k}{k_\star} &
			-\EulerGamma - \frac{1}{4} + 6\omega - \ln \frac{2k}{k_\star} &
			-\EulerGamma - \frac{1}{4} + 6\omega - \ln \frac{2k}{k_\star}
			\\

			&
			1.14139\tmark{a} &
			0.794645\tmark{a} &
			1.48013\tmark{a} &
			0.794645\tmark{a} &
			0.794645\tmark{a}
			\\

 			\cmidrule{2-6}

			\multirow{2}*{$\kappa_{s|i}$} &
			-49 + 240 \omega &
			-\frac{3}{2} + 24 \omega &
			-2\EulerGamma + \frac{40}{17} + \frac{124}{17}\omega -
				2\ln\frac{2k}{k_\star} &
			-\frac{3}{2} + 24 \omega &
			-\frac{3}{2} + 24 \omega
			\\

			&
			-0.344187\tmark{a} &
			3.36558\tmark{a} &
			3.4882\tmark{a} &
			3.36558\tmark{a} &
			3.36558\tmark{a}
			\\

			\cmidrule{2-6}

			\multirow{2}*{$\kappa_{\varepsilon|i}$} &
			-\EulerGamma - \frac{63}{2} + 158 \omega - \ln \frac{2k}{k_\star} &
			-1 + 16 \omega &
			-\frac{8}{17} + \frac{128}{17} \omega &
			-1 + 16 \omega &
			-1 + 16 \omega
			\\

			&
			0.359993\tmark{a} &
			2.24372\tmark{a} &
			1.05587\tmark{a} &
			2.24372\tmark{a} &
			2.24372\tmark{a}
			\\

 			\bottomrule

 		\end{tabular}
 	}
	\settowidth{\tblw}{\usebox{\tableA}}
	\addtolength{\tblw}{-1em}

	\begin{center}
		\usebox{\tableA}
 	\end{center}

	\renewcommand{\arraystretch}{1}
	
	\sbox{\tableB}{%
		\begin{tabular}{l@{\hspace{1mm}}l}
			\tmark{a} & \parbox[t]{\tblw}{%
				Evaluated at the conventional reference scale
				$k_\star = k_t$}
		\end{tabular}
	}
	
	\begin{center}
		\usebox{\tableB}
	\end{center}

 	\caption{\label{table:fNLequilateral}Equilateral limit of $\fNL$ at
 			 lowest-order and next-order.
 			 The numerical constant $\omega$ satisfies
 			 $\omega = \frac{1}{2} \ln \frac{3}{2} = \coth^{-1} 5$.}
 	
 	\end{sidewaystable}

	\begin{table}
	\small
	\heavyrulewidth=.08em
	\lightrulewidth=.05em
	\cmidrulewidth=.03em
	\belowrulesep=.65ex
	\belowbottomsep=0pt
	\aboverulesep=.4ex
	\abovetopsep=0pt
	\cmidrulesep=\doublerulesep
	\cmidrulekern=.5em
	\defaultaddspace=.5em
	\renewcommand{\arraystretch}{1.6}
	
	\sbox{\tableA}{%
		\begin{tabular}{QqQqQq}
		   	\toprule

			&
			\multicolumn{5}{c}{operator}
			\\

			\cmidrule(l){2-6}

		 	& 
		 	\multicolumn{1}{c}{$\zeta'^3$} &
		 	\multicolumn{1}{c}{$\zeta \zeta'^2$} &
		 	\multicolumn{1}{c}{$\zeta ( \partial \zeta)^2$} &
		 	\multicolumn{1}{c}{$\zeta' \partial_j \zeta \partial_j \partial^{-2} \zeta'$} &
		 	\multicolumn{1}{c}{$\partial^2 \zeta (\partial_j \partial^{-2} \zeta')^2$}
		 	\\

			\midrule

			\left. \fNL \right|_{i0} &
			&
			\frac{5}{24} \frac{g_{1\star}}{\zterm_\star} & 
			\frac{5}{8} \frac{g_{2\star}}{\zterm_\star \csstar^2} & 
			&
			\\[2mm]

			\cmidrule{2-6}

			\multirow{2}*{$\kappa_{h|i}$} &
			&
			\EulerGamma + \ln \frac{2k}{k_\star} &
			\EulerGamma - \frac{4}{3} + \ln \frac{2k}{k_\star} &
			&
			\\

			&
			&
			0.577216\tmark{a} &
			-0.756118\tmark{a} &
			&
			\\

			\cmidrule{2-6}

			\multirow{2}*{$\kappa_{v|i}$} &
			&
			-\EulerGamma + 1 - \ln \frac{2k}{k_\star} &
			-\EulerGamma + \frac{5}{3} - \ln \frac{2k}{k_\star} &
			&
			\\

			&
			&
			0.422784\tmark{a} &
			1.08945\tmark{a} &
			&
			\\

 			\cmidrule{2-6}

			\multirow{2}*{$\kappa_{s|i}$} &
			&
			3 &
			-2\EulerGamma + \frac{11}{3} - 2\ln\frac{2k}{k_\star} &
			&
			\\

			&
			&
			&
			2.51224\tmark{a} &
			&
			\\

			\cmidrule{2-6}

			\multirow{2}*{$\kappa_{\varepsilon|i}$} &
			&
			2 &
			\frac{2}{3} &
			&
			\\
			
			&
			&
			&
			0.666667 &
			&
			\\

 			\bottomrule

 		\end{tabular}
	}
	\settowidth{\tblw}{\usebox{\tableA}}
	\addtolength{\tblw}{-1em}

	\begin{center}
		\usebox{\tableA}
 	\end{center}

	\renewcommand{\arraystretch}{1}
	
	\sbox{\tableB}{%
		\begin{tabular}{l@{\hspace{1mm}}l}
			\tmark{a} & \parbox[t]{\tblw}{Evaluated at the reference scale
				$k_\star = 2 \khard$, where $\khard$ is the common
				hard momentum}
		\end{tabular}
	}
	
	\begin{center}
		\usebox{\tableB}
	\end{center}

 	\caption{\label{table:fNLsqueezed}Squeezed limit of $\fNL$ at
 			 lowest-order and next-order.
 			 The numerical constant $\omega$ satisfies
 			 $\omega = \frac{1}{2} \ln \frac{3}{2} = \coth^{-1} 5$.}
 	
 	\end{table}

	\subsection{Shape dependence}
	\label{sec:shapes}

	Stewart \& Lyth's interest in next-order corrections to the power
	spectrum lay in an accurate estimate of its amplitude.
	In comparison, next-order corrections to the bispectrum
	could be relevant in at least two ways. First, they could change the
	amplitude of three-point correlations, as for the power
	spectrum. Second, they could lead to the appearance of new
	``shapes,'' by which is meant the momentum dependence of
	$B(k_1, k_2, k_3)$ \cite{Babich:2004gb},
	defined in~\eqref{eq:bispectrum-def}. In principle, both
	these effects are measurable.
	
	\subsubsection{Inner product and cosine}
	Babich {\etal} introduced a formal ``cosine'' which may be used
	as a measure of similarity in shape between different bispectra
	\cite{Babich:2004gb}.
	Adopting Eq.~\eqref{eq:power-spectrum-def} for the power
	spectrum $P(k)$, one defines an inner product between two
	bispectra $B_1$ and $B_2$ as
	\begin{equation}
		B_1 \cdot B_2 \equiv \; \sum_{\mathclap{\text{triangles}}} \;
			\frac{B_1(k_1, k_2, k_3) B_2(k_1, k_2, k_3)}{P(k_1) P(k_2) P(k_3)} ,
		\label{eq:inner-prod-def}
	\end{equation}
	where the sum is to be taken over all triangular configurations of the
	$\vect{k}_i$. The cosine between $B_1$ and $B_2$ is
	\begin{equation}
		\cos (B_1, B_2) \equiv \frac{B_1 \cdot B_2}
			{(B_1 \cdot B_1)^{1/2} (B_2 \cdot B_2)^{1/2}} .
		\label{eq:cosine-def}
	\end{equation}
	These expressions require some care.
	In certain cases the result
	may be infinite, requiring the summation to be regulated.
	
	\para{Inner product}%
	We define the sum over triangles as an integral over triangular
	configurations in a flat measure,
	so $\sum \rightarrow \int \d^3 k_1 \, \d^3 k_2 \, \d^3 k_3
	\; \delta(\vect{k}_1 + \vect{k}_2 + \vect{k}_3)$.
	It is sometimes useful to introduce
	a more complicated measure, perhaps to model observational effects
	\cite{Fergusson:2008ra}. In this paper we retain the flat measure
	for simplicity.
	The $\delta$-function can be integrated out immediately,
	leaving a space parametrized by two vectors
	forming a planar triangle which we choose
	to be $\vect{k}_1$ and $\vect{k}_2$. The triangle is invariant under
	a group
	$\mathrm{SO}(2) \times \mathrm{U}(1)$,
	representing arbitrary rotations of $\vect{k}_1$
	combined with azimuthal rotations of $\vect{k}_2$;
	these change
	our representation of the triangle but not its intrinsic geometry.
	The volume of this group may
	be factored out of the measure and discarded.
	Reintroducing $k_3$ in favour of
	the remaining angular integration, we conclude
	\begin{equation}
		B_1 \cdot B_2 = \int \big( \prod_i k_i \, \d k_i \big)
		\frac{B_1(k_1, k_2, k_3) B_2(k_1, k_2, k_3)}{P(k_1) P(k_2) P(k_3)} .
	\end{equation}
	The $k_i$ can be parametrized geometrically in terms of the
	perimeter, $k_t$, and two dimensionless ratios.	
	We adopt the parametrization of Fergusson \& Shellard
	\cite{Fergusson:2008ra},
	\begin{subequations}
		\begin{align}
			\label{eq:fergusson-shellard-a}
			k_1 & = \frac{k_t}{4} ( 1 + \alpha + \beta ) \\
			\label{eq:fergusson-shellard-b}
			k_2 & = \frac{k_t}{4} ( 1 - \alpha + \beta ) \\
			\label{eq:fergusson-shellard-c}
			k_3 & = \frac{k_t}{2} ( 1 - \beta ) ,
		\end{align}
	\end{subequations}
	where $0 \leq \beta \leq 1$ and $\beta - 1 \leq \alpha \leq 1 - \beta$.
	The measure
	$\d k_1 \, \d k_2 \, \d k_3$ is proportional to
	$k_t^2 \, \d k_t \, \d \alpha \, \d \beta$.
	Also,
	on dimensional grounds, each bispectrum $B_i$ scales like
	$\tilde{B}_i k_t^{-6}$, where $\tilde{B}_i$ is dimensionless,
	and each power spectrum $P$ scales like $\tilde{P} k_t^{-3}$
	where $\tilde{P}$ is dimensionless.
	In the special case of scale-invariance, $\tilde{P}$ is constant
	and the $\tilde{B}_i$ depend only on $\alpha$ and $\beta$.
	Therefore
	\begin{equation}
		B_1 \cdot B_2 = N
		\int\limits_{\mathclap{\substack{ 0 \leq \beta \leq 1 \\
			\beta - 1 \leq \alpha \leq 1 - \beta}}}
		\d \alpha \, \d \beta \; (1 - \beta)
		(1 + \alpha + \beta)(1 - \alpha + \beta)
		\tilde{B}_1(\alpha, \beta)
		\tilde{B}_2(\alpha, \beta) ,
		\label{eq:inner-prod-well-defined}
	\end{equation}
	where $N$ is a harmless infinite normalization which can
	be divided out. With this understanding we
	use~\eqref{eq:inner-prod-well-defined} to determine
	the cosine of Eq.~\eqref{eq:cosine-def}.
	In practice, our bispectra are not scale invariant
	and therefore~\eqref{eq:inner-prod-well-defined} does not
	strictly apply.
	However, the violations of scale invariance (to be studied
	in \S\ref{sec:running} below) are small.
	
	\para{Divergences}%
	Eq.~\eqref{eq:inner-prod-well-defined} may be infinite.
	For example, the well-studied local bispectrum
	diverges like $(1 + \alpha + \beta)^{-2}$
	or $(1 - \alpha + \beta)^{-2}$ in the limit
	$\beta \rightarrow 0$, $\alpha \rightarrow \pm 1$,
	or like $(1-\beta)^{-2}$ in the limit $\beta \rightarrow 1$,
	$\alpha \rightarrow 0$ \cite{Babich:2004gb}.
	These correspond to the squeezed limits discussed
	in~\S\ref{subsec:bispectrum}.
	Eq.~\eqref{eq:inner-prod-well-defined}
	therefore exhibits power-law divergences on the boundaries
	of the region of integration, and
	in such cases the integral must be regulated to obtain a finite
	answer. For simplicity, we adopt a sharp cutoff which requires
	$k_i / k_t > \deltamin$.
	As $\deltamin \rightarrow 0$ the cosine~\eqref{eq:cosine-def} may
	converge to a nonzero limit if $B_1 \cdot B_2$,
	$B_1 \cdot B_1$ and $B_2 \cdot B_2$ diverge at the same rate.
	Otherwise, except in finely-tuned cases,
	it converges to zero.
	
	For this reason,
	where divergences exist, the value assigned to $\cos (B_1, B_2)$
	is largely a matter of convention.
	However, to resolve the practical question of
	whether two shapes can be distinguished by observation
	it should be remembered that experiments cannot measure
	arbitrarily small wavenumbers.
	Therefore their ability to distinguish shapes peaking in the squeezed
	limit is limited.
	In this case, to obtain accurate forecasts of
	what can be distinguished, one should restore the $k_t$-dependence
	in~\eqref{eq:inner-prod-well-defined}
	and restrict the integration to observable wavenumbers, yielding
	a manifestly finite answer
	\cite{Fergusson:2008ra,Fergusson:2009nv}.

	\subsubsection{Bispectrum shapes from slow-variation parameters}
	
	In a model with arbitrary $g_i$, the bispectrum is a linear combination
	of the shapes produced by the five operators
	in~\eqref{eq:s3tau}.
	Of these, $\zeta \zeta'^2$ and $\zeta(\partial \zeta)^2$ are
	predominantly correlated with the local template and the remainder
	correlate strongly with the equilateral template.
	The $\zeta'^3$ operator has some overlap with the enfolded template,
	yielding a cosine of order $0.75$.
	With generic values of the slow-variation parameters the
	situation at next-order is similar, and each next-order
	shape is largely correlated with its parent lowest-order shape.
	
	\para{Lowest-order shapes}%
	A $P(X,\phi)$
	is not generic in this sense,
	but imposes strong correlations among the $g_i$.
	At lowest order $g_4$ and $g_5$ do not contribute.
	We focus on a model with small sound speed,
	in which next-order corrections are most likely to be observable,
	and retain only contributions enhanced by $\cs^{-2}$.
	The remaining three operators organize themselves
	into a family of shapes
	of the form
	$S_1 + \alphalambda S_2$,
	where $S_2$ arises only from $\zeta'^3$
	but $S_1$ is a linear combination of the shapes produced
	by $\zeta'^3$, $\zeta \zeta'^2$ and $\zeta (\partial \zeta)^2$.
	The parameter $\alphalambda$ is the enhanced
	part of $\lambda/\Sigma$, that is
	\begin{equation}
		\frac{\lambda}{\Sigma} = \frac{\alphalambda}{\cs^2}
		+ \Or(1)
		\quad
		\text{as $\cs \rightarrow 0$} .
	\end{equation}
	In the DBI model $\alphalambda = 1/2$.
	We plot the shapes $S_1$ and $S_2$ in
	Table~\ref{table:lowest-order-shapes}.
	Note that although $S_1$ involves a linear combination of the
	local-shape operators $\zeta \zeta'^2$
	and $\zeta (\partial \zeta)^2$, the
	$P(X,\phi)$ Lagrangian correlates their amplitudes in such a way
	that there is no divergence in the squeezed limit. Both $S_1$
	and $S_2$
	are strongly correlated with the equilateral template.
	They are similar to the $M_1$- and $M_2$-shapes studied
	in a Galileon theory by Creminelli {\etal}
	\cite{Creminelli:2010qf}.

	\parafootnote{Next-order shapes}{We thank Xingang Chen
		and S\'{e}bastien Renaux-Petel
		for helpful discussions relating to the
		material in this section.}%
	At next-order, more shapes are available.
	Na\"{\i}vely,
	the family of enhanced bispectra is
	labelled by $\varepsilon$, $\eta$, $s$ and also $\ell$
	(following Chen {\etal} we define
	$\ell = \dot{\lambda} / H \lambda$
	\cite{Chen:2006nt}).
	In practice there is some degeneracy, because the shapes
	corresponding to these independent parameters may be strongly
	correlated.
	We will see these degeneracies emerge naturally from our analysis.
	
	The $\cs^{-2}$-enhanced next-order shape can be written
	as a linear-combination
	of shapes proportional to the $\varepsilon$,
	$\eta$, $s$ and $\ell$ parameters, modulated by $\alphalambda$,
	\begin{equation}
		\varepsilon S_\varepsilon
		+ \eta S_\eta
		+ s S_s
		+ \alphalambda \left(
			\varepsilon S'_\varepsilon
			+ \eta S'_\eta
			+ s S'_s
			+ \ell S'_\ell
		\right) .
		\label{eq:next-order-shape}
	\end{equation}
	We give overlap cosines of the $S_i$, $S'_i$ with
	the standard templates in Table~\ref{table:slow-roll-shapes}
	and plot their shapes in Table~\ref{table:next-order-shapes}.
	Because the cutoff dependence complicates comparison
	between different analyses
	we list 
	the cosines between templates
	in Table~\ref{table:cos-templates},
	computed using the same conventions.
	Generally speaking, these shapes have strong overlaps with the
	equilateral template. However, two
	are quite different in appearance
	and have a slightly smaller cosine $\sim 0.85$
	with this mode:
	these are $S'_\varepsilon$ and $S_s$.
	We fix two coefficients
	in~\eqref{eq:next-order-shape} by
	choosing a linear combination
	orthogonal
	to both $S_1$ and $S_2$.
	Without loss of generality we can choose these
	to be $\eta$ and $s$. We find
	the required combination to be
	approximately
	\begin{subequations}
	\begin{align}
		\eta &
		\approx
			\frac{0.12 \alphalambda \ell
				(\alphalambda + 0.72)
				(\alphalambda + 1.82)
				-
				0.88 \varepsilon
				(\alphalambda - 9.15)
				(\alphalambda - 0.22)
				(\alphalambda + 1.82)}
				{(\alphalambda - 10.24)
				(\alphalambda - 0.23)
				(\alphalambda + 1.82)}
			\\
		s &
		\approx
			\frac{\alphalambda \ell
				(3.88 - 0.12 \alphalambda)
				- 1.12 \varepsilon (\alphalambda - 8.51)
				(\alphalambda - 0.08)}
				{(\alphalambda - 10.24)
				(\alphalambda - 0.23)}
	\end{align}
	\end{subequations}
	It is possible this procedure is stronger than necessary.
	Both $S_1$ and $S_2$ are correlated with the equilateral template,
	and it may be sufficient to find a linear combination orthogonal
	to that.
	In what follows, however, we insist on orthogonality with
	$S_1$ and $S_2$ and defer generalizations to future work.
	For certain values of $\alphalambda$ the denominator
	of both $\eta$ and $s$ may simultaneously
	vanish, making the required
	$\eta$ and $s$ very large. This implies that, near these
	values of $\alphalambda$, no shape orthogonal
	to both $S_1$ and $S_2$ can be found within the
	validity of next-order perturbation theory.
	Therefore
	we restrict attention to those $\alphalambda$
	which allow acceptably small $\eta$ and $s$.
	
	This process leaves two linear combinations
	proportional to $\varepsilon$ and $\ell$.
	In principle these can be diagonalized, yielding a pair of shapes
	orthogonal to each other and $\{ S_1, S_2 \}$.
	However, the $2\times 2$ matrix of inner products between
	these linear combinations is degenerate. Therefore, only one member
	of this pair is physical and can be realized in a $P(X, \phi)$
	model. The other is not: it has zero inner product
	with~\eqref{eq:next-order-shape}, and is impossible to realize
	because of enforced correlations between
	coefficients.
	We denote the physical orthogonal combination $O$.
	It has a vanishing component proportional to $\ell$.
	This was expected, because
	the shape $S'_\ell$ is the same as $S_2$.
	For this reason, Chen {\etal}
	\cite{Chen:2006nt}
	absorbed $\ell$ into a redefined
	$\lambda/\Sigma$.
	It is indistinguishable from the lowest-order prediction
	and could never be observed separately, which is the origin
	of the degeneracy. We could have arrived at the same
	$O$ by excluding $S'_\ell$ from~\eqref{eq:next-order-shape}.
	Demanding the inner product with $S_1$ and $S_2$ be zero
	reproduces
	the physical linear combination obtained from diagonalization.
	
	We plot the shape of $O$ in Table~\ref{table:orthogonal-plots}.
	Its dependence
	on $\alphalambda$ is modest.
	As a function of the $k_i$
	there are multiple peaks, and therefore $O$
	is not maximized on a unique type of triangle.
	In Table~\ref{table:orthogonal-cosines} we give
	the overlap cosine with
	common templates.
	The lowest-order shapes $S_1$ and $S_2$ are strongly
	correlated with the equilateral template, and since $O$
	is orthogonal to these by construction it also has
	small cosines with the equilateral template, of order $10^{-2}$.
	There is a moderate cosine with the local template
	of order $\sim 0.3$ -- $0.4$.
	The precise value depends on our choice of
	$\deltamin$, but the dependence is not dramatic.
	In Table~\ref{table:orthogonal-cosines} we have used our
	convention $\deltamin = 10^{-3}$. For $\deltamin = 10^{-5}$
	the local cosines change by roughly $25\%$. Overlaps with the
	remaining templates are stable under changes of $\deltamin$.
	There is a cosine of order $0.35$ -- $0.40$ with the orthogonal
	template, and of order $0.30$ -- $0.35$ with the enfolded
	template.
	We conclude that $O$ is not strongly correlated with
	any of the standard templates used in CMB analysis.
	To find subleading effects in the data, it will
	probably be
	necessary to develop a dedicated template for the purpose.

	The $O$-shape is very similar to a highly orthogonal shape
	constructed by
	Creminelli {\etal} \cite{Creminelli:2010qf}
	in a Galileon model,
	although $O$
	contains marginally more fine structure.
	For comparison,
	we plot the Creminelli {\etal}
	shape in Table~\ref{table:creminelli-shape}
	and
	include its cosine with $O$ in
	Table~\ref{table:orthogonal-cosines}.
	For varying
	$\alphalambda$
	we find a cosine in the range $0.8$ -- $0.9$,
	which indicates it would be difficult
	to distinguish these shapes observationally.
	In particular,
	even if a bispectrum with this shape were to be detected,
	further information would be required to distinguish between
	candidate $P(X, \phi)$ or Galileon models for its origin.

	\begin{table}

	\small
	\heavyrulewidth=.08em
	\lightrulewidth=.05em
	\cmidrulewidth=.03em
	\belowrulesep=.65ex
	\belowbottomsep=0pt
	\aboverulesep=.4ex
	\abovetopsep=0pt
	\cmidrulesep=\doublerulesep
	\cmidrulekern=.5em
	\defaultaddspace=.5em
	\renewcommand{\arraystretch}{1.6}

	\sbox{\tableA}{%
		\begin{tabular}{QsSsSsSs}

			\toprule

			&
			\multicolumn{1}{c}{$S_\varepsilon$} &
			\multicolumn{1}{c}{$S'_\varepsilon$} &
			\multicolumn{1}{c}{$S_\eta$} &
			\multicolumn{1}{c}{$S'_\eta$} &
			\multicolumn{1}{c}{$S_s$} &
			\multicolumn{1}{c}{$S'_s$} &
			\multicolumn{1}{c}{$S'_\ell$}
			\\

			\cmidrule{2-8}

			\multicolumn{1}{l}{local\tmark{a}} &
				0.38\tmark{e} &
				0.50\tmark{e} &
				0.37\tmark{e} &
				0.43\tmark{e} &
				0.54\tmark{e} &
				0.39\tmark{e} &
				0.42\tmark{e}
				\\	

			\multicolumn{1}{l}{equilateral\tmark{b}} &
				0.99 &
				0.87 &
				1.00 &
				0.93 &
				0.80 &
				0.94 &
				0.94
				\\

			\multicolumn{1}{l}{orthogonal\tmark{c}} &
				0.084 &
				0.46 &
				0.065 &
				0.31 &
				0.52 &
				0.25 &
				0.29
				\\

			\multicolumn{1}{l}{enfolded\tmark{d}} &
				0.60 &
				0.86 &
				0.59 &
				0.77 &
				0.87 &
				0.72 &
				0.75
				\\

			\bottomrule
		\end{tabular}
	}
	\settowidth{\tblw}{\usebox{\tableA}}
	\addtolength{\tblw}{-1em}
	
	\begin{center}
		\usebox{\tableA}
	\end{center}
	
	\renewcommand{\arraystretch}{1.0}
	\tiny
	
	\sbox{\tableB}{%
		\begin{tabular}{l@{\hspace{1mm}}l}	
			\tmark{a} &
				\parbox[t]{\tblw}{
				See Komatsu \& Spergel~\cite{Komatsu:2001rj}
				and Babich {\etal}~\cite{Babich:2004gb}.} \\
		
			\tmark{b} &
				\parbox[t]{\tblw}{
				See Babich {\etal}~\cite{Babich:2004gb}.} \\
	
			\tmark{c} &
				\parbox[t]{\tblw}{
				See Senatore {\etal}~\cite{Senatore:2009gt}.} \\
	
			\tmark{d} &
				\parbox[t]{\tblw}{See Meerburg {\etal}~\cite{Meerburg:2009ys}
				and Senatore {\etal}~\cite{Senatore:2009gt}.} \\
	
			\tmark{e} & \parbox[t]{\tblw}{
				The local template, and the operators
				$\zeta \zeta'^2$ and $\zeta (\partial \zeta)^2$, are strongly
				peaked in the ``squeezed'' limit where one momentum becomes much
				softer than the other two. For these shapes the inner product
				which defines the cosine is divergent, and must be regulated.
				The resulting cosines are almost entirely
				regulator-dependent. See the discussion
				in \S\ref{sec:shapes}.
		
				The values we quote are meaningful only for our choice
				of regulator. For the values quoted above we have
				used $\deltamin = k/k_t = 10^{-3}$, where
				$\deltamin$ was defined in the main text.}
		\end{tabular}
	}
	
	\begin{center}
		\usebox{\tableB}
	\end{center}

	\caption{\label{table:slow-roll-shapes}Overlap cosines
	for the bispectrum shape proportional to each slow-variation parameter.
	Sign information has been discarded.}

	\end{table}

	\begin{table}
	\small
	\heavyrulewidth=.08em
	\lightrulewidth=.05em
	\cmidrulewidth=.03em
	\belowrulesep=.65ex
	\belowbottomsep=0pt
	\aboverulesep=.4ex
	\abovetopsep=0pt
	\cmidrulesep=\doublerulesep
	\cmidrulekern=.5em
	\defaultaddspace=.5em
	\renewcommand{\arraystretch}{1.6}

	\sbox{\tableA}{%
		\begin{tabular}{lTcTc}

			\toprule
			
			&
				\multicolumn{1}{c}{local} &
				\multicolumn{1}{c}{equilateral} &
				\multicolumn{1}{c}{orthogonal} &
				\multicolumn{1}{c}{enfolded}
			\\
			\cmidrule{2-5}
			
			local &
				1.00 &
				&
				&
			\\
			
			equilateral &
				0.34 &
				1.00 &
				&
			\\
			
			orthogonal &
				0.49 &
				0.03 &
				1.00 &
			\\
			
			enfolded &
				0.60 &
				0.51 &
				0.85 &
				1.00
			\\
			
			\bottomrule
		
		\end{tabular}
	}
	\settowidth{\tblw}{\usebox{\tableA}}
	\addtolength{\tblw}{-1em}
	
	\begin{center}
		\usebox{\tableA}
	\end{center}

	\caption{\label{table:cos-templates}Overlap cosines between common
	templates, defined in Table~\ref{table:slow-roll-shapes}.}

	\end{table}
	
	\begin{table}
	
	\small
	\heavyrulewidth=.08em
	\lightrulewidth=.05em
	\cmidrulewidth=.03em
	\belowrulesep=.65ex
	\belowbottomsep=0pt
	\aboverulesep=.4ex
	\abovetopsep=0pt
	\cmidrulesep=\doublerulesep
	\cmidrulekern=.5em
	\defaultaddspace=.5em
	\renewcommand{\arraystretch}{1.6}

    \sbox{\boxplot}{%
    	\includegraphics[scale=0.15]{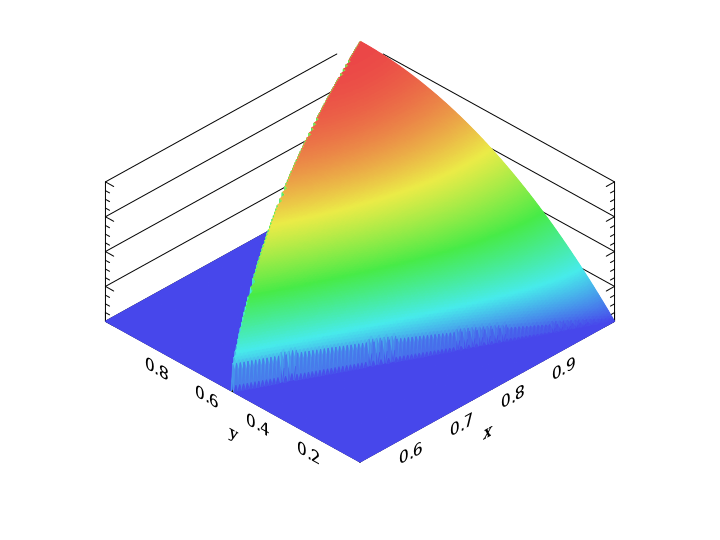}
    }
    \settowidth{\plotw}{\usebox{\boxplot}}
    
    \sbox{\tableA}{%
		\begin{tabular}{ccc}

			\toprule

			&
			Babich\tmark{a} {\etal} &
			Fergusson \& Shellard\tmark{b}
			\\

			\cmidrule(r){2-2}
			\cmidrule(l){3-3}
			
			Shape 1 &
			\parbox[c]{\plotw}{\includegraphics[scale=0.15]{Plots/Babich/3D/ShapeLO/S1}} &
			\parbox[c]{\plotw}{\includegraphics[scale=0.15]{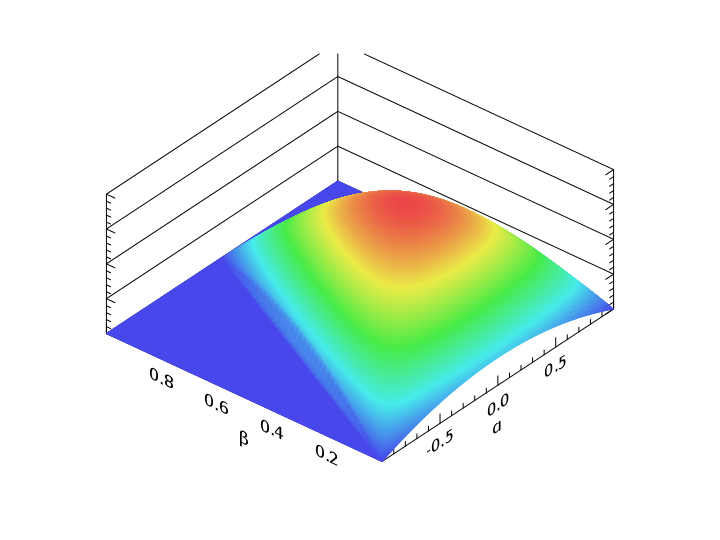}}
			\\

			Shape 2 &
			\parbox[c]{\plotw}{\includegraphics[scale=0.15]{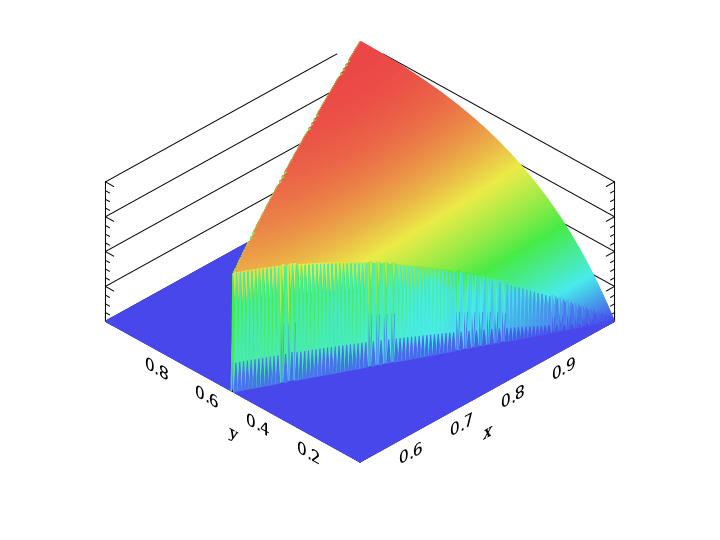}} &
			\parbox[c]{\plotw}{\includegraphics[scale=0.15]{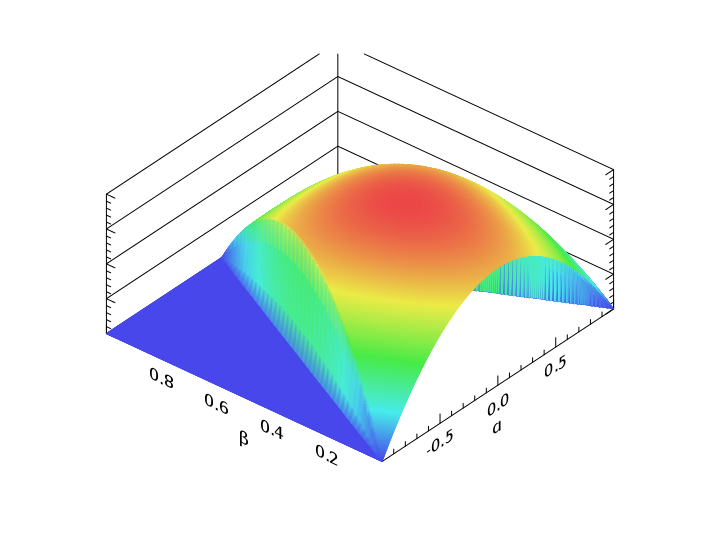}}
			\\

			\bottomrule

		\end{tabular}
	}
	\settowidth{\tblw}{\usebox{\tableA}}
	\addtolength{\tblw}{-1em}

	\begin{center}
		\usebox{\tableA}
	\end{center}
			
	\renewcommand{\arraystretch}{1}
	\tiny
	
	\sbox{\tableB}{%
		\begin{tabular}{l@{\hspace{1mm}}l}
			\tmark{a} & \parbox[t]{\tblw}{%
				See Babich {\etal}~\cite{Babich:2004gb}.
				The plotted quantity is
				$x^2 y^2 B(x, y, 1)$,
				where $x = k_1 / k_3$,
				$y = k_2 / k_3$ and $B$ is the
				bispectrum,
				and normalized to unity at the equilateral
				point $x = y = 1$.} \\
		
	 		\tmark{b} & \parbox[t]{\tblw}{%
		 		See Fergusson \& Shellard~\cite{Fergusson:2008ra}.
		 		The plotted quantity is
		 		$k_1^2 k_2^2 k_3^2 B(k_1, k_2, k_3)$
		 		as a function of the $\alpha$ and
		 		$\beta$ parameters defined in~\eqref{eq:fergusson-shellard-a}--%
		 		\eqref{eq:fergusson-shellard-c}.}
		\end{tabular}
	}
	
	\begin{center}
		\usebox{\tableB}
	\end{center}
	
	\caption{\label{table:lowest-order-shapes}Lowest-order
	bispectrum shapes
	enhanced by $\cs^{-2}$ in $P(X,\phi)$ models.}

	\end{table}

	\begin{table}
	
	\small
	\heavyrulewidth=.08em
	\lightrulewidth=.05em
	\cmidrulewidth=.03em
	\belowrulesep=.65ex
	\belowbottomsep=0pt
	\aboverulesep=.4ex
	\abovetopsep=0pt
	\cmidrulesep=\doublerulesep
	\cmidrulekern=.5em
	\defaultaddspace=.5em
	\renewcommand{\arraystretch}{1.6}

    \sbox{\boxplot}{%
    	\includegraphics[scale=0.15]{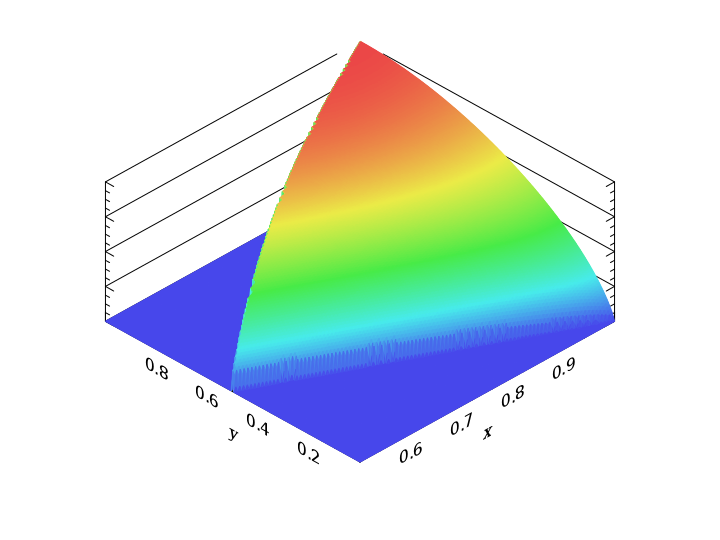}
    }
    \settowidth{\plotw}{\usebox{\boxplot}}
    \sbox{\boxplota}{%
    	\includegraphics[scale=0.1]{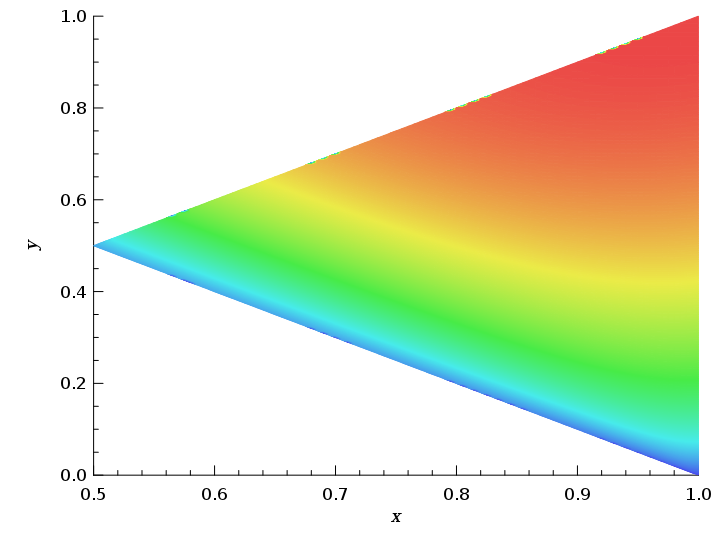}
    }
    \settowidth{\plotwa}{\usebox{\boxplota}}
    
    \sbox{\tableA}{%
		\begin{tabular}{ccccc}

			\toprule

			&
			\multicolumn{2}{c}{Babich {\etal}} &
			\multicolumn{2}{c}{Fergusson \& Shellard}
			\\
			
			\cmidrule(r){2-3}
			\cmidrule(l){4-5}
			
			$S_\varepsilon$ &
			\parbox[c]{\plotwa}{\includegraphics[scale=0.1]{Plots/Babich/2D/ShapeNLO/epsilon}} &
			\parbox[c]{\plotw}{\includegraphics[scale=0.15]{Plots/Babich/3D/ShapeNLO/epsilon}} &
			\parbox[c]{\plotwa}{\includegraphics[scale=0.1]{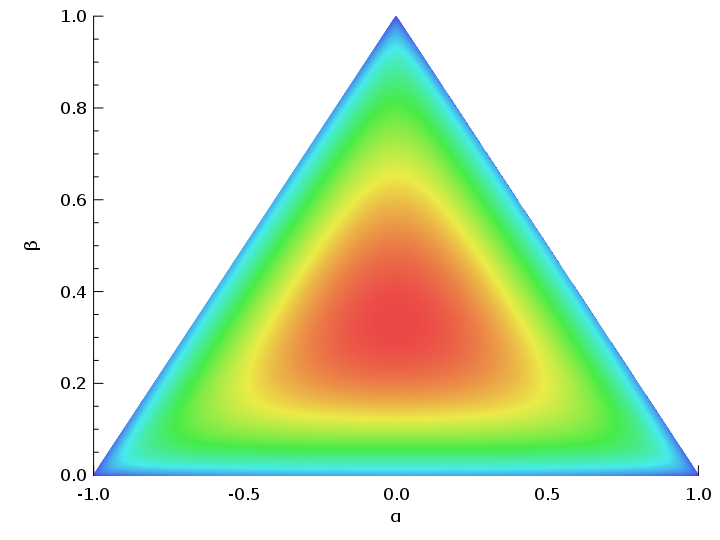}} &
			\parbox[c]{\plotw}{\includegraphics[scale=0.15]{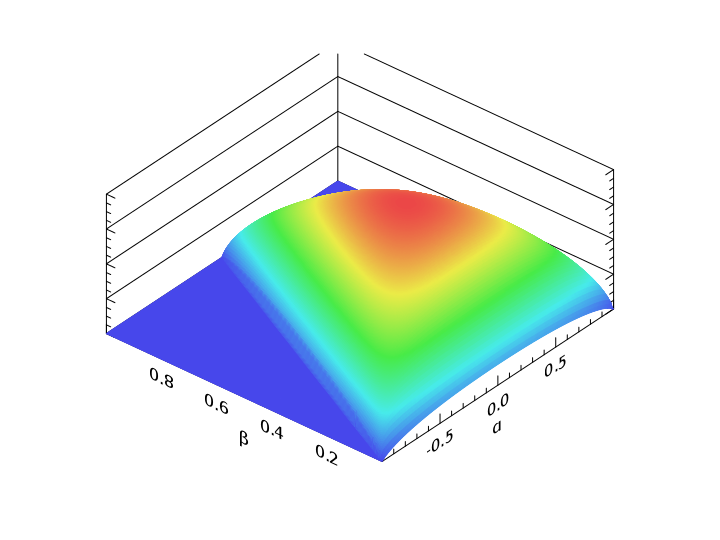}}
			\\

			$S'_\varepsilon$ &
			\parbox[c]{\plotwa}{\includegraphics[scale=0.1]{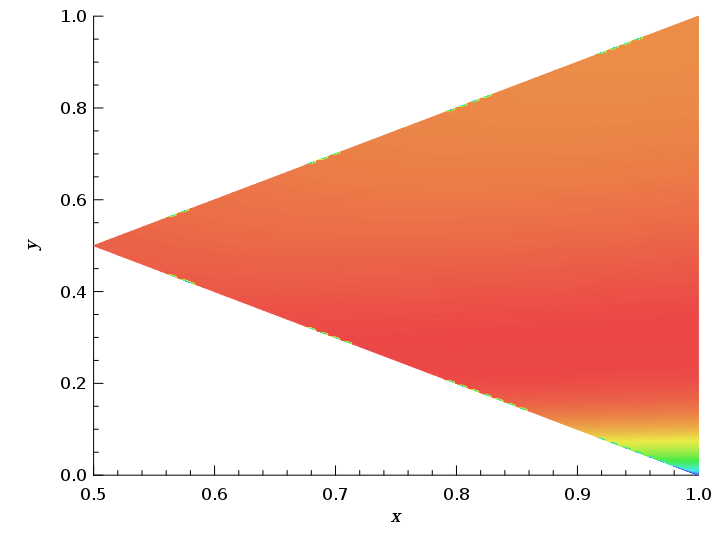}} &
			\parbox[c]{\plotw}{\includegraphics[scale=0.15]{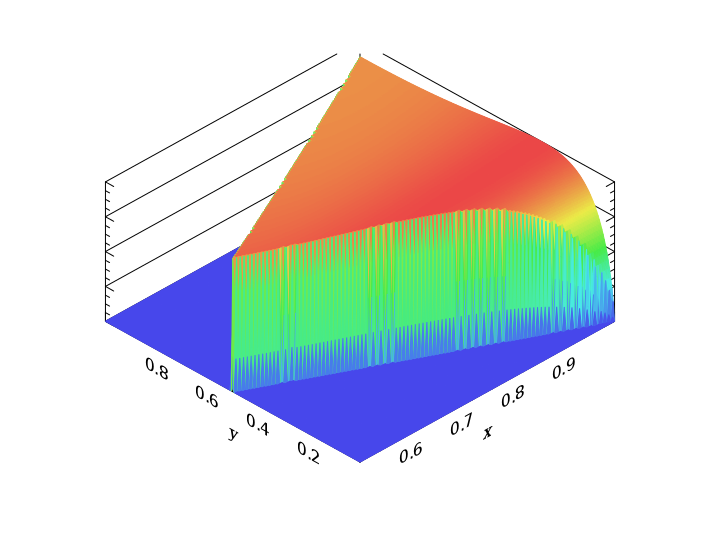}} &
			\parbox[c]{\plotwa}{\includegraphics[scale=0.1]{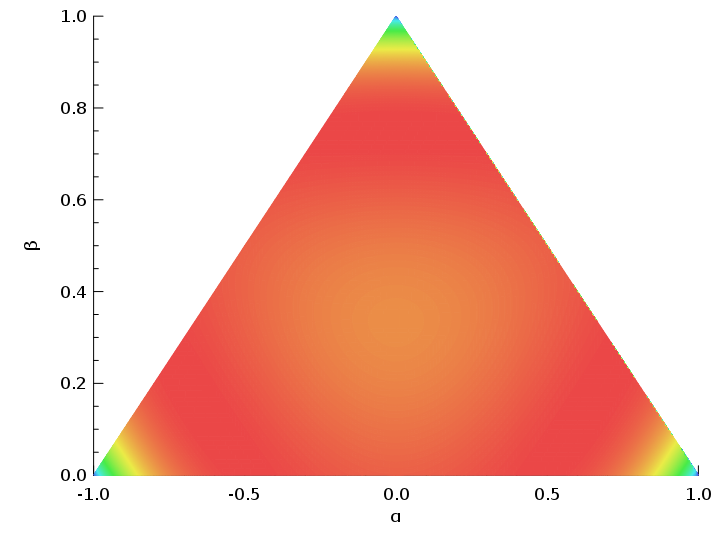}} &
			\parbox[c]{\plotw}{\includegraphics[scale=0.15]{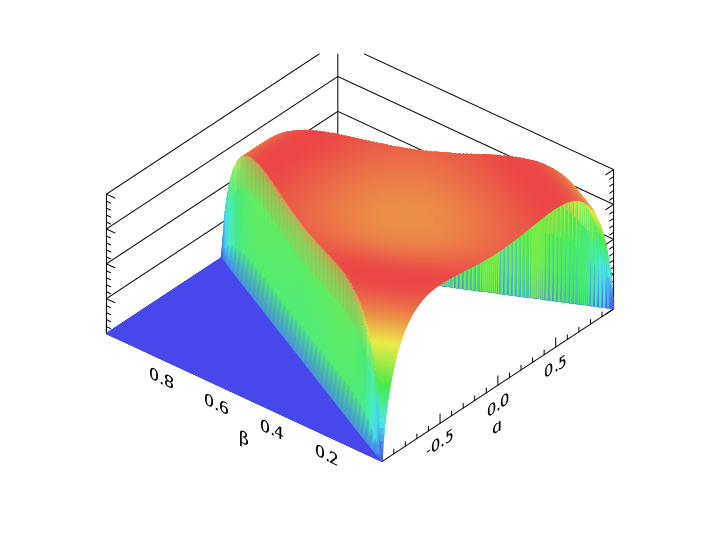}}
			\\

			$S_\eta$ &
			\parbox[c]{\plotwa}{\includegraphics[scale=0.1]{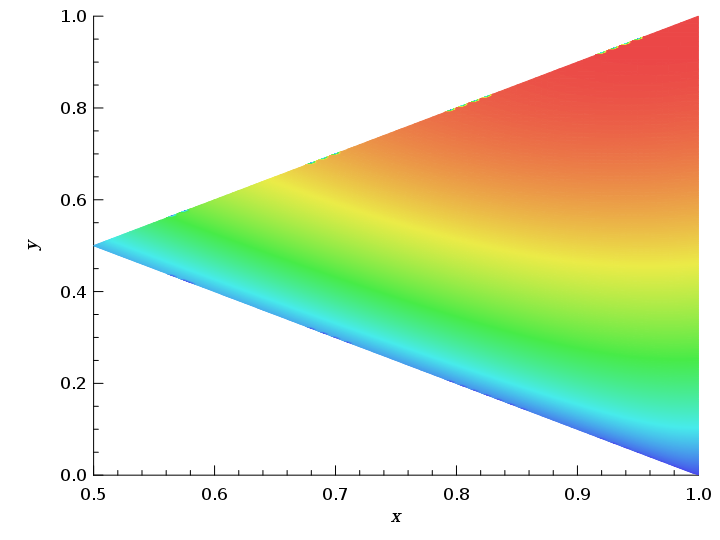}} &
			\parbox[c]{\plotw}{\includegraphics[scale=0.15]{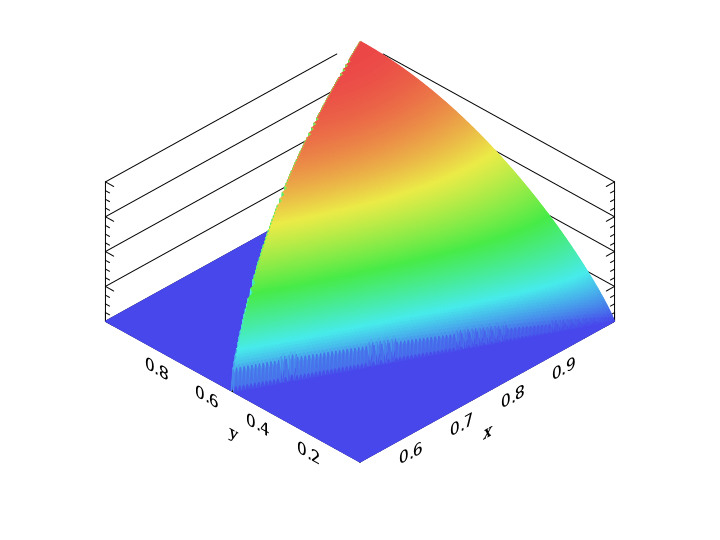}} &
			\parbox[c]{\plotwa}{\includegraphics[scale=0.1]{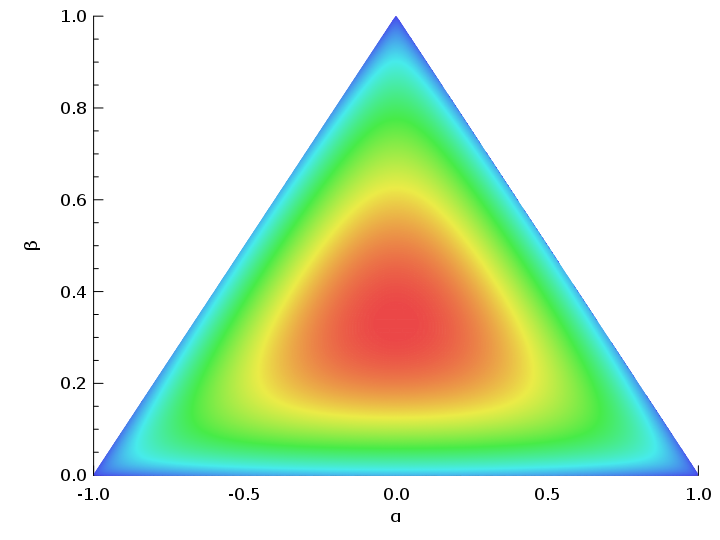}} &
			\parbox[c]{\plotw}{\includegraphics[scale=0.15]{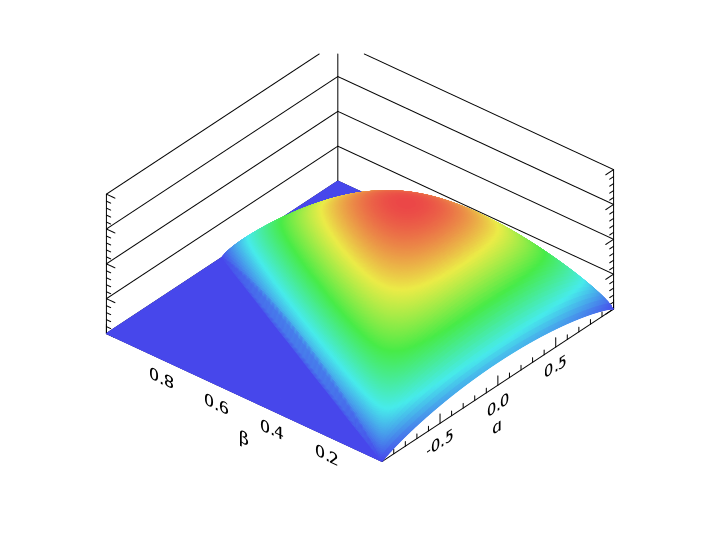}}
			\\

			$S'_\eta$ &
			\parbox[c]{\plotwa}{\includegraphics[scale=0.1]{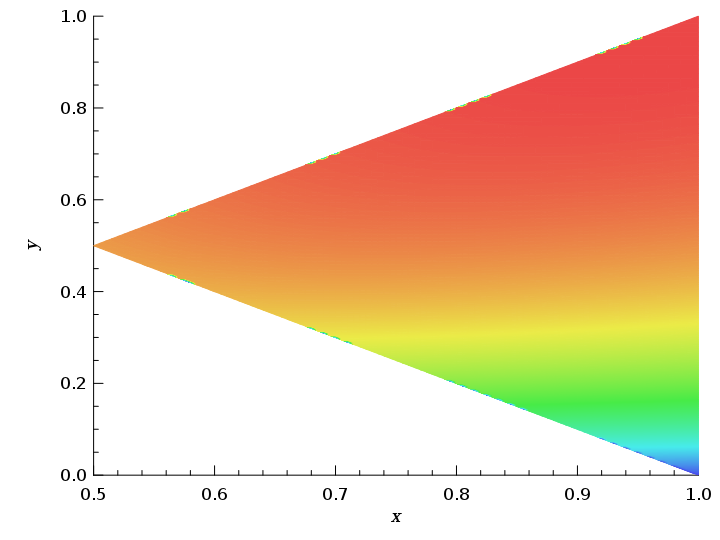}} &
			\parbox[c]{\plotw}{\includegraphics[scale=0.15]{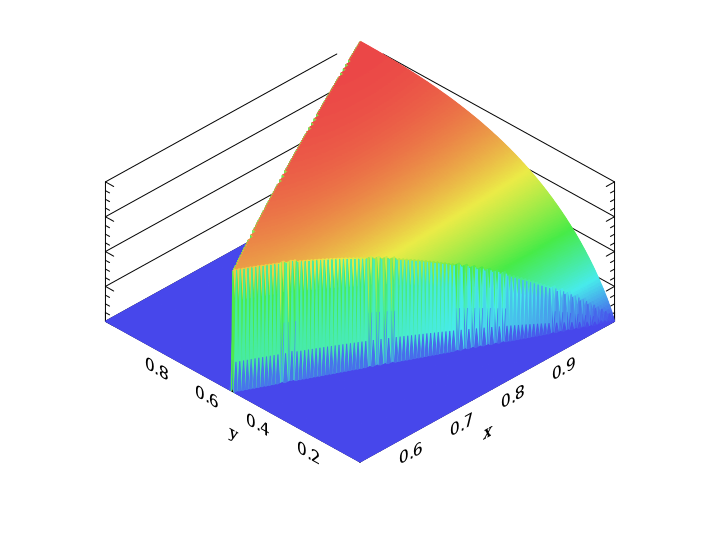}} &
			\parbox[c]{\plotwa}{\includegraphics[scale=0.1]{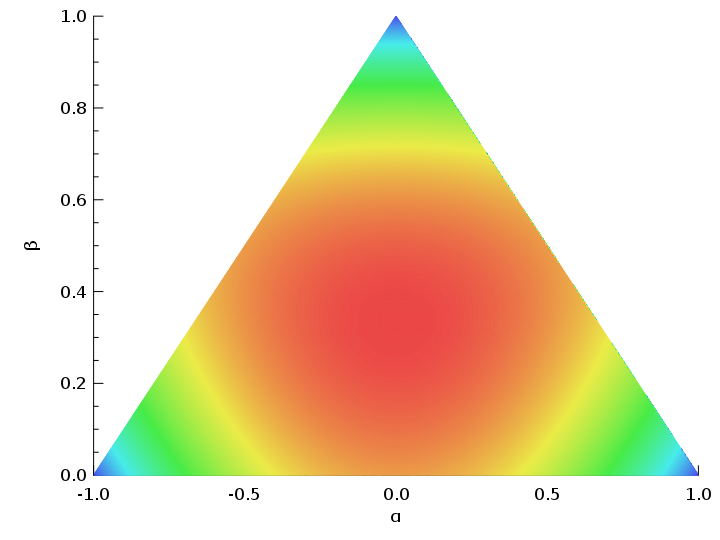}} &
			\parbox[c]{\plotw}{\includegraphics[scale=0.15]{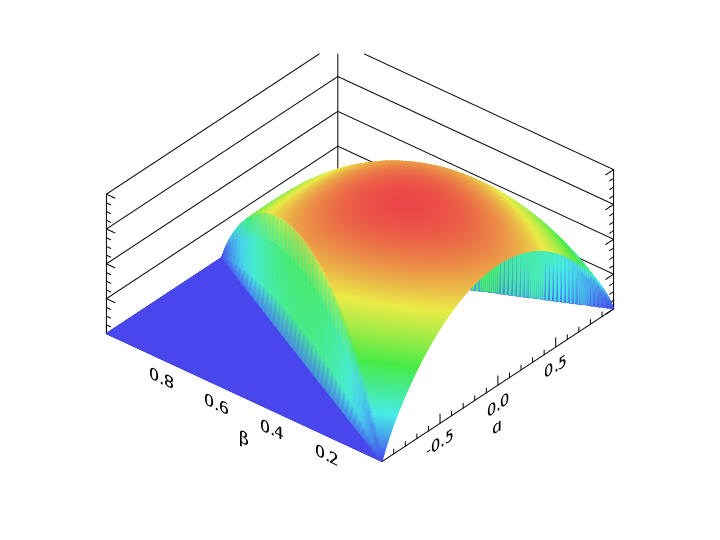}}
			\\

			$S_s$ &
			\parbox[c]{\plotwa}{\includegraphics[scale=0.1]{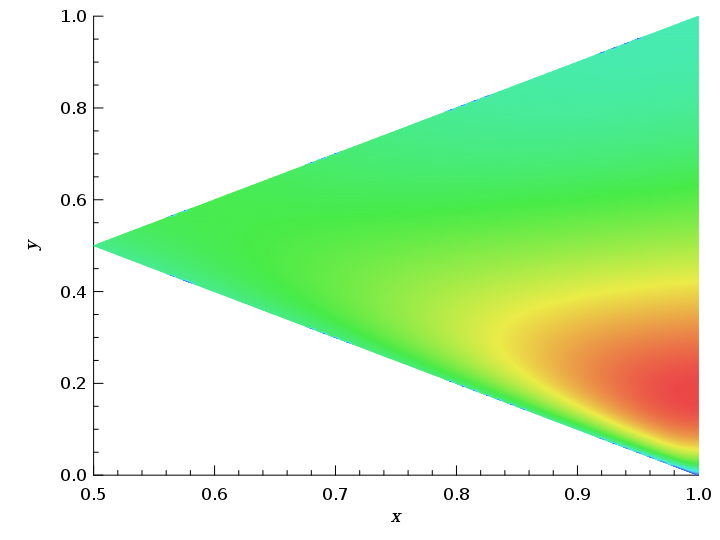}} &
			\parbox[c]{\plotw}{\includegraphics[scale=0.15]{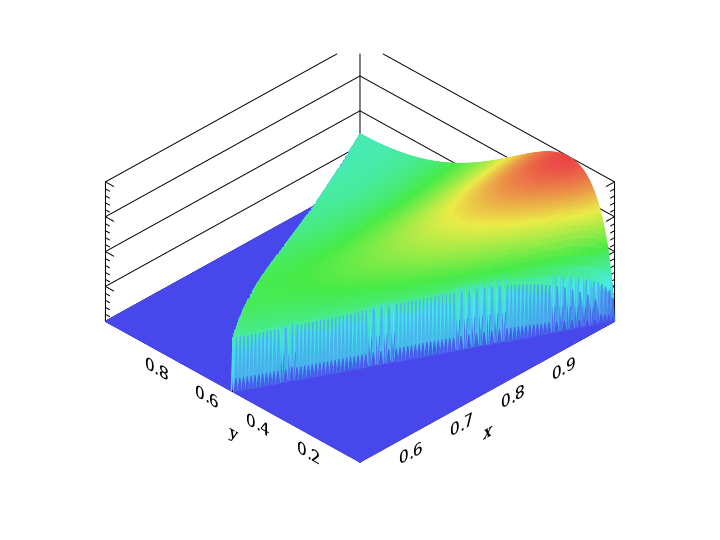}} &
			\parbox[c]{\plotwa}{\includegraphics[scale=0.1]{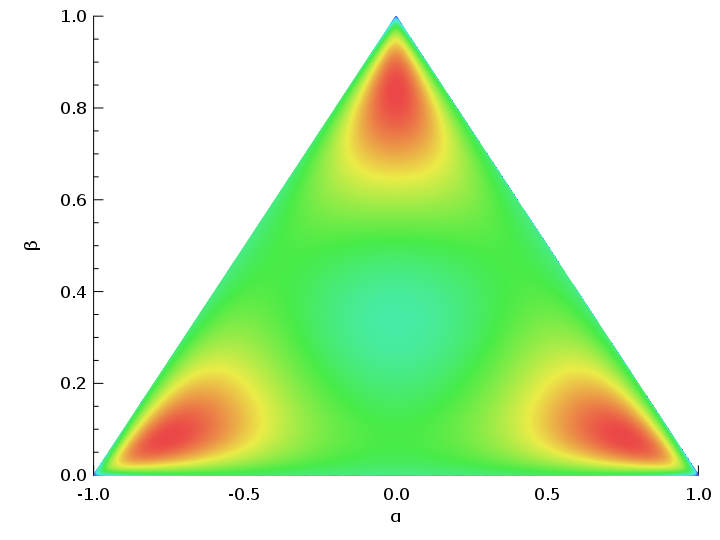}} &
			\parbox[c]{\plotw}{\includegraphics[scale=0.15]{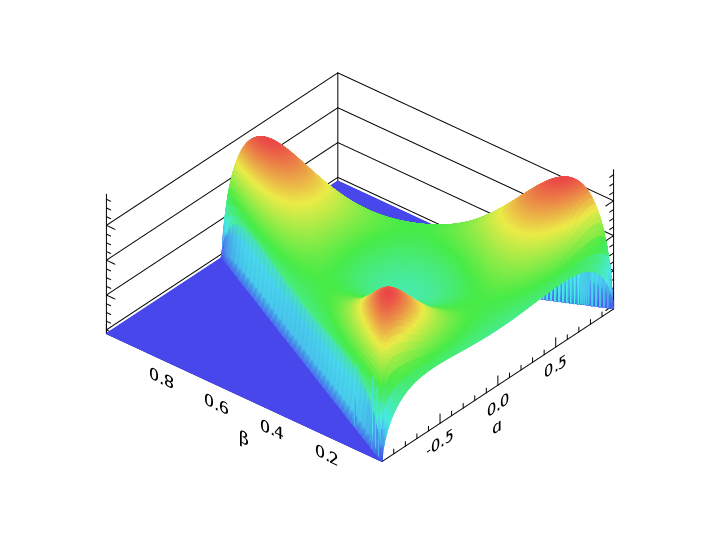}}
			\\

			$S'_s$ &
			\parbox[c]{\plotwa}{\includegraphics[scale=0.1]{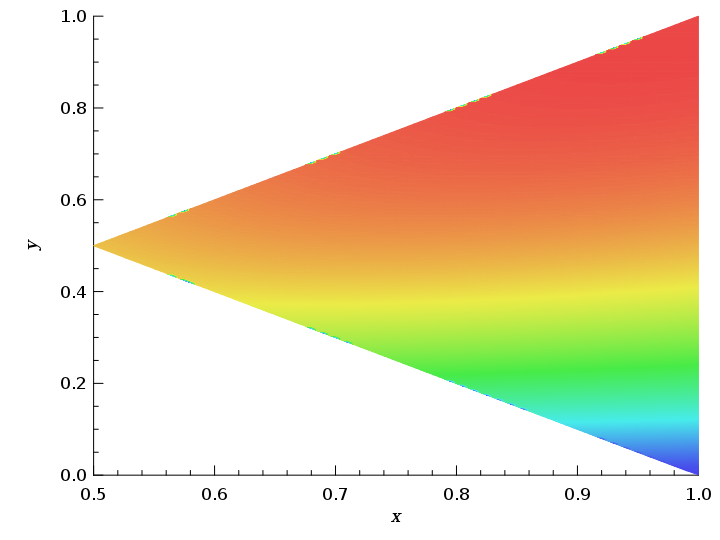}} &
			\parbox[c]{\plotw}{\includegraphics[scale=0.15]{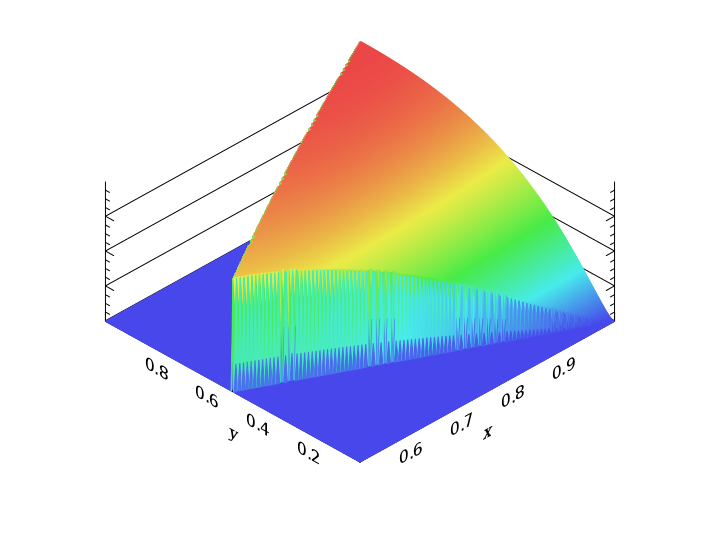}} &
			\parbox[c]{\plotwa}{\includegraphics[scale=0.1]{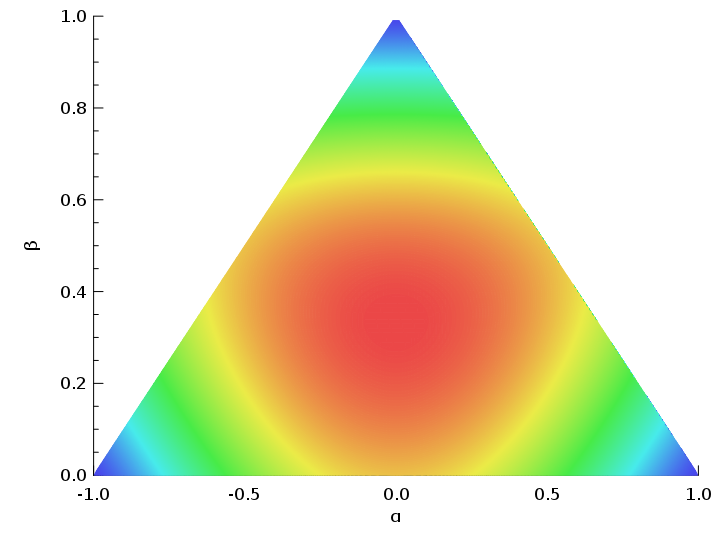}} &
			\parbox[c]{\plotw}{\includegraphics[scale=0.15]{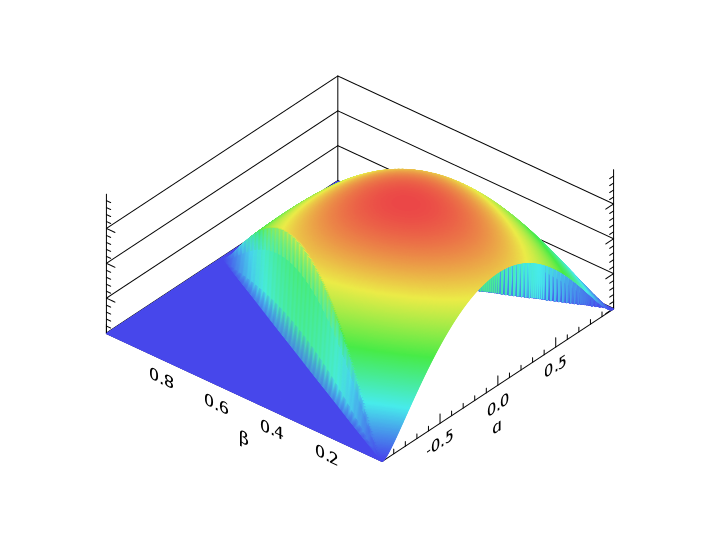}}
			\\

			$S'_\ell$ &
			\parbox[c]{\plotwa}{\includegraphics[scale=0.1]{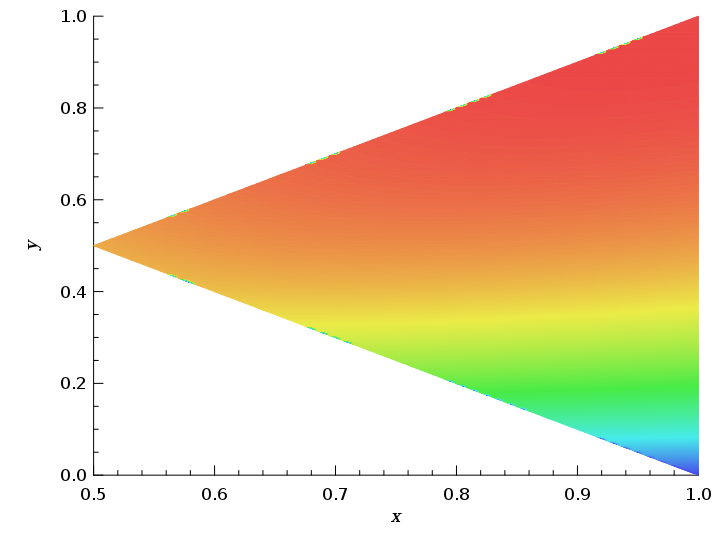}} &
			\parbox[c]{\plotw}{\includegraphics[scale=0.15]{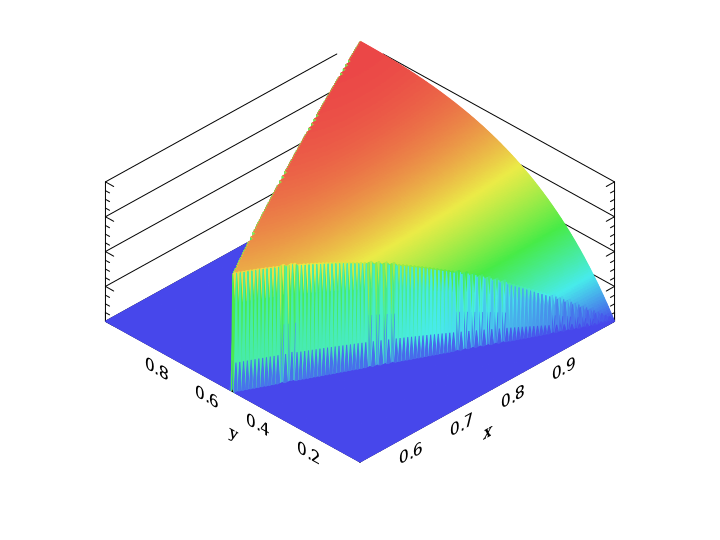}} &
			\parbox[c]{\plotwa}{\includegraphics[scale=0.1]{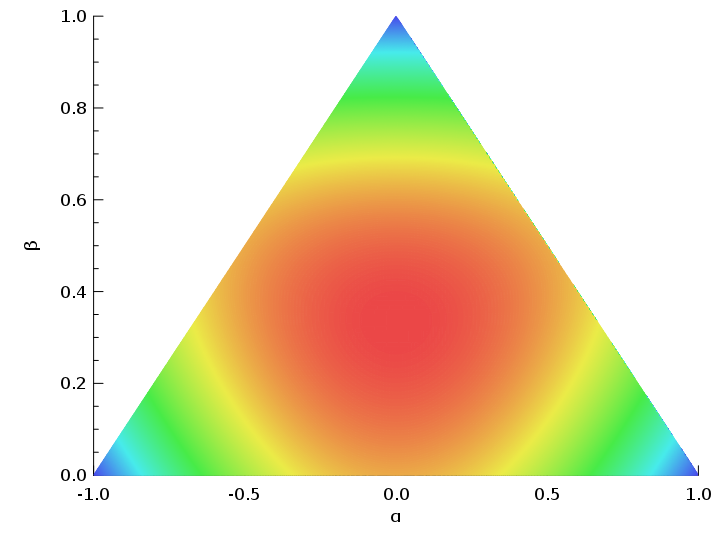}} &
			\parbox[c]{\plotw}{\includegraphics[scale=0.15]{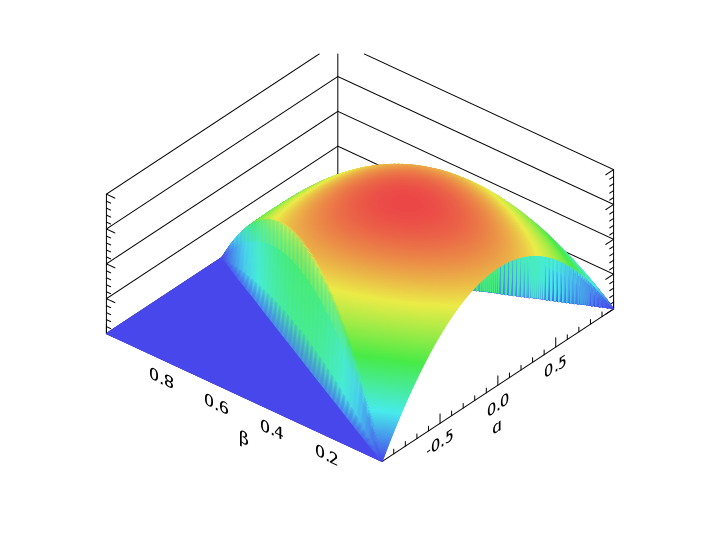}}
			\\

			\bottomrule

		\end{tabular}
	}
	\settowidth{\tblw}{\usebox{\tableA}}
	\addtolength{\tblw}{-1em}

	\begin{center}
		\usebox{\tableA}
	\end{center}
			
	\caption{\label{table:next-order-shapes}Bispectrum shapes enhanced
	by $\cs^{-2}$ at next-order in a $P(X,\phi)$ model.
	The Babich {\etal} and Fergusson--Shellard
	plots are defined in Table~\ref{table:lowest-order-shapes}.}

	\end{table}

	\begin{table}
	
	\small
	\heavyrulewidth=.08em
	\lightrulewidth=.05em
	\cmidrulewidth=.03em
	\belowrulesep=.65ex
	\belowbottomsep=0pt
	\aboverulesep=.4ex
	\abovetopsep=0pt
	\cmidrulesep=\doublerulesep
	\cmidrulekern=.5em
	\defaultaddspace=.5em
	\renewcommand{\arraystretch}{1.6}

    \sbox{\boxplot}{%
    	\includegraphics[scale=0.15]{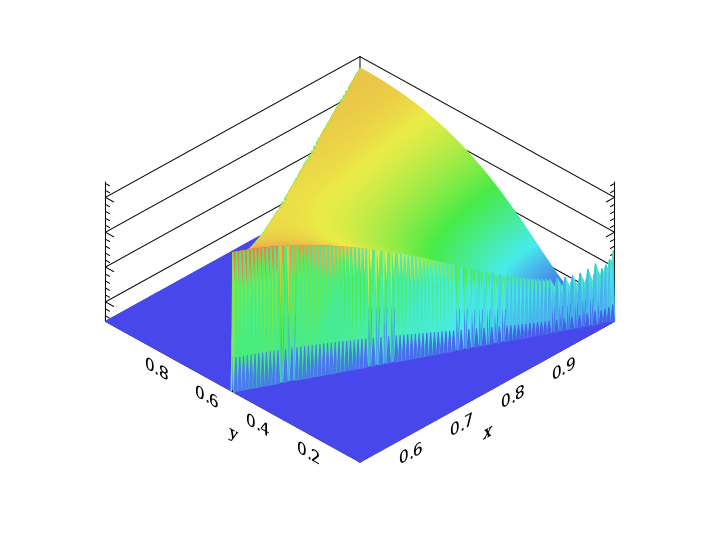}
    }
    \settowidth{\plotw}{\usebox{\boxplot}}
    \sbox{\boxplota}{%
    	\includegraphics[scale=0.1]{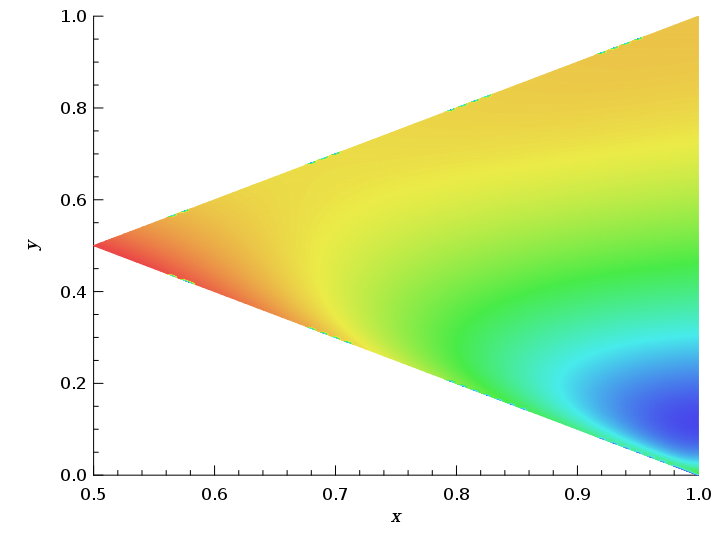}
    }
    \settowidth{\plotwa}{\usebox{\boxplota}}
    
    \sbox{\tableA}{%
		\begin{tabular}{cccc}

			\toprule

			\multicolumn{2}{c}{Babich {\etal}} &
			\multicolumn{2}{c}{Fergusson \& Shellard}
			\\
			
			\cmidrule(r){1-2}
			\cmidrule(l){3-4}
			
			\multicolumn{4}{c}{$\alphalambda = 10^{-3}$}
			\\
			
			\parbox[c]{\plotwa}{\includegraphics[scale=0.1]{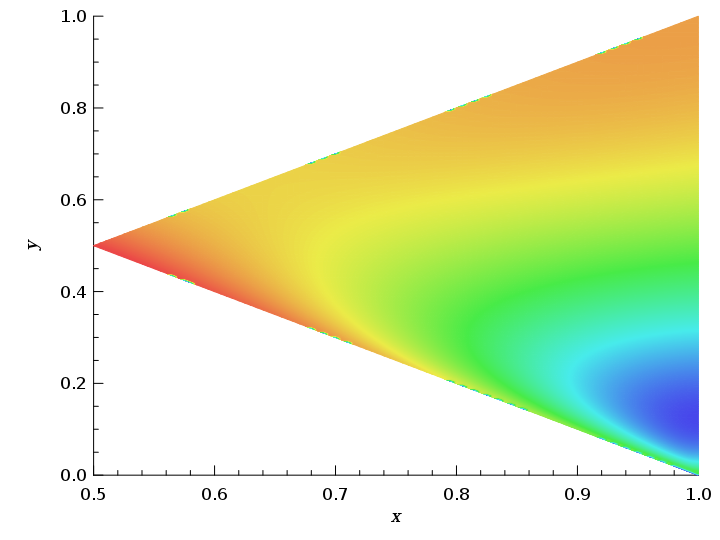}} &
			\parbox[c]{\plotw}{\includegraphics[scale=0.15]{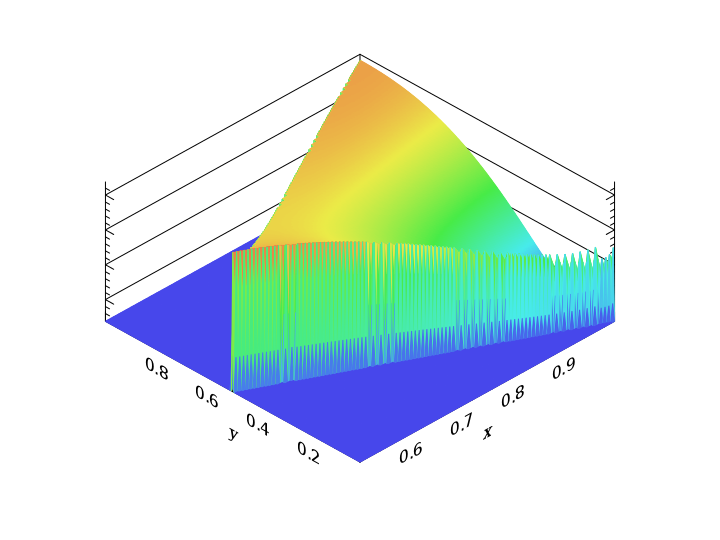}} &
			\parbox[c]{\plotwa}{\includegraphics[scale=0.1]{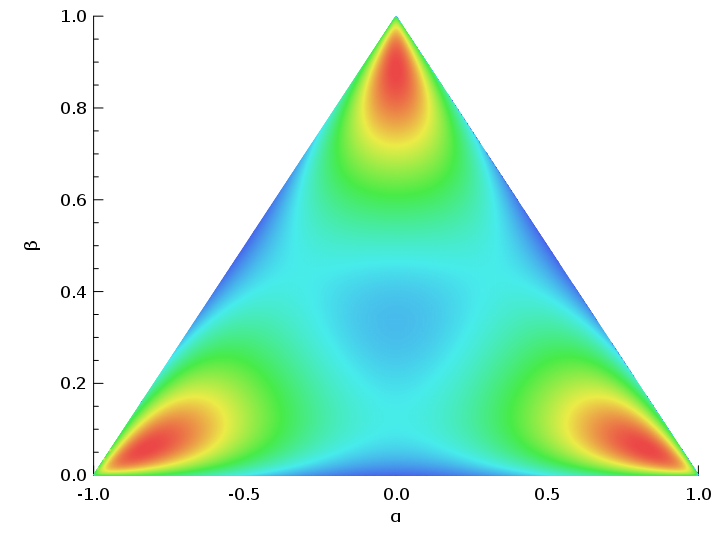}} &
			\parbox[c]{\plotw}{\includegraphics[scale=0.15]{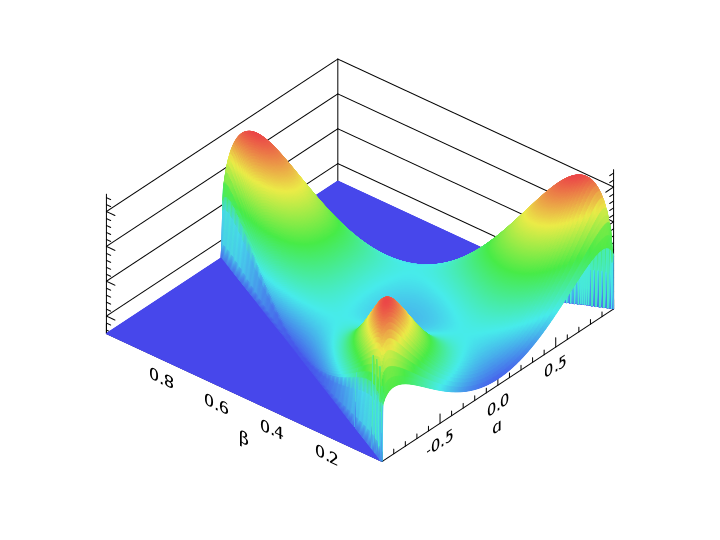}}
			\\
			
			\midrule
			
			\multicolumn{4}{c}{$\alphalambda = 1$}
			\\
			
			\parbox[c]{\plotwa}{\includegraphics[scale=0.1]{Plots/Orthogonal/alpha1/Babich/2D/O}} &
			\parbox[c]{\plotw}{\includegraphics[scale=0.15]{Plots/Orthogonal/alpha1/Babich/3D/O}} &
			\parbox[c]{\plotwa}{\includegraphics[scale=0.1]{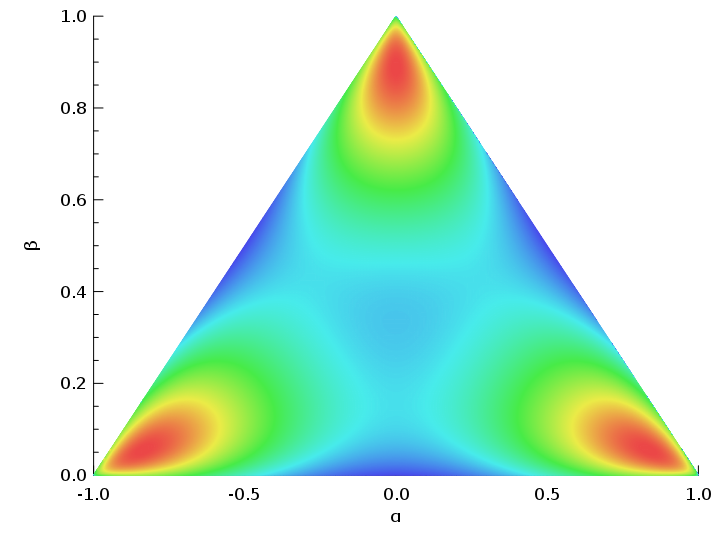}} &
			\parbox[c]{\plotw}{\includegraphics[scale=0.15]{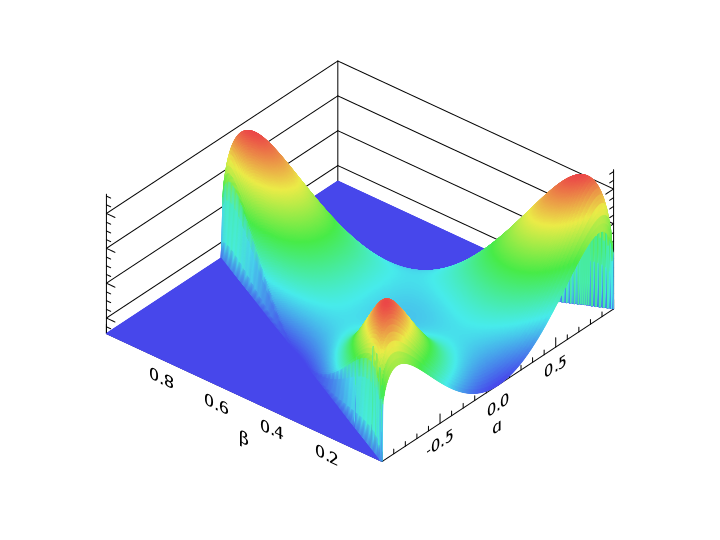}}
			\\

			\midrule
			
			\multicolumn{4}{c}{$\alphalambda = 10$}
			\\
			
			\parbox[c]{\plotwa}{\includegraphics[scale=0.1]{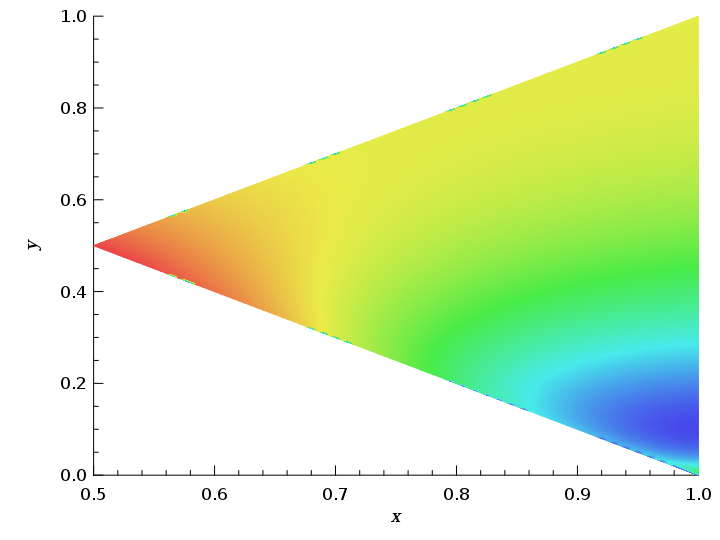}} &
			\parbox[c]{\plotw}{\includegraphics[scale=0.15]{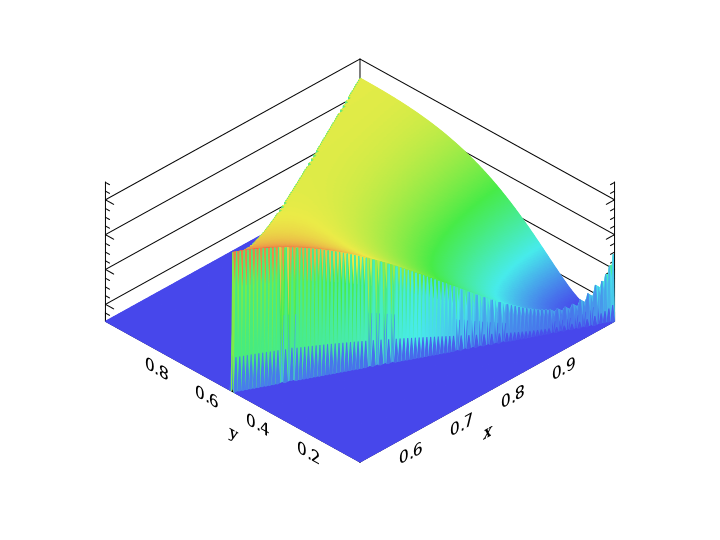}} &
			\parbox[c]{\plotwa}{\includegraphics[scale=0.1]{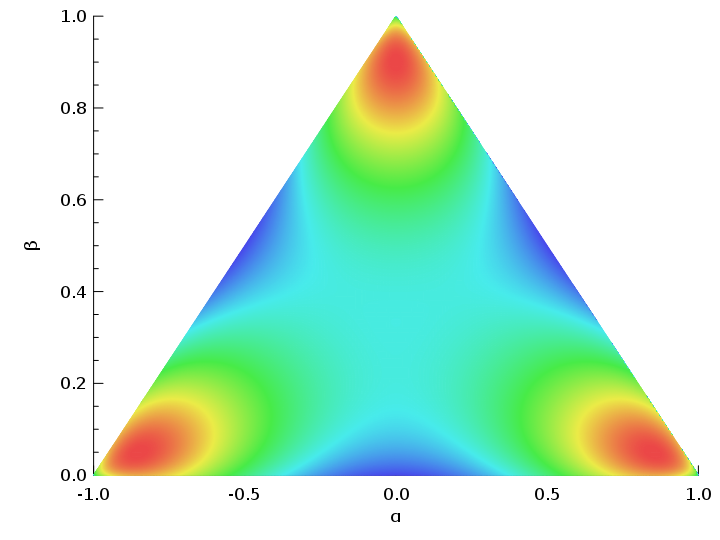}} &
			\parbox[c]{\plotw}{\includegraphics[scale=0.15]{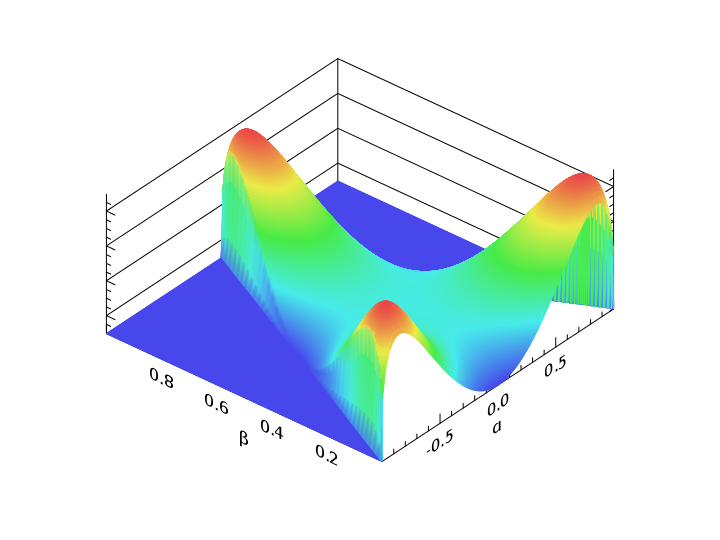}}
			\\

			\bottomrule

		\end{tabular}
	}
	\settowidth{\tblw}{\usebox{\tableA}}
	\addtolength{\tblw}{-1em}

	\begin{center}
		\usebox{\tableA}
	\end{center}
			
	\caption{\label{table:orthogonal-plots}
	The orthogonal bispectrum shape $O$ has zero overlap
	with both the lowest-order possibilities $S_1$ and $S_2$.
	The Babich {\etal} and Fergusson--Shellard plots
	are defined in Table~\ref{table:lowest-order-shapes}.}

	\end{table}

	\begin{table}
	
	\small
	\heavyrulewidth=.08em
	\lightrulewidth=.05em
	\cmidrulewidth=.03em
	\belowrulesep=.65ex
	\belowbottomsep=0pt
	\aboverulesep=.4ex
	\abovetopsep=0pt
	\cmidrulesep=\doublerulesep
	\cmidrulekern=.5em
	\defaultaddspace=.5em
	\renewcommand{\arraystretch}{1.6}

    \sbox{\boxplot}{%
    	\includegraphics[scale=0.1]{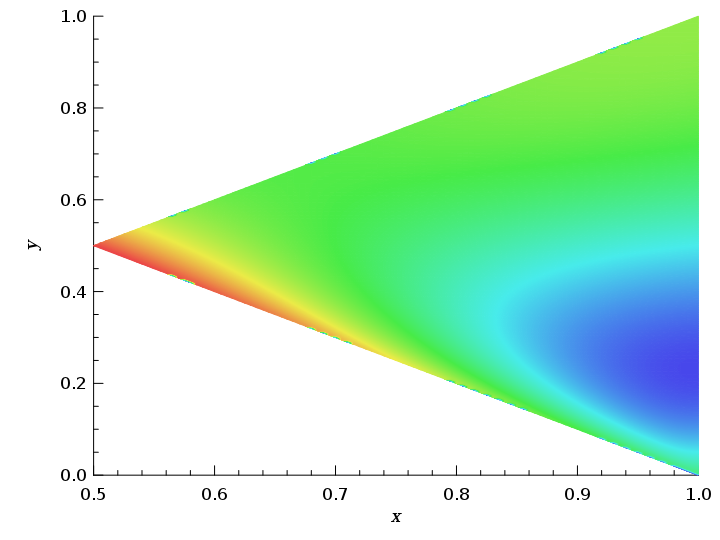}
    }
    \settowidth{\plotw}{\usebox{\boxplot}}
    \sbox{\boxplota}{%
    	\includegraphics[scale=0.15]{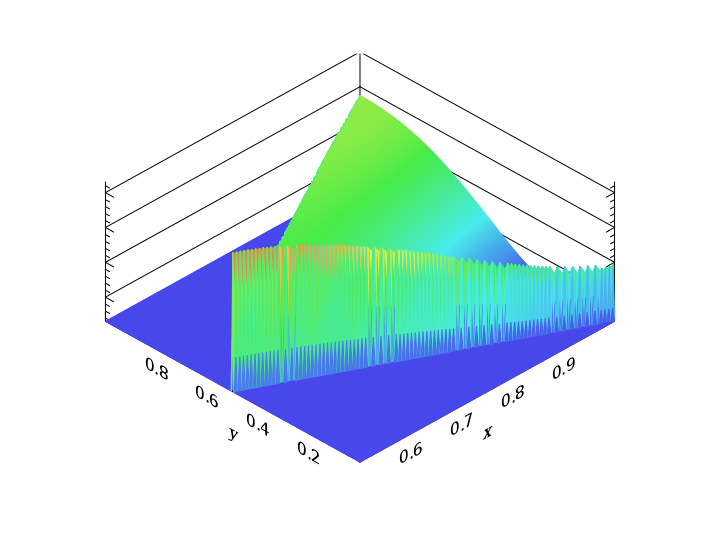}
    }
    \settowidth{\plotwa}{\usebox{\boxplota}}
    
    \sbox{\tableA}{%
		\begin{tabular}{cccc}

			\toprule

			\multicolumn{2}{c}{Babich {\etal}} &
			\multicolumn{2}{c}{Fergusson \& Shellard}
			\\

			\cmidrule(r){1-2}
			\cmidrule(l){3-4}
			
			\parbox[c]{\plotw}{\includegraphics[scale=0.1]{Plots/Creminelli/Babich/2D}} &
			\parbox[c]{\plotwa}{\includegraphics[scale=0.15]{Plots/Creminelli/Babich/3D}} &
			\parbox[c]{\plotw}{\includegraphics[scale=0.1]{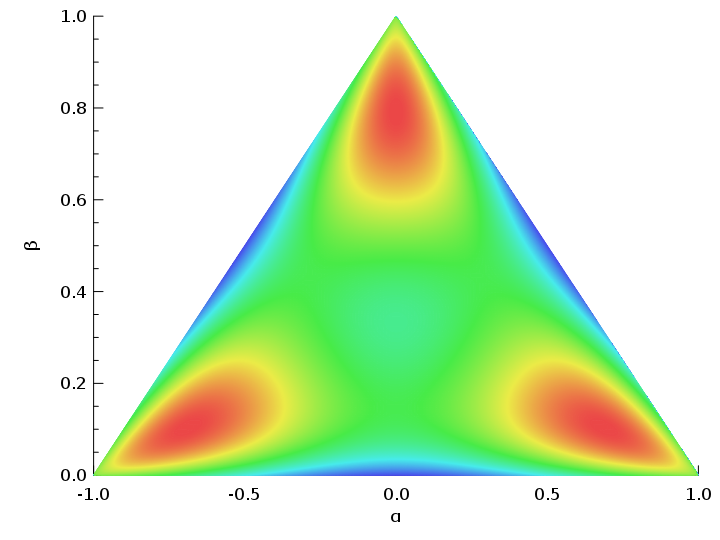}} &
			\parbox[c]{\plotwa}{\includegraphics[scale=0.15]{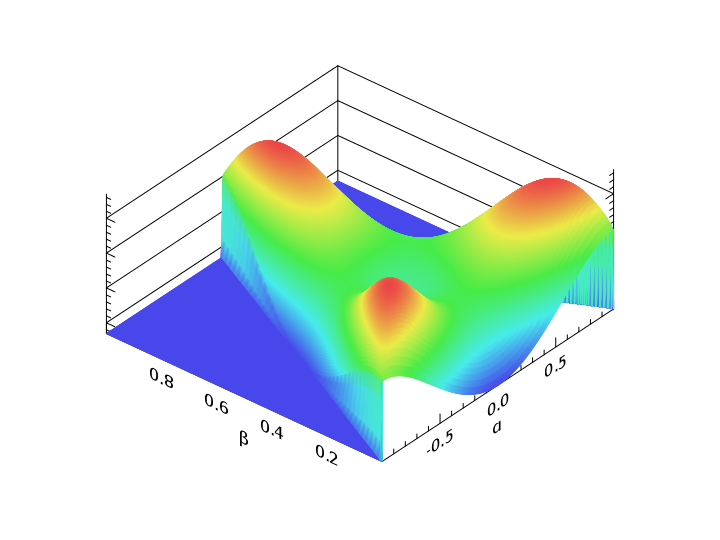}}
			\\

			\bottomrule

		\end{tabular}
	}
	\settowidth{\tblw}{\usebox{\tableA}}
	\addtolength{\tblw}{-1em}

	\begin{center}
		\usebox{\tableA}
	\end{center}
	
	\caption{\label{table:creminelli-shape}Highly orthogonal shape
	constructed by Creminelli {\etal}~\cite{Creminelli:2010qf}.
	The Babich {\etal} and Fergusson--Shellard plots
	were defined in Table~\ref{table:lowest-order-shapes}.}

	\end{table}

	\begin{table}
	\small
	\heavyrulewidth=.08em
	\lightrulewidth=.05em
	\cmidrulewidth=.03em
	\belowrulesep=.65ex
	\belowbottomsep=0pt
	\aboverulesep=.4ex
	\abovetopsep=0pt
	\cmidrulesep=\doublerulesep
	\cmidrulekern=.5em
	\defaultaddspace=.5em
	\renewcommand{\arraystretch}{1.6}

	\sbox{\tableA}{%
		\begin{tabular}{cTcTcT}

			\toprule
			
			$\alphalambda$ &
				\multicolumn{1}{c}{local} &
				\multicolumn{1}{c}{equilateral} &
				\multicolumn{1}{c}{orthogonal} &
				\multicolumn{1}{c}{enfolded} &
				\multicolumn{1}{c}{Creminelli {\etal}\tmark{b}}
			\\
			\cmidrule(r){1-1}
			\cmidrule{2-6}

			$10^{-3}$ &			
				0.35\tmark{a} &
				0.012 &
				0.36 &
				0.32 &
				0.89
			\\

			$1$ &			
				0.38\tmark{a} &
				0.012 &
				0.38 &
				0.33 &
				0.86
			\\
			
			$10$ &
				0.41\tmark{a} &
				0.011 &
				0.40 &
				0.35 &
				0.81
			\\
			
			\bottomrule
		
		\end{tabular}
	}
	\settowidth{\tblw}{\usebox{\tableA}}
	\addtolength{\tblw}{-1em}
	
	\begin{center}
		\usebox{\tableA}
	\end{center}
	
	\renewcommand{\arraystretch}{1.0}
	\tiny
	
	\sbox{\tableB}{%
		\begin{tabular}{l@{\hspace{1mm}}l}	
			\tmark{a} & \parbox[t]{\tblw}{%
				The local template
				is divergent,
				and the values we quote are meaningful only for our choice
				of regulator. For the values quoted above we have
				used $\deltamin = k/k_t = 10^{-3}$, where
				$\deltamin$ was defined in
				\S\ref{sec:shapes}.}
				\\
			
			\tmark{b} & \parbox[t]{\tblw}{%
				For the Creminelli {\etal}
				shape, see Table~\ref{table:creminelli-shape}.}
		\end{tabular}
	}
	
	\begin{center}
		\usebox{\tableB}
	\end{center}

	\caption{\label{table:orthogonal-cosines}Overlap cosines
	between the orthogonal shape $O$ and
	common
	templates, defined in Table~\ref{table:slow-roll-shapes}.
	Sign information has been discarded.}

	\end{table}

	\subsection{Scale dependence}
	\label{sec:running}

	In this section we use the logarithms $\ln k_i / k_\star$
	and $\ln k_t / k_\star$ to study the scale-dependence
	of the three-point function.
	In the squeezed limit this is determined
	by~\eqref{eq:maldacena-condition}.
	The only scale which survives is the common hard momentum
	$\khard$, and the variation of $\fNL$ with this scale is determined by
	the variation of $n_s - 1$. This is typically called the
	\emph{running} of the scalar spectral index \cite{Kosowsky:1995aa},
	and leads to a further consistency relation inherited from
	Maldacena's---and, in general, a hierarchy of such
	consistency equations generated by taking an arbitrary number
	of derivatives.
	In the case of single-field canonical inflation,
	discussed in \S\ref{sec:vanilla} below, we are able to verify this
	explicitly.
	
	Away from the squeezed limit, deformations of
	the momentum triangle may change either its shape or scale.
	Scale dependence occurs in even the simplest models
	for the same reason that the spectrum $\Ps$ and spectral index
	$n_s$ depend on scale
	\cite{Lyth:2005fi,Maldacena:2002vr,Seery:2005wm,Seery:2005gb,Chen:2006nt}.
	Chen introduced a `tilt,' $\nfNL$,
	defined by%
		\footnote{Chen implicitly worked in the equilateral limit
		$k_i  = k$,
		where $k_t = 3k$
		and $\d \ln k_t = \d \ln k$.
		We are defining $\nfNL$ to be the variation of
		$\fNL$ with perimeter for an arbitrary triangle
		if the shape is kept fixed.}
	\cite{Chen:2005fe}
	\begin{equation}
		\nfNL \equiv \frac{\d \fNL}{\d \ln k_t} .
	\end{equation}
	For a fixed triangular shape, this measures changes in $\fNL$
	as the perimeter varies.
	Scale dependence of this type
	was subsequently studied by several authors
	\cite{LoVerde:2007ri,Leblond:2008gg}.
	Observational constraints have been determined
	by Sefusatti {\etal} \cite{Sefusatti:2009xu}.
	Byrnes {\etal}
	performed a similar analysis in the special case of
	multiple-field models producing a local bispectrum
	\cite{Byrnes:2009pe,Byrnes:2010ft}.
	They
	allowed for deformations of the momentum triangle
	including a change of shape, but found
	these to be less important than rescalings of $k_t$.
	
	Shape dependence is often substantially more
	complicated than
	scale dependence.
	Eq.~\eqref{eq:deffnl} makes $\fNL$ dimensionless, but
	contains both powers and logarithms of the $k_i$.
	The powers occur as dimensionless ratios in which $k_t$
	divides out, but the shape dependence remains.
	The argument of each logarithm is also a dimensionless ratio,
	but an extra scale is available:
	the reference scale $k_\star$. When present, this gives rise to the scaling
	logarithms $\ln k_i / k_\star$ and $\ln k_t / k_\star$
	described above, which depend on $k_t$ as well as the shape.
	It follows that a simple way to track the $k_t$-dependence of
	$\fNL$ is to study the $k_\star$-logarithms,
	yielding the identity
	\begin{equation}
		\nfNL = - \frac{\d \fNL}{\d \ln k_\star} .
	\end{equation}

	For a general model, $\nfNL$ can be written
	\begin{equation}
		\begin{split}
		\nfNL =
		\frac{5}{24 z_\star}
		\bigg(
			&
			g_{1\star} H_\star
			\big\{ h_{1\star} -\varepsilon_\star - v_\star   \big\} f_1(k_i)
			+ g_{2\star}
			\big\{ h_{2\star}-v_\star \big\}f_2(k_i)
			\\ & \mbox{}
			+ \frac{ g_{3\star}} {\csstar^2}
			\big\{ h_{3\star}- v_\star - 2 s_\star  \big\}f_3(k_i)
			+ \frac{ g_{4\star}}{ \csstar^2}
			\big\{ h_{4\star} - v_\star \big\}f_4(k_i)
			\\ & \mbox{}
			+ \frac{ g_{5\star}}{ \csstar^2}
			\big\{ h_{5\star} - v_\star \big\}f_5(k_i)
		\bigg)
		\end{split}
	\end{equation}
	where the $f_i(k_i)$ functions are dimensionless ratios of polynomials in
	the $k_i$ which are listed in Table \ref{table:nfNL}.

	\begin{sidewaystable}

	\heavyrulewidth=.08em
	\lightrulewidth=.05em
	\cmidrulewidth=.03em
	\belowrulesep=.65ex
	\belowbottomsep=0pt
	\aboverulesep=.4ex
	\abovetopsep=0pt
	\cmidrulesep=\doublerulesep
	\cmidrulekern=.5em
	\defaultaddspace=.5em
	\renewcommand{\arraystretch}{3}
	\begin{center}
		\small
		\begin{tabular}{SS}

			\toprule
			f_1(k_i) &
			\frac{24k_1^2k_2^2k_3^3}{k_t^3(k_1^3+k_2^3+k_3^3)}
			\\[2mm]

			\rowcolor[gray]{0.9}
			f_2(k_i) &
			\frac{4[k_1^2k_2^2(k_1+k_2)+2k_1^2k_2^2k_3+(k_1+k_2)(k_1^2+k_1k_2+k_2^2)k_3^2+(k_1^2+k_2^2)k_3^3]}{k_t^2(k_1^3+k_2^3+k_3^3)}
	 		\\[2mm]

			f_3(k_i) &
			\frac{2(k_1^2+k_2^2+k_3^2)[k_1^3+2k_1^2(k-2+k_3)+2k_1(k_2^2+k_2k_3+k_3^2)+(k-2+k_3)(k_2^2+k_2k_3+k_3^2)]}{k_t^2(k_1^3+k_2^3+k_3^3)}
			\\[2mm]

			\rowcolor[gray]{0.9} f_4(k_i) &
			\frac{2k_1^5+3k_1^4(k_2+k_3)+(k_2-k_3)^2(k_2+k_3)(2k_2+k_3)(k_2+2k_3)+3k_1(k_2^2-k_3^2)^2-5k_1^3(k_2^2+k_3^2)-k_1^2(k_2+k_3)(5k_2^2+k_2k_3+5k_3^2)}{k_t^2(k_1^3+k_2^3+k_3^3)}
			\\[2mm]

			f_5(k_i) &
			\frac{2k_1^5+k_1^4(k_2+k_3)+k_1(k_2^2-k_3^2)^2-3k_1^3(k_2^2+k_3^2)+(k_2-k_3)^2(k_2+k_3)(2k_2^2+3k_2k_3+2k_3^2)}{k_t^2(k_1^3+k_2^3+k_3^3)}
			\\[2mm]
						
 			\bottomrule
	
		\end{tabular}
	\end{center}
	\caption{Functions determining the momentum dependence of the running of $\fNL$.
	\label{table:nfNL}}
	\end{sidewaystable}	
	
	\para{Squeezed and equilateral limits}%
	In the equilateral limit, we find
	\begin{equation}
		\begin{split}
		\nfNL
		\rightarrow
		- \frac{5}{81 z_\star}
		\bigg(
			&
			g_{1\star} H
			\big\{ \varepsilon_\star + v_\star - h_{1\star} \big\}
			+ 3 g_{2\star}
			\big\{ v_\star - h_{2\star} \big\}
			+ \frac{51 g_{3\star}}{4 \csstar^2}
			\big\{ v_\star + 2 s_\star - h_{3\star} \big\}
			\\ & \mbox{}
			+ \frac{12 g_{4\star}}{4 \csstar^2}
			\big\{ h_{4\star} - v_\star \big\}
			+ \frac{12 g_{5\star}}{4 \csstar^2}
			\big\{ h_{5\star} - v_\star \big\}
		\bigg)
		\end{split}
		\label{eq:nfnl-equilateral}
	\end{equation}
	The squeezed limit gives a simple result,
	\begin{equation}
		\nfNL \rightarrow \frac{5}{24 z_\star} \bigg(
			g_{2\star} \big\{ h_{2\star} - v_\star \big\}
			+ \frac{3 g_{3\star}}{\csstar^2}
			\big\{ h_{3\star} - 2 s_{\star} - v_\star \big\}
		\bigg) .
		\label{eq:nfnl-squeezed}
	\end{equation}
	We define the running of the spectral index, $\alpha_s$, by
	\cite{Kosowsky:1995aa}
	\begin{equation}
		\alpha_s = \frac{\d (n_s - 1)}{\d \ln k} .
		\label{eq:spectral-running}
	\end{equation}
	Compatibility with~\eqref{eq:maldacena-condition}
	in the squeezed limit
	requires
	\begin{equation}
		\nfNL \rightarrow - \frac{5}{12} \left. \alpha_s \right|_{\khard} .
		\label{eq:running-consistency}
	\end{equation}
	In~\S\ref{sec:vanilla}
	we will verify this relation in
	the special case of canonical single-field inflation.

	\section{Tensor modes}
	\label{sec:tensor}

	Inflation will inevitably produce tensor fluctuations to
	accompany the scalar fluctuation $\zeta$.
	Detection of the B-mode polarization signal produced by
	these fluctuations is a major aim of the \emph{Planck}
	satellite and future CMB experiments.
	If it can be measured,
	this signal will provide important constraints on the energy
	scale of inflation.

	In certain models the tensor sector provides sufficient observables
	to allow one or more quantities, such as $\fNL$,
	to be written in terms of other observables.
	In the inflationary literature such relationships are typically
	known as consistency relations,
	and were introduced by Copeland {\etal}
	\cite{Copeland:1993jj,Copeland:1993zn}.
	In the language of particle physics they are
	``observables in terms of observables''---predictions which
	are independent of how we parametrize the theory,
	and which
	the renormalization programme has taught us represent
	the physical content of any quantum field theory.
	Such consistency relations represent important tests of entire
	classes of models.
	To be used effectively with the next-order results of this paper
	we will require next-order predictions for the tensor modes.
	These were obtained by Stewart \& Lyth
	\cite{Stewart:1993bc},
	and are unchanged by the noncanonical action~\eqref{eq:startingaction}.

	In this section our aim is to obtain the next-order
	consistency relation, Eq.~\eqref{eq:next-order-consistency}.
	The tensor fluctuation is a propagating spin-2 mode which belongs
	to the ADM field $h_{ij}$ of~\eqref{eq:adm}. We write
	$h_{ij} = a^2 \e{2\zeta} (\e{\gamma})_{ij}$,
	where $\tr \gamma_{ij} = 0$.
	At quadratic order, the action is
	\cite{Grishchuk:1974ny}
	\begin{equation}
		S_2
		=
		\frac{1}{8}
		\int \d^3x \, \d\tau \; a^2 \bigg[
			\gamma'_{ij} \gamma'_{ij} 
			- \partial_k \gamma_{ij} \partial_k \gamma_{ij}
		\bigg] .
	\end{equation}
	There are two polarizations, traditionally denoted `$+$' and `$\times$,'
	making $\gamma_{ij}$ transverse in the sense
	$\partial_i \gamma_{ij} = 0$.
	Introducing a reference scale $k_\star$ and
	adding the power in each polarization incoherently,
	the resulting dimensionless spectrum can be written
	\cite{Stewart:1993bc}
	\begin{equation}
		\Pstensor
		=
		\frac{2H_{\star}^2}{\pi^2}
		\left[
			1
			+ 2\varepsilon_{\star} \left(
				1
				- \EulerGamma
				- \ln\frac{2k}{k_{\star}}
			\right)
		\right] .
	\end{equation}
	This is the sum of
	two copies of the power
	spectrum for a massless scalar field with $\cs = 1$, and
	is conserved on superhorizon scales.
	Including next-order corrections, the scale dependence of $\Pstensor$
	is measured by the tilt $n_t$,
	\begin{equation}
		\ntstar
		\equiv
		\dfrac{\d \ln \Pstensor}{\d \ln k}
		=
		-2 \varepsilon_{\star} \left[
			1
			+ \varepsilon_{\star}
			- \eta_{\star} \left(
				1-\EulerGamma - \ln \frac{2k}{k_\star}
			\right)
		\right] .
		\label{eq:nt}
	\end{equation}
	It is conventional to measure the amplitude of tensor
	fluctuations relative to $\zeta$.
	One defines the tensor-to-scalar ratio, $r$, by the rule
	\cite{Copeland:1993zn}
	\begin{equation}
		r
		\equiv
		\dfrac{\Pstensor}{\Ps} , 
	\end{equation}
	where $\Ps$ is the dimensionless version of
	Eq.~\eqref{eq:powerspectrum}. We find
	\begin{equation}
		r_\star
		\simeq
		16 \varepsilon_{\star} \csstar \left[
			1
			- 2 \eta_\star
			+ ( s_\star + \eta_\star )
			\left( \EulerGamma + \ln \frac{2k}{k_\star}
			\right)
		\right] .
	\end{equation}
	In canonical models, $r$ can be written purely in terms of observable
	quantities. In the noncanonical case this is not automatically
	possible without the addition of new observables.
	In general,
	\begin{equation}
		r_\star = - 8 \ntstar \csstar
			\left[
				1
				- \varepsilon_\star
				- \eta_\star
				+ s_\star \left(
					\EulerGamma + \ln \frac{2k}{k_\star}
				\right)
			\right] .
		\label{eq:next-order-consistency}
	\end{equation}
	One may use the lowest-order result for $\nt$ to eliminate $\varepsilon$.
	To eliminate $\eta$ would require the scalar spectral index,
	$n_s$.
	It is possible to use $\fNL$ to rewrite $\cs$ in the prefactor
	\cite{Lidsey:2006ia},
	but in doing so one introduces dependence on the parameter $\ell$.
	Therefore at least two extra observables would be required to eliminate the
	dependence on $s$ and $\ell$.
	If these depend on $t$, $\xi$ or similar parameters, then further
	observables could be required.
	We conclude that at next-order, for a general
	$P(X,\phi)$ Lagrangian, the observables $\{ r, n_s, n_t, \fNL \}$ do not
	form a closed set.
	
	\section{Canonical single-field inflation}
	\label{sec:vanilla}

	The simplest model of inflation comprises a single scalar field
	with canonical kinetic terms.
	Maldacena showed that
	the fluctuations in this model are almost
	Gaussian,
	with $\fNL$ of order
	$r$ \cite{Maldacena:2002vr}.
	This is unobservably small.
	In a canonical model, $\cs = 1$.
	
	\para{Nonlinearity parameter}%
	To calculate $\fNL$ we require the flow parameters
	$h_i$, which measure time dependence
	in the vertex factors $g_i$.
	These are
	\begin{equation}
		\begin{aligned}
		h_1 &
		=
		0
		&
		h_2 &
		=
		\frac{\eta ( 2 \epsilon - \eta - \xi )}{\epsilon - \eta}
		&
		&
		\\
		h_3 &
		=
		\frac{\eta ( 2 \epsilon + \eta + \xi )}{\epsilon + \eta}
		&
		h_4 &
		=
		\frac{\eta ( 8 - \epsilon )}{4}
		&
		h_5
		&
		=
		3\eta .
		\end{aligned}
		\quad
		\Bigg\}
		\label{eq:h_i}
	\end{equation}
	In this model
	the time-dependence of $z$ is described by $v=\eta$.
	Collecting contributions from
	Table~\ref{table:fNLequilateral},
	the equilateral limit of $\fNL$ can be
	written
	\begin{equation}
		\begin{split}
			\fNL \rightarrow
			\frac{5}{36} \bigg[
				11 \varepsilon_\star
				+ 3 \eta_\star
				&
				+ \frac{35 \varepsilon_\star^2}{216} \bigg\{
					768 \omega - 54
				\bigg\}
				+ \frac{35 \eta_\star \xi_\star}{36} \bigg\{
					3 \EulerGamma 
					- 8
					+ 3 \ln \frac{3k}{k_\star}
				\bigg\}
				\\ & \mbox{}
				+ \frac{35 \varepsilon_\star \eta_\star}{36} \bigg\{
					11 \EulerGamma
					- 14
					+ 64 \omega
					+ 11 \ln \frac{3k}{k_\star}
				\bigg\}
			\bigg] ,
		\end{split}
		\label{eq:fnlevanilla}
	\end{equation}
	where we have used the numerical constant $\omega = \coth^{-1} 5$,
	and $k$ should be regarded as the common momentum scale,
	$k_i = k$.
	The squeezed limit may be recovered from Table~\ref{table:fNLsqueezed}.
	We find
	\begin{equation}
		\begin{split}
			\fNL
			\rightarrow
			\frac{5}{12} \bigg[
				2 \varepsilon_\star
				+ \eta_\star
				+ 2 \varepsilon_\star^2
				+ \eta_\star \xi_\star \bigg\{
					\EulerGamma
					- 2
					+ \ln \frac{2 k}{k_\star}
				\bigg\}
				+ \varepsilon_\star \eta_\star \bigg\{
					2 \EulerGamma
					- 1
					+ 2 \ln \frac{2 k}{k_\star}
				\bigg\}	
			\bigg] ,
		\end{split}
		\label{eq:fnlsvanilla}
	\end{equation}
	where $k$ should now be regarded as the scale of the hard
	momenta in the correlation function.
	In \S\ref{subsec:fnl} and \S\ref{sec:running}
	we emphasized that $\fNL$ should
	be finite in this limit, containing
	no large logarithms, because these factorize into
	the power spectrum and are subtracted.
	The remaining logarithms [the $\ln 2k / k_\star$
	terms in~\eqref{eq:fnlsvanilla}]	
	track the
	dependence of $\fNL$ on the hard scale, and will be studied below.
	Using~\eqref{eq:maldacena-condition} and comparing with
	the spectral indices quoted in
	Table~\ref{table:spectral-index},
	it is easy to check that our formula correctly reproduces
	the Maldacena limit.
	We note that, strictly, one should regard agreement in this limit as an
	accident which happens because the simple slow-roll model
	contains no other scale which could interfere with factorization
	of the correlation function.
	
	\para{Scale dependence of $\fNL$}%
	Specializing to the equilateral limit of $\nfNL$, we find
	\begin{equation}
		\nfNL \rightarrow \frac{5}{216} \eta_\star
			( 66 \varepsilon_\star + 18 \xi_\star ) .
	\end{equation}
	In the squeezed limit one obtains
	\begin{equation}
		\nfNL \rightarrow \frac{5}{12} \eta_\star
			( 2 \varepsilon_\star + \xi_\star ) 
			= - \frac{5}{12} \alpha_{s \star} ,
		\label{eq:runningsvanilla}
	\end{equation}
	which correctly describes the running of the scalar
	spectral index, $\alpha_{s\star}$,
	in agreement with~\eqref{eq:running-consistency}.
	The consistency
	conditions~\eqref{eq:fnlsvanilla}
	and~\eqref{eq:runningsvanilla} represent a
	nontrivial check on the correctness of our calculation.
	In particular,
	throughout the calculation
	we have cleanly separated the conceptually different
	scales $k_t$ and $k_\star$. Therefore
	the correct formula~\eqref{eq:runningsvanilla} is \emph{not}
	simply
	a consequence of obtaining the correct lowest-order terms
	in~\eqref{eq:fnlsvanilla}---although,
	as described in \S\ref{subsec:second},
	it can be obtained from these
	by differentiating with respect to $\ln k_{\star}$
	after setting $k_\star = k_t$.
	
	\section{Non-canonical single-field inflation}
	\label{sec:noncanonical}
		
	For the noncanonical action~\eqref{eq:startingaction},
	the $h_i$ can be written
	\begin{equation}
		\begin{aligned}
			h_1 &
			= \varepsilon + \eta - 2s +
				\frac{\frac{2\lambda}{\Sigma}
					(\eta - 2 \epsilon - 2 s - \ell) - \frac{2}{\cs^2} s}
					{\frac{1-\cs^2}{\cs^2} - \frac{2\lambda}{\Sigma}}
			&
			h_2 &
			=
			\eta - 4s +
				\frac{\eta ( \epsilon - \xi ) + 6 \cs^2 s}
					{\eta - \epsilon - 3 ( 1 - \cs^2 )}
			\\
			h_3 &
			=
			\eta - 2s +
				\frac{\eta ( \epsilon + \xi ) - 2s ( t - \cs^2 )}
					{\epsilon + \eta - 2s + ( 1 - \cs^2 )}
			&
			h_4 &
			=
			2 \eta - 4s -
				\frac{\eta \epsilon}{4 - \epsilon}
			\\
			h_5 &
			=
			3 \eta - 4 s ,
		\end{aligned}
		\quad
		\Bigg\}
		\label{eq:general-hi}
	\end{equation}
	where we have defined $\xi = \dot{\eta} / H \eta$,
	$t = \dot{s} / H s$.
	
	In the canonical case, it was possible to verify Maldacena's
	consistency condition to next-order.
	In the noncanonical case this is not possible without a next-next-order
	calculation, because
	for $\cs \neq 1$ the leading contribution to $\fNL$ is $\Or(1)$
	in the slow-variation expansion. Therefore our calculation of subleading
	corrections produces a result valid to $\Or(\varepsilon)$, which is
	short of the $\Or(\varepsilon^2)$ accuracy required to verify the
	consistency condition at next-order.
	Chen {\etal} gave the subleading corrections in terms
	of undetermined integrals \cite{Chen:2006nt}. Expanding these
	asymptotically,
	they argued that the consistency relation would be satisfied
	at lowest-order.
	More recently, Renaux-Petel \cite{RenauxPetel:2010ty}
	gave an equivalent demonstration. Here, we have knowledge of the
	full bispectrum to subleading order. Using~\eqref{eq:general-hi},
	it can be verified that in the squeezed limit,
	and expanding around a reference scale $k_\star$,
	\begin{equation}
		\fNL \rightarrow \frac{5}{12} \left(
			2 \varepsilon_\star + \eta_\star + s_\star
		\right) .
	\end{equation}
	One may check that this agrees with
	Eq.~\eqref{eq:maldacena-condition} and Table~\ref{table:spectral-index}.
	We expect $\nfNL = \Or(\varepsilon^2)$, and
	therefore a next-next-order calculation is required
	to estimate the running of $\fNL$ in noncanonical models.
	
	\subsection{Asymptotically power-law models}
	\label{sec:power-law}
	
	Power-law inflationary models were introduced by
	Lucchin \& Matarrese \cite{Lucchin:1984yf,Lucchin:1985wy}, who
	studied potentials producing an expansion history of the
	form $a(t) \propto t^{1/\varepsilon}$.
	The exponent $1/\varepsilon$ is the usual parameter
	$\varepsilon = - \dot{H}/H^2$.
	It need not be small, but should be taken as constant
	which makes $\eta = \xi = 0$.
	The solution is inflating provided $\varepsilon < 1$.
	Exact solutions can be found in the canonical case,
	which form the basis of the next-order calculation
	\cite{Lidsey:1995np}.
	
	In this section we study two examples which
	are asymptotically described by noncanonical power-law
	inflation at late times.
	The first is Dirac--Born--Infeld (``DBI'') inflation,
	which produces a scale-invariant power spectrum
	at lowest-order. Departures from scale invariance
	appear at next-order.
	These properties imply that we can compare our results
	to a formula of Khoury \& Piazza
	which was obtained without invoking the
	slow-roll approximation \cite{Khoury:2008wj}.
	Our second example is $k$-inflation, for which the power
	spectrum is not scale invariant at lowest-order,
	and to which Khoury \& Piazza's result does not apply.

	\subsubsection{Dirac--Born--Infeld inflation}
	\label{sec:dbi}
	The DBI action is a low-energy effective theory
	which
	describes a D3-brane moving in a warped throat.
	It was
	proposed as a model of inflation by Silverstein \& Tong
	\cite{Silverstein:2003hf}, and subsequently developed by
	Alishahiha {\etal} \cite{Alishahiha:2004eh}.
	The action is of the form~\eqref{eq:startingaction},
	with $P(X, \phi)$ satisfying
	\begin{equation}
 		P(X, \phi)
 		=
 		-\dfrac{1}{f(\phi)}
 		\left[
 			\sqrt{1-f(\phi) X} -1
 		\right]
 		-
 		V(\phi)
		\label{eq:dbiaction}
	\end{equation} 
	where $f$ is an arbitrary function of $\phi$
	known as the \emph{warp factor},
	and $V(\phi)$ is a potential arising from couplings between
	the brane and other degrees of freedom.
	The DBI Lagrangian is algebraically special
	\cite{Lidsey:2006ia,deRham:2010eu,Leblond:2008gg}
	and enjoys a number of remarkable properties, including
	a form of nonrenormalization theorem
	\cite{Shmakova:1999ai,DeGiovanni:1999hr,Tseytlin:1999dj}.
	In principle non-minimal curvature couplings
	can be present, of the form $R\phi^2$,
	which spoil inflation
	\cite{Easson:2009kk}.
	This gives a form of the $\eta$-problem,
	and
	we assume such terms to be negligible.

	Eq.~\eqref{eq:dbiaction}
	makes $2\lambda / \Sigma = (1-\cs^2)/\cs^2$,
	which requires $g_1 \rightarrow 0$ but
	causes the denominator of $h_1$ in~\eqref{eq:general-hi}
	to diverge.
	Only the finite combination $g_1 h_1$ appears in physical quantities,
	and it can be checked that $g_1 h_1 \rightarrow 0$ as required.
	The square root in~\eqref{eq:dbiaction} must be real,
	giving a dynamical speed limit for $\phi$.
	It is conventional to define a
	Lorentz factor
	\begin{equation}
		\gamma
		\equiv
		(1-f \dot{\phi}^2)^{-1/2} .
	\end{equation}
	When $\gamma \sim 1$ the motion is nonrelativistic.
	When $\gamma \gg 1$, the brane is moving close to the speed limit.
	The Lorentz factor is related to the speed of sound by
	$\cs = \gamma^{-1}$.
	
	Silverstein \& Tong \cite{Silverstein:2003hf} argued
	that~\eqref{eq:dbiaction} supported attractor solutions
	described at late times by power-law inflation.
	In this limit,
	the slow-variation parameters $\varepsilon$ and $s$ are
	constant, with $\eta = \xi = t = 0$ but $\ell$ not zero.
	Variation of the sound speed gives $s = - 2\varepsilon$, making
	$\varpi = 0$ and yielding 
	scale-invariant fluctuations at
	lowest-order [cf.~\eqref{eq:rge}].
	In the equilateral limit,%
		\footnote{Recall that the equilateral limit is
		$\fNL(k,k,k)$ and is not the quantity constrained by
		experiment.
		See footnote~\ref{footnote:equilateral}.}
	we find that $\fNL$ satisfies
	\begin{equation}
		\fNL \rightarrow
		- \frac{35}{108} ( \gamma^2_\star - 1) \left[
			1 - \frac{\gamma_\star^2}{\gamma_\star^2-1}
			( 3 - 4 \EulerGamma )\varepsilon
			+ \Or(\gamma_\star^{-2})
		\right] .
		\label{eq:dbi-fnl-eq}
	\end{equation}
	In \S\ref{sec:introduction} we estimated the
	relative uncertainty in
	$\fNL$ to be $\sim 14 \varepsilon$,
	working in the limit $\gamma \gg 1$, based on
	$\Or(\varepsilon)$ terms from the vertices only.
	Eq.~\eqref{eq:dbi-fnl-eq} shows that,
	due to an apparently fortuitous cancellation,
	this large contribution is almost
	completely subtracted to leave a small fractional
	correction $\sim 0.69 \varepsilon$.
	
	In the squeezed limit we find
	\begin{equation}
		\fNL \rightarrow
		\frac{10 \gamma^2_\star}{3} \varepsilon^2
		\left(
			4 \EulerGamma
			- 5
			+ 4 \ln \frac{4k}{k_\star}
		\right)
		.
	\end{equation}
	which is $\Or(\varepsilon^2)$, as predicted by the
	Maldacena condition~\eqref{eq:maldacena-condition}
	and the property $\varpi = 0$
	\cite{Alishahiha:2004eh}.
	
	\para{Comparison with previous results}%
	Khoury \& Piazza estimated
	the bispectrum in
	a power-law model
	satisfying $\varpi = 0$
	without invoking an expansion in slow-variation parameters
	\cite{Khoury:2008wj}.%
		\footnote{See also Baumann {\etal} \cite{Baumann:2011dt}.}
	They quoted their
	results in terms of a quantity $f_X$
	which replaces $\lambda$,
	\begin{equation}
		\lambda \equiv \frac{\Sigma}{6} \left(
			\frac{2 f_X +1}{\cs^2} - 1
		\right) .
	\end{equation}
	For the DBI model, $f_X = 1 - \cs^2$.
	They assumed constant $f_X$, making their
	result valid to all orders in $\varepsilon$
	but only lowest-order in the time dependence of $f_X$.
	Working in the
	equilateral limit
	for arbitrary constant $f_X$
	we find
	\begin{equation}
		\begin{split}
			\fNL \rightarrow
			&
			- \frac{5}{972 \csstar^2} \big[
				55 ( 1 - \csstar^2 ) + 8 f_{X}
			\big]
			\\ & \mbox{}
			 +
			 \frac{5 \varepsilon}{972 \csstar^2}
			 \left[
			 	149
			 	- 8 \csstar^2
			 	- 220 \EulerGamma
			 	- 220 \ln \frac{3k}{k_\star}
			 	+ f_X \left\{
			 		40
			 		- 32 \EulerGamma
			 		- 32 \ln \frac{3k}{k_\star}
			 	\right\}
			 \right] .
		\end{split}
		\label{eq:khoury-piazza}
	\end{equation}
	Adopting the evaluation point $k_\star = 3k$, this precisely
	reproduces (8.4) of Khoury \& Piazza \cite{Khoury:2008wj}
	when expanded to order $\varepsilon$.
	Although~\eqref{eq:khoury-piazza}
	does not strictly apply to DBI, where
	$\cs$ is changing,
	it can be checked that effects due to the time dependence
	of $f_X$ do not appear at next-order.
	Indeed, Eq.~\eqref{eq:khoury-piazza} yields~\eqref{eq:dbi-fnl-eq}
	when $f_X = 1 - \gamma^{-2}$.
	
	\para{Generalized DBI inflation}%
	The foregoing analysis was restricted to the
	asymptotic power-law regime, but this is not required.
	Using an arbitrary potential $V(\phi)$ in~\eqref{eq:dbiaction}
	one can construct a generic quasi-de Sitter background.
	Many of their properties, including the attractor behaviour,
	were studied by Franche {\etal} 
	\cite{Franche:2009gk}.
	However,
	in the absence of a controlled calculation of next-order
	terms it has not been possible to estimate
	corrections from the shape of $V(\phi)$ or $f(\phi)$.
	Analogous effects have been computed for Galileon
	inflation~\cite{Burrage:2010cu}, but
	our computation enables us to determine them in
	the DBI scenario for the first time.	
	For $\gamma \gg 1$ the noncanonical structure suppresses
	background dependence
	on details of the potential.
	But small fluctuations around the background
	cannot be shielded from these details,
	which induce three-body interactions
	whether or not they are relevant
	for supporting the quasi-de Sitter epoch.
	These interactions generate relatively unsuppressed
	contributions to the three-point function.
	
	We adapt the notation of Franche {\etal},
	who defined
	quantities measuring the shape of the potential $V$ and the warp
	factor $f$,%
		\footnote{The factor $\sgn( \dot{\phi} f^{1/2} )$ was not
		used by Franche {\etal} \cite{Franche:2009gk}, but is necessary
		here because the relativistic background solution
		requires $f X = 1 + \Or(\gamma^{-2})$
		up to corrections suppressed by
		$\Or(\varepsilon)$ which are higher-order than those we retain.
		Depending on the direction of motion,
		this yields $\dot{\phi} = \pm f^{-1/2} + \Or(\gamma^{-2})$.}
	\begin{equation}
		\epsilonSR
			= \frac{1}{2} \left( \frac{V'}{V} \right)^2 ,
		\quad
		\etaSR
			= \frac{V''}{V} ,
		\quad
		\text{and}
		\quad
		\Delta
			= \sgn(\dot{\phi}f^{1/2} )
			\frac{f'}{f^{3/2}} \frac{1}{3H} .
	\end{equation}
	The same branch of $f^{1/2}$ should be chosen in computing
	$f^{3/2}$ and $\sgn( \dot{\phi} f^{1/2} )$.
	Note that these shape parameters do \emph{not} coincide with
	the global slow-variation parameters
	$\varepsilon$ and $\eta$.
	Franche {\etal} argued that $\Delta \ll 1$ was required
	to obtain attractor solutions, which
	we will assume to be satisfied in what follows.
	In addition,
	we work in the equilateral limit
	and take $\gamma \gg 1$, which is the regime of principal
	interest for observably large $\fNL$.
	We find
	\begin{equation}
		\varepsilon
			= \frac{\epsilonSR}{\gamma},
		\quad
		\eta = 3 \frac{\epsilonSR}{\gamma}
			- \frac{\etaSR}{\gamma}
			- \frac{3}{2} \Delta ,
		\quad
		\text{and}
		\quad
		s = - \frac{\epsilonSR}{\gamma}
			+ \frac{\etaSR}{\gamma}
			- \frac{3}{2} \Delta .
		\label{eq:franche-shape-parameters}
	\end{equation}
	The leading term of~\eqref{eq:dbi-fnl-eq}
	is unchanged,
	but the subleading terms are dominated by
	shape-dependent corrections,
	\begin{equation}
		\begin{split}
			\fNL = - \frac{35\gamma_\star^2}{108}
				\Bigg[
					1
					& \mbox{}
					+ \frac{3 \Delta_\star}{14}
					\Big(
						31
						+ 14 \EulerGamma
						- 228 \omega
						+ 14 \ln \frac{2k}{k_\star}
					\Big)
					+ \frac{\etaSRstar}{7 \gamma_\star}
					\Big(
						3
						- 14 \EulerGamma
						- 14 \ln \frac{3k}{k_\star}
					\Big)
					\\ & \mbox{}
					- \frac{2 \epsilonSRstar}{7 \gamma_\star}
					\Big(
						43
						- 7 \EulerGamma
						- 256 \omega
						- 7 \ln \frac{3k}{k_\star}
					\Big)
					+ \Or(\gamma_\star^{-2})
				\Bigg] .
		\end{split}
	\end{equation}
	These subleading terms
	are more important than those of~\eqref{eq:dbi-fnl-eq},
	which began at relative order $\gamma^{-2}$
	and are therefore strongly suppressed for $\gamma \gg 1$.
	Moreover, inflation can occur even for relatively large values of
	$\epsilonSR$ and $\etaSR$---%
	roughly, whenever $\epsilonSR / \gamma < 1$---%
	so these corrections need not be extremely small.
	For large $|\fNL|$,
	we estimate the relative correction to be
	\begin{equation}
		\frac{\Delta \fNL}{\fNL}
		\approx
		- 2.75  \Delta_\star
		+
		\frac{2.10 \epsilonSRstar - 0.41 \etaSRstar}
		{|\fNL|^{1/2}} .
		\label{eq:dbi-shape-corrections}
	\end{equation}
	For negative equilateral $\fNL$,
	current constraints approximately require
	$|\fNL|^{1/2} \lesssim 12$ \cite{Komatsu:2010fb}.
	Therefore, these corrections can be rather important unless
	the potential is tuned to be flat,
	although some cancellation occurs because $\epsilonSR$
	and $\etaSR$ enter with opposite signs.
	
	To obtain an estimate, suppose that $\Delta_\star$ is negligible.
	Taking the extreme $95\%$-confidence value $\fNL = -151$
	\cite{Komatsu:2010fb} and
	$\epsilonSR \approx |\etaSR| \sim 1$ to obtain an estimate
	for a ``generic'' potential,
	the correction is of order $14\%$ if $\etaSR > 0$
	and $20\%$ if $\etaSR < 0$.
	To reduce these shifts
	one might be prepared to tolerate a small
	tuning, giving perhaps $\epsilonSR \approx |\etaSR| \sim 0.1$
	and suppressing the correction to the percent level.
	However, the corrections grow with decreasing $|\fNL|$.
	Keeping the generic estimate $\epsilonSR \approx |\etaSR| \sim 1$,
	and using
	$|\fNL| \approx 50$, for which $\gamma \approx 10$ and
	the approximation $\gamma \gg 1$ used to
	derive~\eqref{eq:dbi-shape-corrections}
	is at the limit of its applicability,
	we find the corrections to be of order
	$24\%$ for $\etaSR > 0$
	and $36\%$ for $\etaSR < 0$.
	
	Although Franche {\etal} argued that $\Delta_\star$ must be small
	to obtain attractor behaviour, it need not be entirely negligible.
	In such cases it introduces a dependence on the shape of the
	warp factor in addition to the shape of the potential.
	This may be positive or negative.
	If the $\Delta_\star$ and $\varepsilon_\star$ terms add
	constructively, the next-order correction can become rather large.
	
	\para{Infrared model}%
	The 
	DBI scenario can be
	realized in several ways. The original ``ultraviolet'' model
	is now disfavoured by microscopic considerations
	\cite{Baumann:2006cd,Lidsey:2006ia}.
	Chen introduced \cite{Chen:2004gc,Chen:2005ad,Chen:2005fe}
	an alternative ``infrared'' implementation which
	evades these constraints and remains compatible with
	observation
	\cite{Bean:2007eh, Bean:2007hc, Peiris:2007gz, Alabidi:2008ej}.
	In this model
	the warp factor
	$f(\phi)$ is $\lambda/\phi^4$, where $\lambda$ is a dimensionless
	parameter.
	The potential is $V(\phi) = V_0 - \beta^2 H^2 \phi^2 / 2$,
	in which the mass is expressed
	as a fraction $\beta^{1/2}$ of the Hubble scale.
	The constant term $V_0$ is taken to dominate, making
	$\varepsilon$ negligible.
	However the remaining slow-variation parameters need not be small.
	It is convenient to express our results in terms of the number, $N_e$, of
	e-folds to the end of inflation.
	Background quantities evaluated at this time are denoted by a subscript
	`$e$'.
	Computing~\eqref{eq:franche-shape-parameters}
	with these choices of $V$ and $f$
	we find
	$\eta_e \approx 3/N_e$
	and $s_e \approx 1/N_e$.
	The infrared model is an example where $\Delta_e$ is not
	negligible, being also of order $1/N_e$.
	Specializing~\eqref{eq:dbi-shape-corrections} to this case, we find
	\begin{equation}
		\frac{\Delta \fNL}{\fNL}
		\approx
		- \frac{1}{7 N_e} \left(
			65
			+ 14 \EulerGamma
			- 484 \omega
			+ 14 \ln \frac{2k}{k_\star}
		\right)
		\approx
		\frac{4.39}{N_e} ,
	\end{equation}
	where we have chosen $k_\star = 3k$ in the final step. Adopting the
	best-fit value $N_e \approx 38$ suggested by
	Bean {\etal}~\cite{Bean:2007eh}, we find a fractional correction of
	order $12\%$.
	This relatively small correction is a consequence of the
	negligible $\varepsilon$ in this model.
	The analysis of Bean {\etal} gave a reasonable fit for a range of
	$\beta$ of order unity.
	Keeping $N_e \approx 38$
	and using the maximum likelihood value $\beta = 1.77$
	quoted by Bean {\etal}, we find $\Delta \fNL \approx -19$.
	The corresponding shift is from
	$\fNL \approx -163$ without next-order corrections to
	$\fNL \approx -182$ with next-order corrections included.

	\subsubsection{$k$-inflation}
	\label{sec:k-inflation}
	
	The $k$-inflation model of Armend\'{a}riz-Pic\'{o}n {\etal}
	\cite{ArmendarizPicon:1999rj} also
	admits power-law solutions.
	The action is
	\begin{equation}
		P(X, \phi) = \frac{4}{9} \frac{4 - 3 \gamma}{\gamma^2}
			\frac{X^2 -X}{\phi^2} ,
		\label{eq:k-inflation-power-law}
	\end{equation}
	where $\gamma$ is a constant, no longer related to the speed
	of sound by the formula $\cs = \gamma^{-1}$ which applied for DBI.
	Unlike the DBI Lagrangian,
	Eq.~\eqref{eq:k-inflation-power-law} is unlikely to be radiatively
	stable and its microscopic motivation is uncertain. Nevertheless,
	nongaussian properties
	of the inflationary fluctuations in this model
	were studied by Chen {\etal} \cite{Chen:2006nt}.
	There is a solution with
	\begin{equation}
		X = \frac{2 - \gamma}{4 - 3 \gamma} ,
	\end{equation}
	making $\varepsilon = 3 \gamma / 2$ and
	$\cs$ constant. Therefore this model has
	$s = 0$ but $\varepsilon \neq 0$, and is not
	scale-invariant even at lowest-order.
	Inflation occurs if $0 < \gamma < 2/3$.
	The lowest-order contribution to $\fNL$ is of order
	$1/\gamma$, making the next-order term of order unity.
	A next-next-order calculation would be required to accurately
	estimate the term of order $\gamma$.
	
	In the equilateral limit,
	Chen {\etal} quoted the lowest-order result
	$\fNL \approx - 170 / 81 \gamma$.
	Proceeding as in \S\ref{sec:introduction}, one can estimate the
	fractional
	theoretical uncertainty in this prediction
	to be $\sim 9 \gamma$, or roughly $\pm 20$.
	This is comparable to the
	\emph{Planck} error bar, and is likely to exceed the error bar
	achieved by a subsequent CMB satellite.
	Still working in the equilateral limit, we find
	\begin{equation}
		\fNL \rightarrow - \frac{170}{81 \gamma} \left[
			1 - \frac{\gamma}{34} \left( 61 - 192 \ln \frac{3}{2} \right)
			+ \Or(\gamma^2)
		\right] .
		\label{eq:k-inflation-fnl-equilateral}
	\end{equation}
	As for DBI inflation, a fortuitous cancellation brings the
	fractional correction down from our estimate
	$\sim 9 \gamma$ to $\sim 0.5 \gamma$.
	It was not necessary to choose a reference scale $k_\star$ in order
	to evaluate~\eqref{eq:k-inflation-fnl-equilateral}.
	In DBI inflation, to the accuracy of our calculation,
	the power spectrum is scale invariant but $\fNL$
	is not.
	For the power-law solution of $k$-inflation, with the same
	proviso, it transpires that
	$\fNL$ is scale invariant even though the
	the power spectrum is not.
	
	Because $\varpi \neq 0$ in this model, the analysis of
	Khoury \& Piazza does not apply.
	In a recent preprint, Noller \& Magueijo gave a generalization
	which was intended to be valid for
	small $\varpi$ and constant but otherwise arbitrary
	$\varepsilon$ and $s$
	\cite{Noller:2011hd}.
	Their analysis also assumes constant $f_X$.
	We set the reference scale to be $k_\star = 3k$
	and
	work in the equilateral limit for arbitrary constant
	$f_X$. One finds
	\begin{equation}
		\begin{split}
			\fNL \rightarrow
				& - \frac{5}{972 \csstar^2}
				\big[
					55 (1 - \csstar^2) + 8 f_{X}
				\big]
				\\ & \mbox{}
				+ \frac{5\varepsilon}{972 \csstar^2}
				\left[
					177
					+ 120 \csstar^2
					- 1024 \omega ( 1 - \csstar^2 )
					+ f_X \big\{
						264
						- 1280 \omega
					\big\}
				\right]
				\\ & \mbox{}
				+ \frac{5 s}{486 \csstar^2}
				\left[
					7
					+ 55 \EulerGamma
					+ 32 \csstar^2
					- 256 \omega ( 1 - \csstar^2 )
					+ f_X \big\{
						56
						+ 8 \EulerGamma
						- 320 \omega
					\big\}
				\right] .
		\end{split}
		\label{eq:noller-magueijo}
	\end{equation}
	For $s = - 2\varepsilon$, both~\eqref{eq:noller-magueijo} and
	Noller \& Magueijo's formula (A.15)
	reduce to~\eqref{eq:khoury-piazza}
	evaluated at $k_\star = 3k$.
	For $s \neq - 2\varepsilon$, Eq.~\eqref{eq:noller-magueijo}
	disagrees with Noller \& Magueijo's result.
	This occurs
	partially because they
	approximate the propagator~\eqref{eq:wavefunction}
	in the superhorizon limit
	$|k \cs \tau| \ll 1$
	where details of
	the interference between growing and decaying modes
	around the time of horizon exit
	are lost.
	For example,
	their approximation discards
	the $\Ei$-contributions of
	\eqref{eq:orderderiv}
	although these
	are $\Or(\varpi)$
	and as large as other contributions which are retained.
	But were these terms kept, the
	superhorizon limit $|k \cs \tau| \ll 1$ could not be used to
	estimate them.
	Infrared safety of the $J_i$ integrals in~\eqref{eq:clare-integrals}
	is spoiled if truncated at any finite order,
	causing divergences
	in the squeezed limit $\vartheta_i \rightarrow 0$ and
	a spurious contribution to the bispectrum with a local shape.
	As we explain in Appendices~\ref{appendix:propagator}
	and~\ref{appendix:a},
	it appears that---as a point of principle---%
	if $\varpi \neq 0$
	corrections are kept then the shape of
	the bispectrum can be accurately determined
	only if the full time-dependence
	of each wavefunction around the time of horizon exit is retained.
	
	\section{Conclusions}
	\label{sec:conclusions}
	
	In the near future, we can expect key cosmological observables
	to be determined to high precision.
	For example, the \emph{Planck} satellite may determine the
	scalar spectral index $n_s$ to an accuracy of roughly one part in
	$10^3$ \cite{Colombo:2008ta}.
	In a formerly data-starved science, such precision is startling.
	But it cannot be exploited effectively unless our theoretical
	predictions keep pace.
	
	Almost twenty years ago,
	Stewart \& Lyth
	developed analytic formulae
	for the two-point function
	accurate to next-order in
	the small-parameter $\varepsilon = - \dot{H}/H^2$.
	Subsequent observational developments have restricted attention
	to a region of parameter space where $\varepsilon \ll 1$ is
	a good approximation,
	making the lowest-order prediction
	for the power spectrum an accurate
	match for experiment.
	The same need not be true for three-
	and higher $n$-point correlations,
	where the imminent arrival of data
	is expected to
	improve the observational situation.
	The results of \S\ref{sec:noncanonical} show that
	\emph{Planck}'s observational precision in the equilateral mode
	may be comparable to next-order corrections.
	For a future CMB satellite it is conceivable
	that the data will
	be more precise than a lowest-order estimate.
	In this paper we have reported a next-order calculation of the
	bispectrum in a fairly general class of
	single-field inflationary models:
	those which can be described by a Lagrangian of the form
	$P(X, \phi)$, where $X = g^{ab} \nabla_a \phi \nabla_b \phi$
	is twice the field's kinetic energy.
	Our calculation can be translated to $\fNL$,
	and in many models it provides a much more precise estimate
	than the lowest-order result.
	
	Our major results can be categorized into three groups.
	The first group comprises tests of the accuracy, or improvements in
	the precision, of existing lowest-order calculations.
	The second generate new shapes for the bispectrum at next-order.
	The third involves technical refinements in calculating scalar
	$n$-point functions for $n \geq 3$.
	
	\subsection{Accuracy and precision}
	Except in special cases where exact results are possible---%
	such as the result of Khoury \& Piazza for constant $f_X$
	discussed in \S\ref{sec:noncanonical}---%
	predictions for observable quantities
	come with a ``theory error''
	encapsulating uncertainty due to small
	contributions which were not calculated.
	For inflationary observables the typical scale of the theory error
	is set by the accuracy of the slow-variation approximation,
	where the dimensionless quantities $\varepsilon = - \dot{H} / H^2$,
	$\eta = \dot{\varepsilon} / H \varepsilon$,
	$s = \csdot / H \cs$ (and others)
	are taken to be small.

	\para{Power-law DBI inflation and $k$-inflation}%
	In \S\ref{sec:introduction} and
	\S\S\ref{sec:dbi}--\ref{sec:k-inflation}
	we estimated the \emph{precision}
	which could be ascribed to the lowest-order formula
	for $\fNL$ in the
	absence of a complete calculation of next-order effects.
	To do so one may use any convenient---but hopefully
	representative---subset of next-order terms,
	estimating the remainder to be of comparable magnitude but
	uncertain sign.
	Using the next-order contributions from the vertex coefficients
	$g_i$, which can be obtained without detailed calculation,
	we estimated the fractional uncertainty to be of order $14\varepsilon$
	for DBI and $9\gamma$ for $k$-inflation.
	The next-order terms neglected in
	this estimate come from corrections to the propagator, and
	from the time-dependence of each vertex.
	The prospect of such large uncertainties 
	implies one has no option but to carry out the
	full computation of all next-order terms.
	
	In the power-law DBI and
	$k$-inflation scenarios, we find that the terms omitted from these
	estimates generate large cancellations,
	in each case
	reducing the next-order contribution by
	roughly $95\%$.
	Similar large cancellations were observed by
	Gong \& Stewart in their calculation of
	next-next-order corrections to the power spectrum
	\cite{Gong:2001he}.
	After the fact,
	it seems reasonable to infer that 
	the contributions
	from $g_i$ systematically overpredict the next-order
	terms. But this could not have been deduced
	without a calculation of all next-order effects.
	Therefore,
	for power-law DBI and $k$-inflation models we conclude that
	the lowest-order calculation is surprisingly accurate.
	
	For DBI inflation the status of the power-law solution is
	unclear, being under pressure from both observational and theoretical
	considerations.
	More interest is attached to the generalized case to be discussed
	presently.
	For $k$-inflation, taking present-day constraints on the
	spectral index into account,
	the next-order correction is of order $1\%$.
	Estimating the contribution of
	next-next-order terms using all available contributions
	from our calculation
	gives a fractional uncertainty---measured
	with respect to the lowest-order term---of
	order $22\gamma^2$.
	If similar
	large cancellations occur with terms not included in this estimate,
	the next-order result could be rather more precise
	than this would suggest.%
		\footnote{As suggested in the introduction, these terms could be
		calculated using the next-next-order propagator
		corrections provided by
		Gong \& Stewart \cite{Gong:2001he}.}
	Without knowledge of such cancellations, however, we conclude that
	the uncertainty in the prediction for $\fNL$ has diminished
	to $\sim 250\gamma\% \lesssim 10\%$
	of the uncertainty before calculating next-order terms.
	In the case of power-law DBI inflation,
	the same method yields an estimate of next-next-order corrections
	at $\sim 40 \varepsilon^2$, also measured from the dominant
	lowest-order term in the limit $\gamma \gg 1$.
	This reduces the uncertainty to $\sim 300\varepsilon\% \lesssim
	15\%$ of its prior value.

	\para{Generalized DBI inflation}%
	The situation is different
	for a generalized DBI model with arbitrary potential $V(\phi)$
	and warp factor $f(\phi)$.
	The largest next-order corrections measure a qualitatively new
	effect, not included in the power-law solution, arising from the
	shape of $V$ and $f$.
	The fractional shift was quoted in~\eqref{eq:dbi-shape-corrections}
	and can be large, because the DBI action supports inflation
	on relatively steep potentials.
	Indeed, if one were to tune the potential to be flat
	in the sense $\epsilonSR \sim |\etaSR| \lesssim 10^{-2}$
	then
	much of the motivation
	for a higher-derivative model would have been lost.
	Even for rather large values of $|\fNL|$ the correction can be
	several tens of percent for an ``untuned'' potential with
	$\epsilonSR \sim |\etaSR| \sim 1$.
	For slightly
	smaller $|\fNL|$ the correction is increasingly significant,
	perhaps growing to $\sim 35\%$.
	The formulas quoted in \S\ref{sec:dbi} assume $\gamma \gg 1$
	and would require modification for very small $\fNL$
	where $\Or(\gamma^{-1})$ corrections need not be negligible.
	If desired, these can be obtained from our full formulae
	tabulated in \S\ref{subsec:bispectrum}.

	In a concrete model---the infrared DBI scenario
	proposed by Chen~\cite{Chen:2004gc,Chen:2005ad,Chen:2005fe}---we
	find the correction to be
	$\sim 12\%$ for parameter values currently favoured by observation,
	which translates to
	reasonably large shifts in $\fNL$.
	For the maximum-likelihood mass suggested by
	the analysis of Bean {\etal}, we find next-order corrections
	increase the
	magnitude of $\fNL$ by a shift $|\Delta \fNL| \approx 19$.
	This is a little smaller than the error bar which \emph{Planck}
	is expected to achieve, but nevertheless of comparable
	magnitude.
	We conclude that a next-order calculation will be adequate for
	\emph{Planck}, but if the model is not subsequently ruled out
	a next-next-order calculation may be desirable for
	a
	\emph{CMBPol}-
	or \emph{CoRE}-type satellite.

	For models producing small $\fNL$, such as the ``powerlike''
	Lagrangian discussed in {\S}VI.B of Franche {\etal}
	\cite{Franche:2009gk},
	we find similar conclusions. However, in such models $\fNL$ is unlikely
	to be observable
	and the subleading corrections are of less interest.
	
	\subsection{New bispectrum shapes}
	Because
	our final bispectra capture the shape dependence in the squeezed limit,
	we are able to determine the relationship between the
	lowest-order and next-order shapes,
	discussed in \S\ref{sec:shapes}.

	Working in a $P(X, \phi)$ model, the enhanced part of the lowest-order
	bispectrum is well-known to correlate with the
	equilateral template.
	Only two shapes are available,
	plotted in
	Table~\ref{table:lowest-order-shapes},
	and the bispectrum is a linear
	combination of these.
	The next-order bispectrum is a linear combination of \emph{seven}
	different shapes, although these cannot be varied independently:
	strong correlations among their coefficients are imposed by the
	$P(X,\phi)$ Lagrangian. Many of these shapes also correlate with
	the equilateral mode, but two of them are different:
	in the language of \S\ref{sec:shapes}, these are
	$S'_\varepsilon$ and $S_s$.
	For typical values of $\alphalambda$
	(which is the $\cs^{-2}$-enhanced part of $\lambda / \Sigma$)
	one can obtain
	a linear combination of the next-order shapes
	orthogonal to both lowest-order shapes.
	This is the orthogonal shape $O$ appearing in
	Table~\ref{table:orthogonal-plots}.
	
	This shape represents a distinctive prediction of the
	next-order theory, which cannot be reproduced at lowest-order.
	It is very similar to a new shape constructed by
	Creminelli {\etal} \cite{Creminelli:2010qf},
	working in an
	entirely different model---a Galilean-shift invariant Lagrangian
	with at least two derivatives applied to each field.
	We conclude that even a clear detection of a bispectrum with this
	shape will not be sufficient, on its own,
	to distinguish between models of $P(X, \phi)$- and Galileon-type.
	In practice, a network of interlocking predictions is likely to be
	required. For example, this shape cannot be \emph{dominant}
	in a $P(X, \phi)$ model. It must be accompanied by
	a mixture of $S_1$- and $S_2$-shapes of larger amplitude.
	
	The orthogonal
	shape is not guaranteed to be present at an observable
	level in every model,
	even in models where the next-order corrections are detectable.
	This may happen
	because $\alphalambda$ takes a value for
	which no bispectrum orthogonal to $S_1$ and $S_2$ can be
	generated perturbatively, or because
	the slow-variation parameters accidentally conspire to suppress
	its amplitude.
	Nevertheless, it will be present in many models.
	If detected, it would play a role
	similar to the
	well-known consistency relation between $r$ and $n_t$.
	Searching for the $O$-shape in real data is likely to require
	a dedicated template, and until this is constructed it is not
	possible to estimate the 
	signal-to-noise and therefore the amplitude
	detectable by \emph{Planck} or a subsequent experiment.
	It would be interesting to determine the precise amplitude
	to be expected in motivated models, but we
	have not attempted such an estimate in this paper.

	\subsection{Technical results}
	Our calculation includes a number of more technical results.
		
	\para{Treatment of boundary terms}%
	In \S\ref{sec:third-order-action} we gave a systematic treatment
	of boundary terms in the third-order action.
	Although these terms were properly accounted for in previous
	results~\cite{Maldacena:2002vr,Seery:2005wm,Chen:2006nt},
	these calculations used a field redefinition which was not
	\emph{guaranteed}
	to remove all terms in the boundary action.
	
	\para{Pure shape logarithms}%
	The subleading correction to the propagator contains
	an exponential integral
	contribution [of the form $\Ei(x)$]
	whose time dependence cannot be described by
	elementary functions
	\cite{Gong:2001he,Chen:2006nt,Burrage:2010cu}.
	This term must be handled carefully to avoid unphysical
	infrared divergences
	in the squeezed limit where one momentum goes to zero.
	In Appendix~\ref{appendix:a} we describe how this
	contribution yields
	the $J_i$ functions given in Eq.~\eqref{eq:clare-integrals}.
	These are
	obtained using a resummation and analytic continuation technique
	introduced in Ref.~\cite{Burrage:2010cu}.
	The possibility of spurious divergences in the squeezed limit
	shows that, as a matter of principle,
	one should be cautious when determining the shape of the bispectrum
	generated by
	an approximation to the elementary wavefunctions.
	
	In the present case, the $J_i$ contain `pure' shape logarithms
	which are important in describing factorization of the
	three-point function in the limit $k_i \rightarrow 0$.
	Obtaining the quantitatively correct
	\emph{momentum} behaviour requires all
	details of the interference in
	\emph{time} between growing and decaying modes
	near horizon exit.
	The possibility of such interference effects,
	absent in classical mechanics,
	is a typical feature of
	quantum mechanical processes.
	This interference correctly resolves
	the unwanted divergences in the opposing infrared limit
	$k_j \rightarrow 0$ with $j \neq i$.
	We discuss these issues in more detail in
	Appendices~\ref{appendix:propagator} and~\ref{appendix:a}.
	
	\para{Comparison with known results}%
	We have verified Maldacena's consistency
	condition~\eqref{eq:maldacena-condition} to next-order
	in canonical models
	by explicit calculation of the full bispectrum. This agrees with a recent
	calculation by Renaux-Petel \cite{RenauxPetel:2010ty}.
	In the case of power-law inflation with $\varpi = 0$
	and constant $\varepsilon$ and $s$, we reproduce a known result
	due to Khoury \& Piazza
	\cite{Khoury:2008wj}.
	
	A subset of our corrections were computed by Chen {\etal}
	\cite{Chen:2006nt}. Up to first-order terms in
	powers of slow-roll
	quantities---where our results can be
	compared---we find exact agreement.
	
	\begin{center}
		\emph{Note added}
	\end{center}
	Immediately prior to completion of this paper, a preprint
	by Arroja \& Tanaka appeared \cite{Arroja:2011yj}
	which appears to present
	arguments regarding the role of boundary terms which are
	equivalent to those of \S\ref{subsec:boundaryterms}.
	
	\acknowledgments
	We would like to thank Peter Adshead, Xingang Chen,
	Anne-Christine Davis,
	Eugene Lim,
	Donough Regan
	and S\'{e}bastien Renaux-Petel for helpful discussions.
	We appreciate correspondence with Chris Byrnes, 
	Adam Christopherson, Andrew Liddle, Karim Malik,
	Jo\~{a}o Magueijo, Johannes Noller
	and
	Federico Piazza.
	CB is supported by the SNF,
	and acknowledges hospitality from the University of Sussex.
	RHR is supported by FCT through the grant SFRH/BD/35984/2007.
	DS was supported by the Science and Technology Facilities Council
	[grant number ST/F002858/1],
	and acknowledges hospitality from
	the Deutsches Elektronen-Synchrotron DESY
	during the early stages of this work.

	\clearpage

	\appendix

	\section{Propagator corrections}
	\label{appendix:propagator}
	
	At lowest-order, the wavefunctions can be obtained from 
	Eq.~\eqref{eq:wavefunction} by setting all
	slow-variation parameters 
	to zero. Choosing a reference
	scale $k_\star$, this yields the standard result
	\begin{equation}
		u_k(\tau)
			= \im
				\dfrac{H_{\star}}{2\sqrt{z_{\star}}}
				\dfrac{1}{(k \csstar)^{3/2}}
				(1-\im k\csstar\tau) e^{\im k \csstar\tau} .
		\label{eq:wavefunction_background}
	\end{equation}
	
	Next-order corrections to the propagator were
	discussed by Stewart \& Lyth \cite{Stewart:1993bc},
	who quoted their result in terms of special functions
	and expanded uniformly to next-order after taking a late-time limit.
	The uniform next-order expansion at a generic time was given by
	Gong \& Stewart~\cite{Gong:2001he} for canonical models
	and by Chen~{\etal}~\cite{Chen:2006nt} in the noncanonical case.
	Their result was cast in a more convenient
	form in Ref.~\cite{Burrage:2010cu} whose argument we briefly review. 

	The next-order correction is obtained
	after systematic expansion of each quantity in 
	\eqref{eq:wavefunction} to linear order in the slow-variation parameters.
	Contributions arise from each time-dependent factor
	and from the order of the Hankel function.
	Collecting the formulae quoted in Ref.~\cite{Chen:2006nt},
	one finds
	\begin{equation}
		\left.
		\frac{\partial H^{(2)}_{\nu}(x)}{\partial\nu}
		\right|_{\nu = \frac{3}{2}}
			=
			- \dfrac{\im}{x^{3/2}} \sqrt{\dfrac{2}{\pi}}
			\left[
				\e{\im x} (1-\im x) \Ei(-2 \im x)
				- 2 \e{-\im x}
				- \im \dfrac{\pi}{2} \e{-\im x} (1+\im x)
			\right] ,
		\label{eq:orderderiv}
	\end{equation}
	where $\Ei(x)$ is the exponential integral,
	\begin{equation}
		\Ei(x)
			=
			\int_{-\infty}^{x} \dfrac{\e{t}}{t} \d t
		\qquad
		\text{if $x \in \reals$}.
		\label{eq:exponential-integral}
	\end{equation} 
	This is well-defined for $x<0$.
	For $x\geq 0$
	it should  be understood as a Cauchy principal value.
	For complex argument---required by~\eqref{eq:orderderiv}---%
	one promotes Eq.~\eqref{eq:exponential-integral}
	to a contour integral,
	taken on a path running between $t = x \in \complex$
	and $|t| \rightarrow \infty$ in the half-plane $\Re(t) < 0$.
	Using Cauchy's theorem to rotate this contour onto the
	negative real $t$-axis, one finds
	\begin{equation}
		\Ei(2\im k \cs \tau)
			=
			\lim_{\epsilon \rightarrow 0}
			\int_{-\infty(1+\im \epsilon)}^{\tau}
				\dfrac{\d\xi}{\xi} \; \e{2\im k \cs \xi} .
	\end{equation}
	The next-order correction to~\eqref{eq:wavefunction_background},
	expanded uniformly to $\Or(\varepsilon)$ in slow-variation
	parameters but including the \emph{exact} time-dependence, is
	\begin{equation}
		\begin{split}
			\delta u(k, \tau)
				= \mbox{} &
				\dfrac{\im H_{\star}}{2\sqrt{z_{\star}} (k \csstar)^{3/2}}
				\Bigg\{
					- \varpi_{\star} \e{-\im k \csstar \tau}
						(1+\im k \csstar \tau)
						\int_{-\infty(1+\im \epsilon)}^{\tau}
							\dfrac{\d\xi}{\xi} \;
							\e{2\im k \csstar \xi}
					\\ & \mbox{}
					+ \e{\im k \csstar \tau}
						\Big[
							\mu_{0 \star}
							+ \im \mu_{1 \star} k \csstar \tau
							+ s_{\star} k^2 \csstar^2 \tau^2
							+ \Delta N_{\star} \left(
								\varpi_{\star}
								- \im \varpi_{\star} k \csstar \tau
								- s_{\star} k^2 \csstar^2 \tau^2
							\right)
						\Big]
				\Bigg\} .
		\end{split}
		\label{eq:varyingu}
	\end{equation}
	where
	\begin{equation}
		\mu_0 \equiv \varepsilon +v +2s +\im\dfrac{\pi}{2}\varpi
		\quad
		\textrm{and}
		\quad
		\mu_1 \equiv \varepsilon +s -\im\dfrac{\pi}{2}\varpi ,
	\end{equation}
	and quantities labelled `$\star$' are evaluated at
	the horizon-crossing time for the reference scale $k_\star$.
	We have used $\Delta N_{\star}$ to denote the number of e-folds
	which have elapsed since this time, so $\Delta N_\star =
	\ln |k_\star \csstar \tau|$.
	The limit $\epsilon \rightarrow 0$ should be understood
	after the integration has been performed
	and merely guarantees convergence
	as $\xi \rightarrow -\infty$.

	Setting the integral aside,
	Eq.~\eqref{eq:varyingu} has the appearance of an expansion in
	powers of $k\tau$. In fact, the time dependence is exact.
	Although terms of high-order in $k\tau$ become increasingly
	irrelevant as $|k\tau| \rightarrow \infty$,
	all such terms make comparable contributions to a generic $n$-point
	function at the time of horizon crossing, where $|k\tau| \sim 1$,
	and should not be discarded.
	This reflects interference
	effects between the growing and decaying modes
	at horizon exit.
	Indeed, a high-order term such as $(k\tau)^n$
	typically generates a contribution $\sim n! (k / k_t)^n$.
	Such terms are suppressed by the factor $(k/k_t)^n$ which typically
	varies between $0$ and $1/2$, but are enhanced by the rapidly
	growing factorial.
	Therefore their contribution to the shape
	dependence must usually be retained.
	An infinite series of such terms may converge only for certain
	ratios $k/k_t$, requiring a continuation technique to obtain the
	momentum dependence---and hence the shape of the bispectrum---for
	arbitrary $k_i$. We will see an explicit example
	in Appendix~\ref{appendix:a} below.

	To evaluate the cubic action~\eqref{eq:s3tau} we will require
	the time derivative $\delta u'$,
	where $'$ denotes a derivative with respect to conformal time.
	Using the identity $\mu_0+\mu_1-2\varpi_1=0$, it follows that
	\begin{equation}
		\begin{split}
			\delta u'(k, \tau)
				= \mbox{} &
				\dfrac{\im H_{\star} (k \csstar)^{1/2}}
					{2\sqrt{z_{\star}}}
				\tau
				\Bigg\{
					- \varpi_{1 \star} \e{-\im k \csstar \tau}
					\int_{-\infty(1+\im\epsilon)}^{\tau}
						\dfrac{\d\xi}{\xi} \; \e{2\im k \csstar \xi}
				\\ & \hspace{1cm} \mbox{}
				+ \e{\im k \csstar \tau}
					\Big[
						s_{\star}
						- \mu_{1\star}
						+ \im s_{\star} k \csstar \tau
						+ \Delta N_{\star} \left(
							\varpi_{1\star}
							- 2s_{\star}
							- \im s_{\star} k \csstar \tau
						\right)
					\Big]
				\Bigg\} .
		\end{split}
		\label{eq:varyinguprime}
	\end{equation} 
	
	In order that the power spectrum remains conserved
	on superhorizon scales,
	one should verify that $\delta u$ approaches a constant and
	$\delta u' \rightarrow 0$ as $|k\tau| \rightarrow 0$.
	The explicitly time-dependent term $\Delta N_\star$
	apparently spoils the required behaviour, but
	is compensated by a logarithmic divergence from the integral.
	Indeed, for $\tau \rightarrow 0$ one finds
	\begin{equation}
		\int_{-\infty(1+\im\epsilon)}^{\tau}
			\dfrac{\d\xi}{\xi} \; \e{2\im k \csstar \xi}
		= \ln |2\im k \csstar \tau| + \Or( k \csstar \tau ) ,
	\end{equation}
	which precisely cancels the time dependence arising from $\Delta N_\star$.
	Note the incomplete cancellation of the logarithm, which
	leaves a residual of the form $\varpi_\star \ln 2$
	and is the origin of the $\ln 2$ term in the Stewart--Lyth constant
	$C = -2 + \ln 2 + \EulerGamma$ \cite{Stewart:1993bc,Lidsey:1995np}.
	This is a primitive form of the incomplete cancellation which leads
	to residual $\ln k_i/k_t$ and $\ln k_t / k_\star$ terms
	after cancellation of $\ln \tau$ logarithms in the three-point function.
	
	\section{Integrals involving the exponential integral $\Ei(z)$}
	\label{appendix:a}

	The principal obstruction to evaluation of the next-order
	corrections using standard methods is the $\Ei$-term
	in~\eqref{eq:varyingu}.
	This can not be expressed directly in terms of elementary functions
	whose integrals can be computed in closed form.
	It was explained in Appendix~\ref{appendix:propagator}
	that although $\Ei(2\im k \cs \tau)$
	contains terms of high orders in
	$k\tau$ which become increasingly irrelevant as $|k\tau| \rightarrow 0$,
	these cannot usually be neglected when computing
	$n$-point functions.
	A term of order $(k\tau)^n$ generates a contribution
	$\sim n! (k/k_t)^n$ and
	as part of an infinite series will converge at best slowly for
	certain values of $k/k_t$---typically,
	when the scale $k$ itself is squeezed to zero.
	The ratio $k/k_t$ is maximized when $k$ is left fixed but
	one of the \emph{other} momenta is squeezed.
	To obtain a quantitative description of this limit
	may require continuation.
	Hence,
	in order to understand the shape generated by arbitrary $k_i$,
	it follows that the entire $k\tau$-dependence must be retained.
	It was for this reason that we were careful to expand uniformly
	only in slow-roll parameters in Appendix~\ref{appendix:propagator},
	while keeping all time-dependent terms.

	The integrals we require are of the form
	\cite{Burrage:2010cu}
	\begin{equation}
		I_m(k_3)
			=
			\int_{-\infty}^{\tau}
				\d\zeta	\; \zeta^m \e{\im (k_1+k_2-k_3)\cs \zeta}
				\int_{-\infty}^{\zeta}
					\dfrac{\d\xi}{\xi} \; \e{2\im k_3 \cs \xi} .
		\label{eq:iminitial}
	\end{equation}
	We are again using the convention that,
	although $I_m$ depends on all three $k_i$, only the
	asymmetric momentum is written as an explicit argument.
	In the following discussion we specialize to $I_0$.
	Comparable results for arbitrary $m$ can be found by
	straightforward modifications
	and are given in Ref.~\cite{Burrage:2010cu}.

	We introduce the dimensionless combinations
	\begin{equation}
		\vartheta_3
			=
			1
			- \dfrac{2k_3}{k_t}
		\quad
		\text{and}
		\quad
		\theta_3
			=
			\dfrac{k_3}{k_t-2k_3}
			=
			\frac{1-\vartheta_3}{2 \vartheta_3} .
		\label{eq:theta-def}
	\end{equation}
	After contour rotation and some algebraic simplification,
	we can express $I_0$ in terms of
	$\vartheta_3$ and $\theta_3$ \cite{Burrage:2010cu}
	\begin{equation}
		I_0(k_3)
			=
			- \dfrac{\im}{\vartheta_3 k_t \cs}
				\int_{0}^{\infty}{\d u \; \e{-u}}
				\int_{\infty}^{\theta_3 u}{\dfrac{\d v}{v} \; \e{-2v}} .
		\label{eq:i02}
	\end{equation}
	The $v$-integral has a
	Puiseaux series representation
	\begin{equation}
		\int_{\infty}^{\theta_3 u}
			\dfrac{\d v}{v} \e{-2v}
		=
		\EulerGamma
		+ \ln(2\theta_3 u)
		+ \sum_{n=1}^{\infty}
			\dfrac{(-2\theta_3 u)^n}{n! n}
		\label{eq:puiseaux}
	\end{equation}
	where the sum converges uniformly for all complex $\theta_3 u$.
	The right-hand side has a singularity at $\theta_3 u = 0$,
	and the logarithm generates
	a branch cut along $|\arg (\theta_3 u)| = \pi$.
	The $u$-independent term $\EulerGamma$ is the Euler--Mascheroni
	constant
	$\EulerGamma \equiv \int_{\infty}^{0}{\e{-x}\ln x \, \d x}$
	and is obtained
	by expanding $\e{-2v}$ in series, integrating term-by-term,
	and matching the undetermined constant of integration
	with the left-hand side of~\eqref{eq:puiseaux}
	in the limit $u \rightarrow 0$.

	Because~\eqref{eq:puiseaux} is uniformly convergent we may
	exchange integration and summation,
	evaluating $I_0$ using term-by-term integration.
	We find \cite{Burrage:2010cu}
	\begin{equation}
		I_0(k_3)
			=
			-\dfrac{\im}{\vartheta_3 k_t \cs}
			\left[
				\ln (2\theta_3) +
				\sum_{n=1}^{\infty}
					\frac{(-2\theta_3)^n}{n}
			\right] .
		\label{eq:i03}
	\end{equation}
	This is singular when $\vartheta_3 \rightarrow 0$,
	which corresponds to the squeezed limit where
	either $k_1$ or $k_2 \rightarrow 0$.
	There is nothing unphysical about this arrangement of momenta,
	for which $\theta_3 \rightarrow \infty$,
	but we will encounter
	a power-law divergence unless the bracket $[ \cdots ]$
	vanishes sufficiently rapidly in the same limit.

	The sum converges absolutely if 
	$-1/2 < \Re(\theta_3) <1/2$,
	corresponding to the narrow physical region $0 < k_3 < k_t / 4$
	where $k_3$ is being squeezed.
	For $\theta_3$ satisfying this
	condition the summation can be performed explicitly,
	yielding
	$\ln (1+2\theta_3)^{-1} = \ln \vartheta_3$.
	It can be verified that $I_0(k_3)$ is analytic
	in the half-plane $\Re(\theta_3) > 0$.
	Therefore it is possible to
	analytically continue to $\theta_3 \geq 1/2$,
	allowing the behaviour as $\theta_3 \rightarrow \infty$
	and $\vartheta_3 \rightarrow 0$
	to be studied.
	We find
	\begin{equation}
		I_0(k_3)
			=
			- \frac{\im}{\vartheta_3 k_t \cs} \ln ( 1 - \vartheta_3 )
			\equiv
			- \frac{\im}{k_t \cs} J_0(k_3) ,
		\label{eq:I0final}
	\end{equation}
	where we have introduced a function $J_0$, which is related to
	$I_0$ by a simple change of normalization.
	It is more convenient to express our final bispectra in terms of
	$J_0$ rather than $I_0$.
	In the limit $\vartheta_3 \rightarrow 0$, Eq.~\eqref{eq:I0final}
	is finite with $J_0 \rightarrow -1$.
	The $\theta_3 \rightarrow \infty$ behaviour of the sum has
	subtracted both the power-law divergence of the prefactor
	and the logarithmic divergence of $\ln (2 \theta_3)$.
	To obtain this conclusion required information about the
	entire momentum dependence of~\eqref{eq:iminitial}
	in order to perform a meaningful analytic continuation.%
		\footnote{A precisely analogous situation occurs at finite
		temperature, where one first computes the imaginary time
		Green's function $G(\im \omega_n)$
		on a discrete spectrum of Matsubara frequencies $\{ \omega_n \}$.
		To obtain the real-time Green's function one must
		extend $G$ into the complex plane
		(in the sense that
		one finds a function $G(z)$, where $z \in \mathbb{C}$, which
		coincides with the imaginary time Green's function when
		$z = \im \omega_n$),
		and analytically continue to the real axis.
		Unless one has knowledge of $G$ on the complete set of
		Matsubara frequencies---which may be impossible
		in an effective field theory---it is difficult
		to extract meaningful results
		for the real-space correlation function.
		In the present case,
		the requirement to know the series expansion
		in~\eqref{eq:puiseaux} or~\eqref{eq:i03} for all $n$
		is analogous to the requirement
		to know $G(\im \omega_n)$ for all
		Matsubara frequencies.\label{footnote:matsubara}}
	Had we truncated the $k_3 \zeta$ dependence of~\eqref{eq:iminitial},
	or equivalently the $\theta_3 u$ dependence of~\eqref{eq:puiseaux},
	we would have encountered a spurious divergence in the squeezed
	limit and a misleading prediction of a strong signal in the local
	mode. Several of our predictions, including
	recovery of the Maldacena limit
	described by~\eqref{eq:fnlsvanilla},
	depend on the precise numerical value of $J_0$
	as $\theta_3 \rightarrow \infty$ and therefore constitute tests
	of this procedure.

	We can now evaluate $I_m(k_3)$ for arbitrary $m$,
	although only the cases $m \leq 2$ are required for the calculation
	presented in the main text.
	The necessary expressions are
	\begin{subequations}
		\begin{align}
			I_1(k_3)
				& =
				\dfrac{1}{(\vartheta_3 k_t \cs)^2}
				\left[
					\vartheta_3
					+ \ln(1-\vartheta_3)
				\right]
				\equiv
				\dfrac{1}{(k_t \cs)^2} J_1(k_3)
			\\
			I_2(k_3)
				& =
				\dfrac{\im}{(\vartheta_3 k_t \cs)^3}
				\left[
					\vartheta_3 (2+\vartheta_3)
					+ 2\ln(1-\vartheta_3)
				\right]
				\equiv
				\dfrac{\im}{(k_t \cs)^3} J_2(k_3) .
		\end{align}
	\end{subequations}
	In the limit $\vartheta_3 \rightarrow 0$ one can verify
	that $J_1 \rightarrow -1/2$ and $J_2 \rightarrow -2/3$.

	\section{Useful integrals}
	\label{appendix:b}

	To simplify evaluation of the various integrals which
	arise in computing the bispectrum,
	it is helpful to have available a formula for a master integral,
	$\mathcal{J}$,
	for which the integrals of interest are special cases
	\cite{Burrage:2010cu}.
	We define
	\begin{equation}
		\begin{split}
			\mathcal{J}_{\star}
			=
			\im \int_{-\infty}^{\tau} \d\xi \; \e{\im k_t \csstar\xi}
			\bigg[
				&
				\gamma_0
				+ \im \gamma_1 \csstar \xi
				+ \gamma_2 \csstar^2 \xi^2
				+ \im\gamma_3 \csstar^3 \xi^3
				+ \gamma_4 \csstar^4 \xi^4
			\\ & \mbox{}
				+ N_\star \left(
					\delta_0
					+ \im \delta_1 \csstar \xi
					+ \delta_2 \csstar^2 \xi^2
					+ \im \delta_3 \csstar^3 \xi^3
					+ \delta_4 \csstar^4 \xi^4
				\right)
			\bigg]
		\end{split}
		\label{eq:masterj}
	\end{equation}
	where $N_{\star}$
	represents the number of e-folds elapsed since horizon exit
	of the reference mode $k_\star$, giving
	$N_\star = \ln | k_\star \csstar \xi |$.
	Eq.~\eqref{eq:masterj} is an oscillatory integral,
	for which asymptotic techniques are well-developed,
	and which can be evaluated using repeated integration by parts.
	To evaluate the $N_\star$-dependent terms one requires the
	standard integral
	\begin{equation}
		\lim_{\tau \rightarrow 0} \int_{-\infty}^{\tau} \d\xi
		\; N_{\star} \, \e{\im k_t \csstar} 
		=
		\dfrac{\im}{k_t \csstar}
		\left(
			\EulerGamma +
			\im \dfrac{\pi}{2}
		\right) ,
	\end{equation}
	which can be obtained by contour rotation
	and use of the Euler--Mascheroni constant
	defined below Eq.~\eqref{eq:puiseaux}.
	We conclude
	\begin{equation}
		\begin{split}
			\mathcal{J}_\star
			=
			\frac{1}{k_t \csstar}
			\bigg[
				&
				\gamma_0
				- \frac{\gamma_1 + \delta_1}{k_t}
				- \dfrac{2\gamma_2+3\delta_2}{k_t^2}
				+ \dfrac{6\gamma_3+11\delta_3}{k_t^3}
				+ \dfrac{24\gamma_4+50\delta_4}{k_t^4}
				\\
				& \mbox{}
				- \left(
					\EulerGamma
					+ \ln \frac{k_t}{k_\star}
					+ \im \dfrac{\pi}{2}
				\right)
				\left(
					\delta_0
					- \dfrac{\delta_1}{k_t}
					- 2 \dfrac{\delta_2}{k_t^2}
					+ 6 \dfrac{\delta_3}{k_t^3}
					+ 24\dfrac{\delta_4}{k_t^4}
				\right)
			\bigg]\;.
		\end{split}
		\label{eq:masterjresult}
	\end{equation} 
	We use this result repeatedly in \S\ref{subsec:3pf}
	to evaluate integrals in closed form.
	
	\section{The special functions $R$ and $Q$}
	\label{appendix:compare}
	
	For completeness, we give explicit formulae for the functions
	$Q(k_1, k_2, k_3)$ and $R(k_1, k_2, k_3)$ which were
	used to express a subset
	of the slow-roll corrections in Ref.~\cite{Chen:2006nt}.
	These functions can be written in closed form
	using the integrals
	$J_0$, $J_1$ and $J_2$ defined in Appendix~\ref{appendix:a}.
	To do so, we introduce a quantity
	$\alpha(k_1)$ which satisfies
	\begin{equation}
		\alpha(k_1) \equiv \frac{k_t}{k_1} - 1 .
	\end{equation}
	We are adopting our usual convention in which $\alpha(k_1)$ is a function
	of all $k_i$, but only the asymmetrically occurring momentum is written
	explicitly.
	Also, in this Appendix, the Greek symbols
	$\alpha$, $\beta$ and $\gamma$ are used independently of their
	meaning elsewhere in this paper.
	The combination $\vartheta_1$ can be written
	[\emph{cf.}~\eqref{eq:theta-def}]
	\begin{equation}
		\vartheta_1 = \frac{\alpha(k_1) - 1}{\alpha(k_1) + 1} .
	\end{equation}
	The definitions of $R$ and $Q$ can be found in Eqs.~(B.84)--(B.86)
	of Ref.~\cite{Chen:2006nt}.
	We find
	\begin{equation}
		R = \frac{k_2^2 k_3^2}{k_1}
			\left[
				\frac{2}{[1+\alpha(k_1)]^2}
				-
				\frac{1}{1+\alpha(k_1)}
				\left(
					J_0(k_1) - J_1(k_1)
					+
					\frac{\alpha(k_1) J_2(k_1)}{[1+\alpha(k_1)]^2}
				\right)
			\right]
			+ \text{sym.},
		\label{eq:xingang-r}
	\end{equation}
	where `sym.' denotes the symmetric sum formed by adding the two
	distinct combinations $k_1 \rightarrow k_2$ and $k_1 \rightarrow
	k_3$.
	Eq.~\eqref{eq:xingang-r} is accurate up to terms which vanish
	like powers of $|k \tau|$, and are therefore negligible a few
	e-folds after horizon crossing.
	With the same understanding, we find that $Q$ can be written
	\begin{equation}
	\begin{split}
		Q = \mbox{} & \frac{\beta_0(k_1)}{2}
			\big(
				4 + 2 [ \alpha(k_1) - 1][\EulerGamma + \ln 2]
			\big)
			+
			(k_2^3 + k_3^3)
			\left(
				\ln \frac{k_t}{k_1}
				+
				\EulerGamma
			\right)
			-
			\gamma_1(k_1)
			+
			\frac{\gamma_2(k_1)}{1 + \alpha(k_1)}
			\\
			& \mbox{}
			-
			\delta_0(k_1) \big(
				\ln [ 1 + \alpha(k_1) ]
				+
				\EulerGamma
				-
				1
			\big)
			+
			\frac{1}{1+\alpha(k_1)}
			\left(
				\beta_2(k_1) J_0(k_1)
				-
				\frac{\beta_3(k_1) J_1(k_1)}{1+\alpha(k_1)}
			\right)
			+
			\text{sym.}
	\end{split}
	\label{eq:xingang-q}
	\end{equation}
	The functions appearing here are
	\begin{align}
		\beta_0(k_1) & \equiv
			- \frac{k_1}{6} \sum_i k_i^2 \\
		\beta_2(k_1) & \equiv
			k_1 ( k_2^2 + k_3^2 )
			+
			\frac{k_2^2 k_3^2}{k_1}
			+
			\frac{k_1}{6} \left(
				\frac{k_2 k_3}{k_1^2} - \alpha(k_1)
			\right)
			\sum_i k_i^2 \\
		\beta_3(k_1) & \equiv
			- k_1 k_2 k_3 \alpha(k_1)
			+
			\frac{k_2^2 k_3^2}{k_1}
			+
			\frac{k_2 k_3}{6 k_1}
			\sum_i k_i^2 \\
		\gamma_0(k_1) & \equiv
			k_1 ( k_2^2 + k_3^2 )
			+
			\frac{k_1}{3}
			\sum_i k_i^2 \\
		\gamma_1(k_1) & \equiv
			- k_1 ( k_2^2 + k_3^2 )
			-
			k_1 k_2 k_3 \alpha(k_1)
			-
			\frac{k_1 \alpha(k_1)}{3} \sum_i k_i^2 \\
		\gamma_2(k_1) & \equiv
			- k_1 k_2 k_3 \alpha(k_1)
			-
			2 \frac{k_2^2 k_3^2}{k_1}
			-
			\frac{1}{3} \frac{k_2 k_3}{k_1} \sum_i k_i^2 \\
		\delta_0(k_1) & \equiv
			[\gamma_0(k_1) + \beta_0(k_1)][ 1 + \alpha(k_1) ] + \gamma_1(k_1) .
	\end{align}

	\bibliographystyle{JHEPmodplain}
	\bibliography{paper}

\end{document}